\documentclass[12pt]{article}
\usepackage{amssymb,amscd,array}
\catcode `\@=11
\@addtoreset{equation}{section}

\newtheorem{thm}{Theorem}[section]
\newtheorem{rem}[thm]{Remark}

\newtheorem{prop}[thm]{Proposition}
\newtheorem{cor}[thm]{Corollary}

\def\qed{\blacksquare}
\newcommand{\be}{\begin{equation}}
\newcommand{\ee}{\end{equation}}
\newcommand{\bea}{\begin{eqnarray}}
\newcommand{\eea}{\end{eqnarray}}
\newcommand{\R}{\mathbb{R}}
\newcommand{\N}{\mathbb{N}}
\newcommand{\C}{\mathbb{C}}

\begin{document}
\begin{titlepage}

\begin{center}
{\bf \Large{Cohomological Aspects of Gauge Invariance in the Causal Approach\\}}
\end{center}
\vskip 1.0truecm
\centerline{D. R. Grigore, 
\footnote{e-mail: grigore@theory.nipne.ro}}
\vskip5mm
\centerline{Department of Theoretical Physics, Institute for Physics and Nuclear
Engineering ``Horia Hulubei"}
\centerline{Institute of Atomic Physics}
\centerline{Bucharest-M\u agurele, P. O. Box MG 6, ROM\^ANIA}

\vskip 2cm
\bigskip \nopagebreak
\begin{abstract}
\noindent
Quantum theory of the gauge models in the causal approach leads to some cohomology problems. We investigate these problems in detail.
\end{abstract}
\end{titlepage}

\section{Introduction}

The general framework of perturbation theory consists in the construction of 
the chronological products such that Bogoliubov axioms are verified \cite{BS}, \cite{EG}, \cite{DF}, \cite{ano}; for every set of Wick monomials 
$ 
W_{1}(x_{1}),\dots,W_{n}(x_{n}) 
$
acting in some Fock space
$
{\cal H}
$
one associates the operator
$ 
T^{W_{1},\dots,W_{n}}(x_{1},\dots,x_{n}); 
$  
all these expressions are in fact distribution-valued operators called chronological products. It will be convenient to use another notation: 
$ 
T(W_{1}(x_{1}),\dots,W_{n}(x_{n})). 
$ 
The construction of the chronological products can be done recursively according to Epstein-Glaser prescription \cite{EG}, \cite{Gl} (which reduces the induction procedure to a distribution splitting of some distributions with causal support) or according to Stora prescription \cite{PS} (which reduces the renormalization procedure to the process of extension of distributions). 
These products are not uniquely defined but there are some natural limitation on this arbitrariness. If the arbitrariness does not grow with $n$ we have a renormalizable theory.
An equivalent point of view uses retarded products \cite{St1}.

Gauge theories describe particles of higher spin. Usually such theories are not renormalizable. However, one can save renormalizablility using ghost fields. Such theories are defined in a Fock space
$
{\cal H}
$
with indefinite metric, generated by physical and un-physical fields (called {\it ghost fields}). One selects the physical states assuming the existence of an operator $Q$ called {\it gauge charge} which verifies
$
Q^{2} = 0
$
and such that the {\it physical Hilbert space} is by definition
$
{\cal H}_{\rm phys} \equiv Ker(Q)/Im(Q).
$
The space
$
{\cal H}
$
is endowed with a grading (usually called {\it ghost number}) and by construction the gauge charge is raising the ghost number of a state. Moreover, the space of Wick monomials in
$
{\cal H}
$
is also endowed with a grading which follows by assigning a ghost number to every one of the free fields generating
$
{\cal H}.
$
The graded commutator
$
d_{Q}
$
of the gauge charge with any operator $A$ of fixed ghost number
\be
d_{Q}A = [Q,A]
\ee
is raising the ghost number by a unit. It means that
$
d_{Q}
$
is a co-chain operator in the space of Wick polynomials. From now on
$
[\cdot,\cdot]
$
denotes the graded commutator.
 
A gauge theory assumes also that there exists a Wick polynomial of null ghost number
$
T(x)
$
called {\it the interaction Lagrangian} such that
\be
~[Q, T] = i \partial_{\mu}T^{\mu}
\ee
for some other Wick polynomials
$
T^{\mu}.
$
This relation means that the expression $T$ leaves invariant the physical states, at least in the adiabatic limit. In all known models one finds out that there exist a chain of Wick polynomials
$
T^{\mu},~T^{\mu\nu},~T^{\mu\nu\rho},\dots
$
such that:
\be
~[Q, T] = i \partial_{\mu}T^{\mu}, \quad
[Q, T^{\mu}] = i \partial_{\nu}T^{\mu\nu}, \quad
[Q, T^{\mu\nu}] = i \partial_{\rho}T^{\mu\nu\rho},\dots
\label{descent}
\ee
It so happens that for all these models the expressions
$
T^{\mu\nu},~T^{\mu\nu\rho},\dots
$
are completely antisymmetric in all indices; it follows that the chain of relation stops at the step $4$ (if we work in four dimensions). We can also use a compact notation
$
T^{I}
$
where $I$ is a collection of indices
$
I = \{\nu_{1},\dots,\nu_{p}\}~(p = 0,1,\dots,);
$
all these polynomials have the same canonical dimension
\be
\omega(T^{I}) = \omega_{0},~\forall I
\ee
and because the ghost number of
$
T \equiv T^{\emptyset}
$
is null, then we also have:
\be
gh(T^{I}) = |I|.
\ee
One can write compactly the relations (\ref{descent}) as follows:
\be
d_{Q}T^{I} = i~\partial_{\mu}T^{I\mu}.
\label{descent1}
\ee

For concrete models the equations (\ref{descent}) can stop earlier: for 
instance in the Yang-Mills case we have
$
T^{\mu\nu\rho} = 0
$
and in the case of gravity
$
T^{\mu\nu\rho\sigma} = 0.
$

Now we can construct the chronological products
$$
T^{I_{1},\dots,I_{n}}(x_{1},\dots,x_{n}) \equiv T(T^{I_{1}}(x_{1}),\dots,T^{I_{n}}(x_{n}))
$$
according to the recursive procedure. We say that the theory is gauge invariant in all orders of the perturbation theory if the following set of identities generalizing (\ref{descent1}):
\be
d_{Q}T^{I_{1},\dots,I_{n}} = 
i \sum_{l=1}^{n} (-1)^{s_{l}} {\partial\over \partial x^{\mu}_{l}}
T^{I_{1},\dots,I_{l}\mu,\dots,I_{n}}
\label{gauge}
\ee
are true for all 
$n \in \N$
and all
$
I_{1}, \dots, I_{n}.
$
Here we have defined
\be
s_{l} \equiv \sum_{j=1}^{l-1} |I|_{j}
\ee
(see also \cite{DB}). In particular, the case
$
I_{1} = \dots = I_{n} = \emptyset
$
it is sufficient for the gauge invariance of the scattering matrix, at least
in the adiabatic limit.

Such identities can be usually broken by {\it anomalies} i.e. expressions of the type
$
A^{I_{1},\dots,I_{n}}
$
which are quasi-local and might appear in the right-hand side of the relation (\ref{gauge}). These expressions verify some consistency conditions - the so-called Wess-Zumino equations. One can use these equations in the attempt to eliminate the anomalies by redefining the chronological products. All these operations can be proven to be of cohomological nature and naturally lead to descent equations of the same type as (\ref{descent1}) but for different ghost number and canonical dimension. 

If one can choose the chronological products such that gauge invariance is true then  there is still some freedom left for redefining them. To be able to decide if the theory is renormalizable one needs the general form of such arbitrariness. Again, one can reduce the study of the arbitrariness to descent equations of the type as (\ref{descent1}).

Such type of cohomology problems have been extensively studied in the more popular approach to quantum gauge theory based on functional methods (following from some path integration method). In this setting the co-chain operator is non-linear and makes sense only for classical field theories. On the contrary, in the causal approach the co-chain operator is linear so the cohomology problem makes sense directly in the Hilbert space of the model. 
One needs however a classical field theory machinery to analyze the descent equations more easily.

In this paper we want to give a general description of these methods and we will apply them for Yang-Mills models. In the next Section we remind the axioms verified by the chronological products and consider the particular case of gauge models.  

In Section \ref{WZ} we give some general results about the structure of the anomalies and reduce the proof of (\ref{gauge}) to descent equations. In Section \ref{geom} we provide a convenient geometric setting for our problem. We will prove an algebraic form of the Poincar\'e lemma valid for on-shell fields (The usual Poincar\'e cannot be applied because the homotopy operator of de Rham does not leave invariant the space of on-shell polynomials.) In Section \ref{q} we determine the cohomology of the operator 
$
d_{Q}
$
for Yang-Mills models. Using this cohomology and the algebraic Poincar\'e lemma we can solve the descent equations in various ghost numbers in Section \ref{relative}. We make some comments about higher orders of perturbation theory in Section \ref{yang}. For the case of quantum electro-dynamics we give the shortest proof of gauge invariance in all orders. 

The present paper includes the results of some previous papers \cite{YM}, \cite{standard},\cite{fermi}, \cite{ano} but many the proofs are new and use in an optimal way various cohomological structures. In \cite{Sc1} and \cite{Sc2} one can find similar results but the cohomological methods are not used for the proofs.
\newpage

\section{General Gauge Theories\label{ggt}}
 
We give here the essential ingredients of perturbation theory. 

\subsection{Bogoliubov Axioms}{\label{bogoliubov}}

The chronological products
$ 
T(W_{1}(x_{1}),\dots,W_{n}(x_{n})) \quad n = 1,2,\dots
$
are verifying the following set of axioms:
\begin{itemize}
\item
Skew-symmetry in all arguments
$
W_{1}(x_{1}),\dots,W_{n}(x_{n}):
$
\be
T(\dots,W_{i}(x_{i}),W_{i+1}(x_{i+1}),\dots,) =
(-1)^{f_{i} f_{i+1}} T(\dots,W_{i+1}(x_{i+1}),W_{i}(x_{i}),\dots)
\ee
where
$f_{i}$
is the number of Fermi fields appearing in the Wick monomial
$W_{i}$.
\item
Poincar\'e invariance: for all 
$(a,A) \in inSL(2,\C)$
we have:
\be
U_{a, A} T(W_{1}(x_{1}),\dots,W_{n}(x_{n})) U^{-1}_{a, A} =
T(A\cdot W_{1}(A\cdot x_{1}+a),\dots,A\cdot W_{n}(A\cdot x_{n}+a));
\label{invariance}
\ee

Sometimes it is possible to supplement this axiom by other invariance properties: space and/or time inversion, charge conjugation invariance, global symmetry invariance with respect to some internal symmetry group, supersymmetry, etc.
\item
Causality: if
$x_{i} \geq x_{j}, \quad \forall i \leq k, \quad j \geq k+1$
then we have:
\be
T(W_{1}(x_{1}),\dots,W_{n}(x_{n})) =
T(W_{1}(x_{1}),\dots,W_{k}(x_{k}))~~T(W_{k+1}(x_{k+1}),\dots,W_{n}(x_{n}));
\label{causality}
\ee
\item
Unitarity: We define the {\it anti-chronological products} according to
\be
(-1)^{n} \bar{T}(W_{1}(x_{1}),\dots,W_{n}(x_{n})) \equiv \sum_{r=1}^{n} 
(-1)^{r} \sum_{I_{1},\dots,I_{r} \in Part(\{1,\dots,n\})}
\epsilon~~T_{I_{1}}(X_{1})\cdots T_{I_{r}}(X_{r})
\label{antichrono}
\ee
where the we have used the notation:
\be
T_{\{i_{1},\dots,i_{k}\}}(x_{i_{1}},\dots,x_{i_{k}}) \equiv 
T(W_{i_{1}}(x_{i_{1}}),\dots,W_{i_{k}}(x_{i_{k}}))
\ee
and the sign
$\epsilon$
counts the permutations of the Fermi factors. Then the unitarity axiom is:
\be
\bar{T}(W_{1}(x_{1}),\dots,W_{n}(x_{n}))
= T(W^{*}_{1}(x_{1}),\dots,W^{*}_{n}(x_{n}))
\label{unitarity}
\ee
\item
The ``initial condition"
\be
T(W(x)) = W(x).
\ee
\end{itemize}

It can be proved that this system of axioms can be supplemented with
\bea
T(W_{1}(x_{1}),\dots,W_{n}(x_{n}))
\nonumber \\
= \sum \epsilon \quad
<\Omega, T(W'_{1}(x_{1}),\dots,W'_{n}(x_{n}))\Omega>~~
:W_{1}"(x_{1}),\dots,W_{n}"(x_{n}):
\label{wick-chrono2}
\eea
where
$W'_{i}$
and
$W_{i}"$
are Wick submonomials of
$W_{i}$
such that
$W_{i} = :W'_{i} W_{i}":$
the sign
$\epsilon$
takes care of the permutation of the Fermi fields and
$\Omega$
is the vacuum state. This is called the {\it Wick expansion property}. 

We can also include in the induction hypothesis a limitation on the order of singularity of the vacuum averages of the chronological products associated to arbitrary Wick monomials
$W_{1},\dots,W_{n}$;
explicitly:
\be
\omega(<\Omega, T^{W_{1},\dots,W_{n}}(X)\Omega>) \leq
\sum_{l=1}^{n} \omega(W_{l}) - 4(n-1)
\label{power}
\ee
where by
$\omega(d)$
we mean the order of singularity of the (numerical) distribution $d$ and by
$\omega(W)$
we mean the canonical dimension of the Wick monomial $W$; in particular this means
that we have
\be
T(W_{1}(x_{1}),\dots,W_{n}(x_{n}))
= \sum_{g} t_{g}(x_{1},\dots,x_{n})~W_{g}(x_{1},\dots,x_{n})
\label{generic}
\ee
where
$W_{g}$
are Wick polynomials of fixed canonical dimension and
$t_{g}$
are distributions with the order of singularity bounded by the power counting
theorem \cite{EG}:
\be
\omega(t_{g}) + \omega(W_{g}) \leq
\sum_{j=1}^{n} \omega(W_{j}) - 4 (n - 1)
\label{power1}
\ee
and the sum over $g$ is essentially a sum over Feynman graphs.

Up to now, we have defined the chronological products only for Wick monomials 
$
W_{1},\dots,W_{n}
$
but we can extend the definition for Wick polynomials by linearity.

One can modify the chronological products without destroying the basic property of causality {\it iff} one can make
\bea
T(W_{1}(x_{1}),\dots,W_{n}(x_{n})) \rightarrow
T(W_{1}(x_{1}),\dots,W_{n}(x_{n}))
\nonumber \\
+ R_{W_{1},\dots,W_{n}}(x_{1},\dots,x_{n})
\label{renorm}
\eea
where $R$ are quasi-local expressions; by a {\it quasi-local expression} we mean an expression of the form
\be
R_{W_{1},\dots,W_{n}}(x_{1},\dots,x_{n}) =
\sum_{g} \left[ P_{g}(\partial)\delta(X)\right]
W_{g}(x_{1},\dots,x_{n}) 
\label{renorm1}
\ee
with 
$P_{g}$ 
monomials in the partial derivatives and 
$W_{g}$
are Wick polynomials; here
$
\delta(X)
$
is the $n$-dimensional delta distribution
$
\delta(X) \equiv \delta(x_{1} - x_{n})\cdots \delta(x_{n-1} - x_{n}).
$
Because of the delta function we can consider that   
$P_{g}$
is a monomial only in the derivatives with respect to, say
$
x_{2},\dots,x_{n}.
$
If we want to preserve (\ref{power}) we impose the restriction
\be
deg(P_{g}) + \omega(W_{g}) \leq
\sum_{j=1}^{n} \omega(W_{j}) - 4 (n - 1)
\label{power2}
\ee
and some other restrictions are following from the preservation of Lorentz covariance and unitarity.

The redefinitions of the type (\ref{renorm}) are the so-called {\it finite renormalizations}. Let us note that this arbitrariness, described by the number of independent coefficients of the polynomials
$
P_{g}
$
can grow with $n$ and in this case the theory is called {\it non-renormalizable}. This can happen if some of the Wick monomials
$
W_{j}, j = 1,\dots,n
$
have canonical dimension greater than $4$. If all the monomials have canonical dimension less of equal to $4$ then the arbitrariness is bounded independently of $n$ and the theory is called {\it renormalizable}. However, even in the non-renormalizable case if the theory verifies some additional symmetry properties it could happen that the number of arbitrary coefficients from
$
P_{g}
$
is finite. This seems to be the case for quantum gravity. We will analyze this case in another paper.

It is not hard to prove that any finite renormalization can be rewritten in the form
\be
R(x_{1},\dots,x_{n}) =
\delta(X)~W(x_{1}) + \sum_{j=1}^{n}~{\partial \over \partial x^{\mu}_{l}}R_{l}(X)
\label{renorm2}
\ee
where the expressions
$
R_{l}(X)
$
are also quasi-local. But it is clear that the sum in the above expression is null in the adiabatic limit. This means that we can postulate that the finite renormalizations have a much simpler form, namely
\be
R(x_{1},\dots,x_{n}) = \delta(X)~W(x_{1})
\label{renorm3}
\ee  
where the Wick polynomial $W$ is constrained by
\be
\omega(W) \leq \sum_{j=1}^{n} \omega(W_{j}) - 4 (n - 1).
\label{power3}
\ee

\subsection{Gauge Theories and Anomalies\label{anomalies}}

From now on we consider that we work in the four-dimensional Minkowski space and we have the Wick polynomials
$
T^{I}
$
such that the descent equations (\ref{descent1}) are true and we also have
\be
T^{I}(x_{1})~T^{J}(x_{2}) = (-1)^{|I||J|}~T^{J}(x_{2})~T^{I}(x_{1})
\label{graded-comm}
\ee
for 
$
x_{1} - x_{2}
$
space-like i.e. these expressions causally commute in the graded sense. 

The equation (\ref{descent1}) are called a {\it relative cohomology} problem. The co-boundaries for this problem are of the type
\be
T^{I} = d_{Q}B^{I} + i~\partial_{\mu}B^{I\mu}.
\label{coboundary}
\ee

Next we construct the associated chronological products
$$
T^{I_{1},\dots,I_{n}}(x_{1},\dots,x_{n}) = T(T^{I_{1}}(x_{1}),\dots,T^{I_{n}}(x_{n})).
$$

Because of the previous assumption, it follows from the skew-symmetry axiom that we can choose them such that we have the graded symmetry property:
\be
T(\dots,T^{I_{k}}(x_{k}),T^{I_{k+1}}(x_{k+1}),\dots) = (-1)^{|I_{k}| |I_{k+1}|}~
T(\dots,T^{I_{k+1}}(x_{k+1}),T^{I_{k}}(x_{k}),\dots).
\label{symmetryT}
\ee
We also have
\be
gh(T^{I_{1},\dots,I_{n}}) = \sum_{l=1}^{n} |I_{l}|.
\label{ghT}
\ee

In the case of a gauge theory there are renormalizations of the type (\ref{renorm1}) which call {\it trivial}, namely those of the type
\be
R^{\dots}(X) = d_{Q}B^{\dots}(X) 
+ i~\sum_{l=1}^{n}~{\partial \over \partial x^{\mu}_{l}}B^{l;\dots}(X)
\label{renorm4}
\ee

Indeed, as it was remarked above, any co-boundary operator induces the null operator on the physical Hilbert space. Also any total divergence gives a null contribution in the adiabatic limit.

We now write the gauge invariance condition (\ref{gauge}) in a compact form. We consider the space 
$
{\cal C}_{n}
$
of co-chains of the form 
$
C^{I_{1},\dots,I_{n}}(X)
$
which are distribution-valued operators in the Hilbert space with antisymmetry in all indices from every
$
I_{j},~(j = 1,\dots,n)
$
and also verifying:
\be
C^{\dots,I_{k},I_{k+1},\dots}(\dots,x_{k},x_{k+1},\dots) 
= (-1)^{|I_{k}| |I_{k+1}|} \times
C^{\dots,I_{k+1},I_{k},\dots}(\dots,x_{k+1},x_{k},\dots).
\label{symmetryC}
\ee
Then we can define the operator
$
\delta: {\cal C}_{n} \longrightarrow {\cal C}_{n+1}
$ 
according to:
\be
\delta ~C^{I_{1},\dots,I_{n}} \equiv
\sum_{l=1}^{n} (-1)^{s_{l}} {\partial\over \partial x^{\mu}_{l}}
C^{I_{1},\dots,I_{l}\mu,\dots,I_{n}}
\label{deltaA}.
\ee
It is easy to prove that we have:
\be
\delta^{2} = 0;
\ee
we also note that
$\delta$ 
commutes with
$
d_{Q}.
$
One can now write the equation (\ref{gauge}) in a more compact way:
\be
d_{Q}T^{I_{1},\dots,I_{n}} = i \delta T^{I_{1},\dots,I_{n}}. 
\label{gauge1}
\ee

We now determine the obstructions for the gauge invariance relations (\ref{gauge1}). These relations are true for $n = 1$ according to (\ref{descent1}). If we suppose that they are true up to order
$
n - 1
$
then it follows easily that in order $n$ we must have:
\be
d_{Q}T^{I_{1},\dots,I_{n}} = i \delta T^{I_{1},\dots,I_{n}}
+ A^{I_{1},\dots,I_{n}}
\label{gauge2}
\ee
where the expressions
$
A^{I_{1},\dots,I_{n}}(x_{1},\dots,x_{n})
$
are quasi-local operators and are called {\it anomalies}. It is clear that we have from (\ref{symmetryT}) a similar symmetry for the anomalies: namely we have complete antisymmetry in all indices from every
$
I_{j},~(j = 1,\dots,n) 
$
and
\be
A^{\dots,I_{k},I_{k+1},\dots}(\dots,x_{k},x_{k+1},\dots) 
= (-1)^{|I_{k}| |I_{k+1}|} \times
A^{\dots,I_{k+1},I_{k},\dots}(\dots,x_{k+1},x_{k},\dots).
\label{symmetryA}
\ee
i.e. 
$
A^{I_{1},\dots,I_{n}}(x_{1},\dots,x_{n})
$
are also co-chains. We also have
\be
gh(A^{I_{1},\dots,I_{n}}) = \sum_{l=1}^{n} |I_{l}| + 1.
\label{ghA}
\ee

Let
$
\omega_{0} \equiv \omega(T);
$
then one has:
\be
A^{I_{1},\dots,I_{n}}(X) = 0
\quad {\it iff} \quad \sum_{l=1}^{n} |I_{l}| > n (\omega_{0} - 1) + 4
\label{limit}
\ee

We can write a more precise form for the anomalies, namely:
\bea
A^{I_{1},\dots,I_{n}}(x_{1},\dots,x_{n})
= \sum_{k} \sum_{i_{1},\dots,i_{k} > 1}
[ \partial^{i_{1}}_{\rho_{1}} \dots \partial^{i_{k}}_{\rho_{k}} \delta(X) ]
W^{I_{1},\dots,I_{n};\rho_{1},\dots,\rho_{k}}_{i_{1},\dots,i_{k}}(x_{1})
\label{genericA}
\eea
and in this expression the Wick polynomials
$
W^{I_{1},\dots,I_{n};\rho_{1},\dots,\rho_{k}}_{i_{1},\dots,i_{k}}
$
are uniquely defined. Now from (\ref{power1}) we have
\be
\omega(W^{I_{1},\dots,I_{n};\rho_{1},\dots,\rho_{k}})
\leq n ( \omega_{0} - 4) + 5 - k
\label{power4}
\ee
which gives a bound on $k$ in the previous sum. We also have some consistency conditions on the expressions verified by the anomalies. If one applies the operator
$d_{Q}$
to (\ref{gauge2}) one obtains the so-called {\it Wess-Zumino consistency conditions}:
\be
d_{Q}A^{I_{1},\dots,I_{n}} = - i~\delta A^{I_{1},\dots,I_{n}}.
\label{wz}
\ee

Suppose now that we have fixed the gauge invariance (\ref{gauge1}) and we investigate the renormalizability issue i.e. we make the redefinitions
\be
T(T^{I_{1}}(x_{1}),\dots,T^{I_{n}}(x_{n})) \rightarrow
T(T^{I_{1}}(x_{1}),\dots,T^{I_{n}}(x_{n}))
+ R^{I_{1},\dots,I_{n}}(x_{1},\dots,x_{n})
\label{renorm5}
\ee
where $R$ are  quasi-local expressions. As before we have
\be
R^{\dots,I_{k},I_{k+1},\dots}(\dots,x_{k},x_{k+1},\dots) 
= (-1)^{|I_{k}| |I_{k+1}|} \times
R^{\dots,I_{k+1},I_{k},\dots}(\dots,x_{k+1},x_{k},\dots).
\label{symmetryR}
\ee
We also have
\be
gh(R^{I_{1},\dots,I_{n}}) = \sum_{l=1}^{n} |I_{l}|.
\label{ghR}
\ee
and
\be
R^{I_{1},\dots,I_{n}} = 0,
\quad
\sum_{l=1}^{n} |I_{l}| > n (\omega_{0} - 1) + 4.
\label{limit1}
\ee

If we want to preserve (\ref{gauge}) it is clear that the quasi-local operators
$
R^{I_{1},\dots,I_{n}}
$
should also verify 
\be
d_{Q}R^{I_{1},\dots,I_{n}} = i~\delta R^{I_{1},\dots,I_{n}}
\label{wz3}
\ee
i.e. equations of the type (\ref{wz}). In this case we note that we have more structure;
according to the previous discussion we can impose the structure (\ref{renorm1}):
\be
R^{I_{1},\dots,I_{n}}(x_{1},\dots,x_{n}) 
= \delta(X)~W^{I_{1},\dots,I_{n}}(x_{1})
\ee
and we obviously have:
\be
gh(W^{I_{1},\dots,I_{n}}) = \sum_{l=1}^{n} |I_{l}|
\label{ghR1}
\ee
and
\be
W^{I_{1},\dots,I_{n}} = 0,
\quad
\sum_{l=1}^{n} |I_{l}| > n (\omega_{0} - 1) + 4.
\label{limit2}
\ee

From (\ref{wz3}) we obtain after some computations that there are Wick polynomials
$
R^{I}
$
such that
\be
W^{I_{1},\dots,I_{n}} = (-1)^{s}~R^{I_{1} \cup\dots\cup I_{n}}.
\label{gauge6}
\ee
where
\be
s \equiv \sum_{k <l \leq n} |I_{k}| |I_{l}|.
\ee

Moreover, we have
\be
gh(R^{I}) = |I|
\label{ghR2}
\ee
and
\be
R^{I} = 0,
\quad
|I| > n (\omega_{0} - 1) + 4.
\label{limit3}
\ee

Finally, the following descent equations are true:
\be
d_{Q}R^{I} = i~\partial_{\mu}R^{I\mu}
\label{gauge5}
\ee
and have obtained another relative cohomology problem similar to (\ref{coboundary}) but in another ghost sector and canonical dimensions. The relative Co-boundaries of this problem correspond to the relative Co-boundaries from (\ref{renorm}). 

\newpage
\section{A Particular Case of the Wess-Zumino Consistency Conditions \label{WZ}}

In this Section we consider a particular form of (\ref{gauge2}) and (\ref{wz}) namely the case when all polynomials
$
T^{I}
$
have canonical dimension
$
\omega_{0} =4.
$
In this case (\ref{limit}) becomes:
\be
A^{I_{1},\dots,I_{n}}(X) = 0
\quad {\it iff} \quad \sum_{l=1}^{n} |I_{l}| > 4
\label{limit4}
\ee
and this means that only a finite number of the equations (\ref{gauge2}) can be anomalous. It is convenient to define 
\bea
A_{1} \equiv A^{\emptyset,\dots,\emptyset},~
A_{2}^{\mu} \equiv A^{[\mu],\emptyset,\dots,\emptyset},~
A_{3}^{[\mu\nu]} \equiv A^{[\mu\nu],\emptyset,\dots,\emptyset},
\nonumber \\
A_{4}^{\mu;\nu} \equiv A^{[\mu],[\nu],\emptyset,\dots,\emptyset},~
A_{5}^{[\mu\nu];\rho} \equiv A^{[\mu\nu],\rho,\emptyset,\dots,\emptyset},~
A_{6}^{[\mu\nu];[\rho\sigma]} \equiv A^{[\mu\nu],[\rho\sigma],\emptyset,\dots,\emptyset},
\nonumber \\
A_{7}^{\mu;\nu;\rho} \equiv A^{[\mu],[\nu],[\rho],\emptyset,\dots,\emptyset},~
A_{8}^{[\mu\nu];\rho;\sigma} \equiv A^{[\mu\nu],[\rho],[\sigma],\emptyset,\dots,\emptyset},~
A_{9}^{\mu;\nu;\rho;\sigma} \equiv A^{[\mu],[\nu],[\rho],[\sigma],\emptyset,\dots,\emptyset}
\eea
where we have emphasized the antisymmetry properties with brackets. We have from (\ref{gauge2}) the following anomalous gauge equations:
\bea
d_{Q} T(T(x_{1}),\dots,T(x_{n})) =
\nonumber \\
i \sum_{l=1}^{n} {\partial\over \partial x^{\mu}_{l}} 
T(T(x_{1}),\dots,T^{\mu}(x_{l}),\dots,T(x_{n}))
+ A_{1}(X)
\label{ym1}
\eea
\bea
d_{Q} T(T^{\mu}(x_{1}),T(x_{2}),\dots,T(x_{n})) =
i {\partial\over \partial x^{\mu}_{1}} 
T(T^{\mu\nu}(x_{1}),T(x_{2}),\dots,T(x_{n}))
\nonumber \\
-i \sum_{l=2}^{n} {\partial\over \partial x^{\nu}_{l}} 
T(T^{\mu}(x_{1}),T(x_{2}),\dots,T^{\nu}(x_{l}),\dots,T(x_{n}))
+ A^{\mu}_{2}(X)
\label{ym2}
\eea
\bea
d_{Q} T(T^{\mu\nu}(x_{1}),T(x_{2}),\dots,T(x_{n})) =
\nonumber \\
i \sum_{l=2}^{n} {\partial\over \partial x^{\rho}_{l}} 
T(T^{\mu\nu}(x_{1}),T(x_{2}),\dots,T^{\rho}(x_{l}),\dots,T(x_{n}))
+ A^{[\mu\nu]}_{3}(X)
\label{ym3}
\eea
\bea
d_{Q} T(T^{\mu}(x_{1}),T^{\nu}(x_{2}),T(x_{3}),\dots,T(x_{n})) =
\nonumber \\
i {\partial\over \partial x^{\rho}_{1}} 
T(T^{\mu\rho}(x_{1}),T^{\nu}(x_{2}),T(x_{3}),\dots,T(x_{n}))
- i {\partial\over \partial x^{\rho}_{2}} 
T(T^{\mu}(x_{1}),T^{\nu\rho}(x_{2}),T(x_{3}),\dots,T(x_{n}))
\nonumber \\
+ i \sum_{l=3}^{n} {\partial\over \partial x^{\rho}_{l}} 
T(T^{\mu}(x_{1}),T^{\nu}(x_{2}),T(x_{3}),\dots,T^{\rho}(x_{l}),\dots,
T(x_{n})) 
+ A^{\mu;\nu}_{4}(X)
\label{ym4}
\eea
\bea
d_{Q} T(T^{\mu\nu}(x_{1}),T^{\rho}(x_{2}),T(x_{3}),\dots,T(x_{n})) =
\nonumber \\
i {\partial\over \partial x^{\sigma}_{2}} 
T(T^{\mu\nu}(x_{1}),T^{\rho\sigma}(x_{2}),T(x_{3}),\dots,T(x_{n}))
\nonumber \\
- i \sum_{l=3}^{n} {\partial\over \partial x^{\sigma}_{l}} 
T(T^{\mu\nu}(x_{1}),T^{\rho}(x_{2}),\dots,T^{\sigma}(x_{l}),\dots,T(x_{n}))
+ A^{[\mu\nu];\rho}_{5}(X)
\label{ym5}
\eea
\bea
d_{Q} T(T^{\mu\nu}(x_{1}),T^{\rho\sigma}(x_{2}),T(x_{3}),\dots,T(x_{n})) =
\nonumber \\
i \sum_{l=3}^{n} {\partial\over \partial x^{\lambda}_{l}} 
T(T^{\mu\nu}(x_{1}),T^{\rho\sigma}(x_{2}),T(x_{3}),\dots,T^{\lambda}(x_{l}),
\dots,T(x_{n}))
\nonumber \\
+ A^{[\mu\nu];[\rho\sigma]}_{6}(X)
\label{ym6}
\eea
\bea
d_{Q} T(T^{\mu}(x_{1}),T^{\nu}(x_{2}),T^{\rho}(x_{3}),T(x_{4}),\dots,T(x_{n}))
= \nonumber \\
i {\partial\over \partial x^{\sigma}_{1}} 
T(T^{\mu\sigma}(x_{1}),T^{\nu}(x_{2}),T^{\rho}(x_{3}),T(x_{4}),\dots,T(x_{n}))
\nonumber \\
- i {\partial\over \partial x^{\sigma}_{2}} 
T(T^{\mu}(x_{1}),T^{\nu\sigma}(x_{2}),T^{\rho}(x_{3}),T(x_{4}),\dots,T(x_{n}))
\nonumber \\
+ i {\partial\over \partial x^{\sigma}_{3}} 
T(T^{\mu}(x_{1}),T^{\nu}(x_{2}),T^{\rho\sigma}(x_{3}),T(x_{4}),\dots,T(x_{n}))
\nonumber \\
- i \sum_{l=4}^{n} {\partial\over \partial x^{\sigma}_{l}} 
T(T^{\mu}(x_{1}),T^{\nu}(x_{2}),T^{\rho}(x_{3}),T(x_{4}),\dots,
T^{\sigma}(x_{l}),\dots,T(x_{n})) 
\nonumber \\
+ A^{\mu;\nu;\rho}_{7}(X)
\label{ym7}
\eea
\bea
d_{Q} T(T^{\mu\nu}(x_{1}),T^{\rho}(x_{2}),T^{\sigma}(x_{3}),T(x_{4}),
\dots,T(x_{n})) =
\nonumber \\
i {\partial\over \partial x^{\lambda}_{2}} 
T(T^{\mu\nu}(x_{1}),T^{\rho\lambda}(x_{2}),T^{\sigma}(x_{3}),T(x_{4}),
\dots,T(x_{n}))
\nonumber \\
- i {\partial\over \partial x^{\lambda}_{3}} 
T(T^{\mu\nu}(x_{1}),T^{\rho}(x_{2}),T^{\sigma\lambda}(x_{3}),T(x_{4}),
\dots,T(x_{n}))
\nonumber \\ 
+ i \sum_{l=4}^{n} {\partial\over \partial x^{\lambda}_{l}} 
T(T^{\mu\nu}(x_{1}),T^{\rho}(x_{2}),T^{\sigma}(x_{3}),T(x_{4}),\dots,
T^{\lambda}(x_{l}),\dots,T(x_{n}))
\nonumber \\
+ A^{[\mu\nu];\rho;\sigma}_{8}(X)
\label{ym8}
\eea
\bea
d_{Q} T(T^{\mu}(x_{1}),T^{\nu}(x_{2}),T^{\rho}(x_{3}),T^{\sigma}(x_{4}),\dots,
T(x_{n})) =
\nonumber \\
i {\partial\over \partial x^{\lambda}_{1}} 
T(T^{\mu\lambda}(x_{1}),T^{\nu}(x_{2}),T^{\rho}(x_{3}),T^{\sigma}(x_{4}),
T(x_{5}),\dots,T(x_{n}))
\nonumber \\
- i {\partial\over \partial x^{\lambda}_{2}} 
T(T^{\mu}(x_{1}),T^{\nu\lambda}(x_{2}),T^{\rho}(x_{3}),T^{\sigma}(x_{4}),
T(x_{5}),\dots,T(x_{n}))
\nonumber \\
+ i {\partial\over \partial x^{\lambda}_{3}} 
T(T^{\mu}(x_{1}),T^{\nu}(x_{2}),T^{\rho\lambda}(x_{3}),T^{\sigma}(x_{4}),
T(x_{5}),\dots,T(x_{n}))
\nonumber \\
- i {\partial\over \partial x^{\lambda}_{4}} 
T(T^{\mu}(x_{1}),T^{\nu}(x_{2}),T^{\rho}(x_{3}),T^{\sigma\lambda}(x_{4}),
T(x_{5}),\dots,T(x_{n}))
\nonumber \\
+ i \sum_{l=5}^{n} {\partial\over \partial x^{\lambda}_{l}} 
T(T^{\mu}(x_{1}),T^{\nu}(x_{2}),T^{\rho}(x_{3}),T^{\sigma}(x_{4}),
T(x_{5}),\dots,T^{\lambda}(x_{l}),\dots,T(x_{n})) 
\nonumber \\
+ A^{\mu;\nu;\rho;\sigma}_{9}(X)
\label{ym9}
\eea
where we can assume that:
\bea
A^{\mu;\nu}_{4}(X) = 0, \quad A^{\mu\nu;\rho}_{5} = 0, \quad
A^{\mu\nu;\rho\sigma}_{6} = 0, \quad |X| = 1,
\nonumber \\
A^{\mu;\nu;\rho}_{7}(X) = 0, \quad A^{\mu\nu;\rho;\sigma}_{8} = 0,
\quad |X| \leq 2,
\nonumber \\
A^{\mu;\nu;\rho;\sigma}_{9}(X) = 0, \quad |X| \leq 3
\eea
without losing generality. 

From (\ref{symmetryA}), we get the following symmetry properties:
\be
A_{1}(x_{1},\dots,x_{n}) \quad
{\rm is~ symmetric~ in} \quad x_{1},\dots,x_{n};
\label{sA1}
\ee
\be
A_{2}^{\mu}(x_{1},\dots,x_{n}) \quad
{\rm is~ symmetric~ in} \quad x_{2},\dots,x_{n};
\label{sA2}
\ee
\be
A_{3}^{[\mu\nu]}(x_{1},\dots,x_{n}) \quad
{\rm is~ symmetric~ in} \quad x_{2},\dots,x_{n};
\label{sA3}
\ee
\be
A_{4}^{\mu;\nu}(x_{1},\dots,x_{n}) \quad
{\rm is~ symmetric~ in} \quad x_{3},\dots,x_{n};
\label{sA4}
\ee
\be
A_{5}^{[\mu\nu];\rho}(x_{1},\dots,x_{n}) \quad
{\rm is~ symmetric~ in} \quad x_{3},\dots,x_{n};
\label{sA5}
\ee
\be
A_{6}^{[\mu\nu];[\rho\sigma]}(x_{1},\dots,x_{n}) \quad
{\rm is~ symmetric~ in} \quad x_{3},\dots,x_{n};
\label{sA6}
\ee
\be
A_{7}^{\mu;\nu;\rho}(x_{1},\dots,x_{n}) \quad
{\rm is~ symmetric~ in} \quad x_{4},\dots,x_{n};
\label{sA7}
\ee
\be
A_{8}^{[\mu\nu];\rho;\sigma}(x_{1},\dots,x_{n}) \quad
{\rm is~ symmetric~ in} \quad x_{4},\dots,x_{n};
\label{sA8}
\ee
\be
A_{9}^{\mu;\nu;\rho;\sigma}(x_{1},\dots,x_{n}) \quad
{\rm is~ symmetric~ in} \quad x_{5},\dots,x_{n}
\label{sA9}
\ee
and we also have:
\be
A_{4}^{\mu;\nu}(x_{1},\dots,x_{n}) = - A_{4}^{\nu;\mu}(x_{2},x_{1},x_{3},\dots,x_{n}) ;
\label{s4'}
\ee
\be
A_{6}^{[\mu\nu];[\rho\sigma]}(x_{1},\dots,x_{n}) = 
A_{6}^{[\rho\sigma];[\mu\nu]}(x_{2},x_{1},x_{3},\dots,x_{n});
\label{s6'}
\ee
\be
A_{7}^{\mu;\nu;\rho}(x_{1},\dots,x_{n}) 
= - A_{7}^{\nu;\mu;\rho}(x_{2},x_{1},x_{3},\dots,x_{n})
= - A_{7}^{\mu;\rho;\nu}(x_{1},x_{3},x_{2},x_{4},\dots,x_{n});
\label{s7'}
\ee
\be
A_{8}^{[\mu\nu];\rho;\sigma}(x_{1},x_{2},\dots,x_{n}) =
- A_{8}^{[\mu\nu];\sigma;\rho}(x_{1},x_{3},x_{2},x_{4},\dots,x_{n});
\label{s8'}
\ee
\bea
A_{9}^{\mu;\nu;\rho;\sigma}(x_{1},\dots,x_{n}) 
= - A_{9}^{\nu;\mu;\rho;\sigma}(x_{2},x_{1},x_{3},\dots,x_{n})
\nonumber \\
= - A_{9}^{\mu;\rho;\nu;\sigma}(x_{1},x_{3},x_{2},x_{4},\dots,x_{n})
= - A_{9}^{\mu;\nu;\sigma;\rho}(x_{1},x_{2},x_{4},x_{3},x_{5},\dots,x_{n}).
\label{s9'}
\eea
 
The Wess-Zumino consistency conditions are in this case:
\be
d_{Q} A_{1}(x_{1},\dots,x_{n}) 
= - i \sum_{l=1}^{n} {\partial\over \partial x^{\mu}_{l}} 
A^{\mu}_{2}(x_{l},x_{1},\dots,\hat{x}_{l},\dots,x_{n})
\label{WZ1}
\ee
\be
d_{Q} A^{\mu}_{2}(x_{1},\dots,x_{n})
= - i {\partial\over \partial x^{\nu}_{1}}
A^{[\mu\nu]}_{3}(x_{1},\dots,x_{n})
+ i \sum_{l=2}^{n} {\partial\over \partial x^{\nu}_{l}} 
A^{\mu;\nu}_{4}(x_{1},x_{l},x_{2},\dots,\hat{x}_{l},\dots,x_{n})
\label{WZ2}
\ee
\be
d_{Q} A^{[\mu\nu]}_{3}(x_{1},\dots,x_{n})
= - i \sum_{l=2}^{n} {\partial\over \partial x^{\rho}_{l}} 
A^{[\mu\nu];\rho}_{5}(x_{1},x_{l},x_{2},\dots,\hat{x}_{l},\dots,x_{n})
\label{WZ3}
\ee
\bea
d_{Q} A^{\mu;\nu}_{4}(x_{1},\dots,x_{n})
= - i {\partial\over \partial x^{\rho}_{1}}
A^{[\mu\rho];\nu}_{5}(x_{1},\dots,x_{n})
+ i {\partial\over \partial x^{\rho}_{2}}
A^{[\nu\rho];\mu}_{5}(x_{2},x_{1},x_{3},\dots,x_{n})
\nonumber \\
- i \sum_{l=3}^{n} {\partial\over \partial x^{\rho}_{l}}
A^{\mu;\nu;\rho}_{7}(x_{1},x_{2},x_{l},x_{3},\dots,\hat{x}_{l},\dots,x_{n})
\label{WZ4}
\eea
\bea
d_{Q} A^{[\mu\nu];\rho}_{5}(x_{1},\dots,x_{n})
= - i {\partial\over \partial x^{\sigma}_{2}}
A^{[\mu\nu];[\rho\sigma]}_{6}(x_{1},\dots,x_{n})
\nonumber \\
+ i \sum_{l=3}^{n} {\partial\over \partial x^{\sigma}_{l}}
A^{[\mu\nu];\rho;\sigma}_{8}(x_{1},x_{2},x_{l},x_{3},\dots,\hat{x}_{l},\dots,x_{n})
\label{WZ5}
\eea
\be
d_{Q} A^{[\mu\nu];[\rho\sigma]}_{6}(x_{1},\dots,x_{n}) = 0;
\label{WZ6}
\ee
\bea
d_{Q} A^{\mu;\nu;\rho}_{7}(x_{1},\dots,x_{n})
= - i {\partial\over \partial x^{\sigma}_{1}}
A^{[\mu\sigma];\nu;\rho}_{8}(x_{1},\dots,x_{n})
\nonumber \\
+ i {\partial\over \partial x^{\sigma}_{2}}
A^{[\nu\sigma];\mu;\rho}_{8}(x_{2},x_{1},x_{3},\dots,x_{n})
- i {\partial\over \partial x^{\sigma}_{3}}
A^{[\rho\sigma];\mu;\nu}_{8}(x_{3},x_{1},x_{2},x_{4},\dots,x_{n})
\nonumber \\
+ i \sum_{l=4}^{n} {\partial\over \partial x^{\rho}_{l}}
A^{\mu;\nu;\rho;\sigma}_{9}(x_{1},x_{2},x_{3},x_{l},x_{4},\dots,\hat{x}_{l},
\dots,x_{n})
\label{WZ7}
\eea
\be
d_{Q} A^{[\mu\nu];\rho;\sigma}_{8}(x_{1},\dots,x_{n}) = 0;
\label{WZ8}
\ee
\be
d_{Q} A^{\mu;\nu;\rho;\sigma}_{9}(x_{1},\dots,x_{n}) = 0.
\label{WZ9}
\ee

We recall that the generic form of the anomalies is given by (\ref{genericA}). We propose to simplify this expression using appropriate redefinitions of the chronological products. It is better to work out first the case 
$
n = 2
$
and one will see how to proceed for higher orders. In the case 
$
n = 2
$
we have the following possible anomalous gauge invariance relations:
\be
d_{Q} T(T(x_{1}),T(x_{2})) =
\nonumber \\
i {\partial\over \partial x^{\mu}_{1}}T(T^{\mu}(x_{1}),T(x_{2}))
+ i {\partial\over \partial x^{\mu}_{2}}T(T(x_{1}),T^{\mu}(x_{2}))
+ A_{1}(x_{1},x_{2})
\label{g1}
\ee
\be
d_{Q} T(T^{\mu}(x_{1}),T(x_{2})) =
i {\partial\over \partial x^{\mu}_{1}}T(T^{\mu\nu}(x_{1}),T(x_{2}))
-i {\partial\over \partial x^{\nu}_{2}}T(T^{\mu}(x_{1}),T^{\nu}(x_{2}))
+ A^{\mu}_{2}(x_{1},x_{2})
\label{g2}
\ee
\be
d_{Q} T(T^{\mu\nu}(x_{1}),T(x_{2})) =
i {\partial\over \partial x^{\rho}_{2}}T(T^{\mu\nu}(x_{1}),T^{\rho}(x_{2}))
+ A^{[\mu\nu]}_{3}(x_{1},x_{2})
\label{g3}
\ee
\be
d_{Q} T(T^{\mu}(x_{1}),T^{\nu}(x_{2})) =
i {\partial\over \partial x^{\rho}_{1}}T(T^{\mu\rho}(x_{1}),T^{\nu}(x_{2}))
- i {\partial\over \partial x^{\rho}_{2}}T(T^{\mu}(x_{1}),T^{\nu\rho}(x_{2}))
+ A^{\mu;\nu}_{4}(x_{1},x_{2})
\label{g4}
\ee
\be
d_{Q} T(T^{\mu\nu}(x_{1}),T^{\rho}(x_{2})) =
i {\partial\over \partial x^{\sigma}_{2}}T(T^{\mu\nu}(x_{1}),T^{\rho\sigma}(x_{2}))
+ A^{[\mu\nu];\rho}_{5}(x_{1},x_{2})
\label{g5}
\ee
\be
d_{Q} T(T^{\mu\nu}(x_{1}),T^{\rho\sigma}(x_{2})) =
A^{[\mu\nu];[\rho\sigma]}_{6}(x_{1},x_{2}).
\label{g6}
\ee

We have the following result:
\begin{thm}

One can redefine the chronological products such that
\bea
A_{1}(x_{1},x_{2}) = \delta(x_{1} - x_{2})~W(x_{1}), \qquad
A^{\mu}_{2}(x_{1},x_{2}) = \delta(x_{1} - x_{2})~W^{\mu}(x_{1})
\nonumber \\
A^{[\mu\nu]}_{3}(x_{1},x_{2}) = \delta(x_{1} - x_{2})~W^{[\mu\nu]}(x_{1}), \qquad
A^{\mu;\nu}_{4}(x_{1},x_{2}) = - \delta(x_{1} - x_{2})~W^{[\mu\nu]}(x_{1}),
\nonumber \\
A^{[\mu\nu];\rho}_{5}(x_{1},x_{2}) = 0, \qquad 
A^{[\mu\nu];[\rho\sigma]}_{6}(x_{1},x_{2}) = 0.
\eea
Moreover one has the following descent equations:
\be
d_{Q}W = - i~\partial_{\mu}W^{\mu},\qquad
d_{Q}W^{\mu} = i~\partial_{\nu}W^{[\mu\nu]},\qquad
d_{Q}W^{[\mu\nu]} = 0.
\label{W}
\ee

The expressions $W$ and
$
W^{\mu}
$
are relative co-cyles and are determined up to relative co-boundaries. The expression
$
W^{[\mu\nu]}
$
is a cocycle and it is determined up to a co-boundary.
\end{thm}

{\bf Proof:}
The symmetry properties are in this case 
\be
A_{1}(x_{1},x_{n}) = A_{1}(x_{2},x_{1})
\label{sA1-2}
\ee
\be
A_{4}^{\mu;\nu}(x_{1},x_{2}) = - A_{4}^{\nu;\mu}(x_{2},x_{1}) ;
\label{s4'-2}
\ee
\be
A_{6}^{[\mu\nu];[\rho\sigma]}(x_{1},x_{2}) = 
A_{6}^{[\rho\sigma];[\mu\nu]}(x_{2},x_{1})
\label{s6'-2}
\ee
and the corresponding Wess-Zumino consistency conditions
\be
d_{Q} A_{1}(x_{1},x_{2}) 
= - i {\partial\over \partial x^{\mu}_{1}}A^{\mu}_{2}(x_{1},x_{2})
- i {\partial\over \partial x^{\mu}_{2}}A^{\mu}_{2}(x_{2},x_{1})
\label{WZ1-2}
\ee
\be
d_{Q} A^{\mu}_{2}(x_{1},x_{2})
= - i {\partial\over \partial x^{\nu}_{1}}A^{\mu\nu}_{3}(x_{1},x_{2})
+ i {\partial\over \partial x^{\nu}_{2}}A^{\mu;\nu}_{4}(x_{1},x_{2})
\label{WZ2-2}
\ee
\be
d_{Q} A^{[\mu\nu]}_{3}(x_{1},x_{2})
= - i {\partial\over \partial x^{\rho}_{2}}A^{[\mu\nu];\rho}_{5}(x_{1},x_{2})
\label{WZ3-2}
\ee
\be
d_{Q} A^{\mu;\nu}_{4}(x_{1},x_{2})
= - i {\partial\over \partial x^{\rho}_{1}}A^{[\mu\rho];\nu}_{5}(x_{1},x_{2})
+ i {\partial\over \partial x^{\rho}_{2}}A^{[\nu\rho];\mu}_{5}(x_{2},x_{1})
\label{WZ4-2}
\ee
\be
d_{Q} A^{[\mu\nu];\rho}_{5}(x_{1},x_{2})
= - i {\partial\over \partial x^{\sigma}_{2}}A^{[\mu\nu];[\rho\sigma]}_{6}(x_{1},x_{2})
\label{WZ5-2}
\ee
\be
d_{Q} A^{[\mu\nu];[\rho\sigma]}_{6}(x_{1},x_{2}) = 0
\label{WZ6-2}
\ee
will be enough to obtain the result from the statement.

(i) From (\ref{genericA}) we have:
\be
A_{1}(x_{1},x_{2})
= \sum_{k \leq 4} 
\partial_{\mu_{1}} \dots \partial_{\mu_{k}} \delta(x_{2} - x_{1})
W^{\{\mu_{1},\dots,\mu_{k}\}}_{1}(x_{1})
\label{A1-2}
\ee
where we have emphasized the symmetry properties by curly brackets. We have the restrictions 
\be
\omega(W^{\{\mu_{1},\dots,\mu_{k}\}}_{1}) \leq 5 - k, \qquad
gh(W^{\{\mu_{1},\dots,\mu_{k}\}}_{1}) = 1
\ee
for all 
$
k = 0, \dots, 4.
$
We perform the finite renormalization:
\be
T(T^{\mu}(x_{1}),T(x_{2})) \rightarrow T(T^{\mu}(x_{1}),T(x_{2})) 
+ \partial_{\nu}\partial_{\rho}\partial_{\sigma}~\delta(x_{2} - x_{1})
U^{\mu;\{\nu,\rho,\sigma\}}_{2}(x_{1})
\label{R2a-2}
\ee
and it is easy to see that if we choose
$
U^{\mu;\{\nu,\rho,\sigma\}}_{2} = - {i\over 2}~W^{\{\mu,\nu,\rho,\sigma\}}_{1}
$
then we obtain a new expression (\ref{A1-2}) for the anomaly 
$
A_{1}
$
where the sum goes only up to 
$
k = 3.
$
(Although the monomials
$
W^{\{\mu_{1},\dots,\mu_{k}\}}_{1}
$
will be changed after this finite renormalization we keep the same notation.) Now we impose the symmetry property (\ref{sA1-2}) and consider only the terms with three derivatives on 
$\delta$;
it easily follows that
$
W^{\{\mu,\nu,\rho\}}_{1} = 0
$
i.e. in the expression (\ref{A1-2}) for the anomaly 
$
A_{1}
$
the sum goes only up to 
$
k = 2.
$

Next we perform the finite renormalization:
\be
T(T^{\mu}(x_{1}),T(x_{2})) \rightarrow T(T^{\mu}(x_{1}),T(x_{2})) 
+ \partial_{\nu}\delta(x_{2} - x_{1})
U^{\mu;\nu}_{2}(x_{1})
\label{R2b-2}
\ee
and it is easy to see that if we choose
$
U^{\mu;\nu}_{2} = - {i\over 2}~W^{\{\mu,\nu\}}_{1}
$
then we obtain a new expression (\ref{A1-2}) for the anomaly 
$
A_{1}
$
where the sum goes only up to 
$
k = 1.
$
Again we impose the symmetry property (\ref{sA1-2}) and consider only the terms with one derivative on 
$\delta$;
it easily follows that
$
W^{\mu}_{1} = 0
$
i.e. the expression (\ref{A1-2}) has the form from the statement.

(ii) From (\ref{genericA}) we have:
\be
A^{\mu}_{2}(x_{1},x_{2})
= \sum_{k \leq 3} 
\partial_{\rho_{1}} \dots \partial_{\rho_{k}} \delta(x_{2} - x_{1})
W^{\mu;\{\rho_{1},\dots,\rho_{k}\}}_{2}(x_{1})
\label{A2-2}
\ee
and we have the restrictions 
\be
\omega(W^{\mu;\{\rho_{1},\dots,\rho_{k}\}}_{2}) \leq 5 - k, \quad
gh(W^{\mu;\{\rho_{1},\dots,\rho_{k}\}}_{2}) = 2
\ee
for all 
$
k = 0, \dots, 3.
$
We use Wess-Zumino consistency condition (\ref{WZ1-2}); if we consider only the terms with four derivatives on 
$\delta$
we obtain that the completely symmetric part of
$
W^{\mu;\{\nu,\rho,\sigma\}}_{2}
$
is null:
$
W^{\{\mu;\nu,\rho,\sigma\}}_{2} = 0.
$
In this case it is easy to prove that one can write 
$
W^{\mu;\{\nu,\rho,\sigma\}}_{2}
$
in the following form:
\be
W^{\mu;\{\nu,\rho,\sigma\}}_{2} = 
{1 \over 3}~(\tilde{W}^{[\mu\nu];\{\rho\sigma\}}_{2}
+ \tilde{W}^{[\mu\rho];\{\nu\sigma\}}_{2}
+ \tilde{W}^{[\mu\sigma];\{\nu\rho\}}_{2})
\ee 
with 
\be
\tilde{W}^{[\mu\nu];\{\rho\sigma\}}_{2} \equiv 
{3 \over 4}~W^{\mu;\{\nu,\rho,\sigma\}}_{2} - (\mu \leftrightarrow \nu).
\ee

We perform the finite renormalization
\be
T(T^{[\mu\nu]}(x_{1}),T(x_{2})) \rightarrow T(T^{[\mu\nu]}(x_{1}),T(x_{2})) 
+ \partial_{\rho}\partial_{\sigma}\delta(x_{2} - x_{1})
~U_{3}^{[\mu\nu];\{\rho\sigma\}}(x_{1})
\label{R3a-2}
\ee
with
$
U_{3}^{[\mu\nu];\{\rho\sigma\}} = - i~\tilde{W}^{[\mu\nu];\{\rho\sigma\}}_{2}
$
and we eliminate the contributions corresponding to 
$k = 3$
from (\ref{A2-2}). Now we consider the contribution corresponding to
$k = 2$; 
again we use the Wess-Zumino consistency condition (\ref{WZ1-2}); if we consider only the terms with three derivatives on 
$\delta$
we obtain that the completely symmetric part of
$
W^{\mu;\{\nu,\rho\}}_{2}
$
is null
$
W^{\{\mu;\nu,\rho\}}_{2} = 0
$
and write:
\be
W_{2}^{\mu;\{\nu\rho\}} = 
{1\over 2}~(\tilde{W}_{2}^{[\mu\nu];\rho} + \tilde{W}_{2}^{[\mu\rho];\nu})
\ee
with 
\be
\tilde{W}_{2}^{[\mu\nu];\rho} = {2\over 3}~W_{2}^{\mu;\{\nu\rho\}} 
- (\mu \leftrightarrow \nu).
\ee

Now we consider the finite renormalization
\be
T(T^{[\mu\nu]}(x_{1}),T(x_{2})) \rightarrow T(T^{[\mu\nu]}(x_{1}),T(x_{2})) 
+ \partial_{\rho}\delta(x_{2} - x_{1})~U_{3}^{[\mu\nu];\rho}(x_{1})
\label{R3b-2}
\ee
with 
$
U_{3}^{[\mu\nu];\rho} = i~\tilde{W}_{2}^{[\mu\nu];\rho} 
$
and we get a new expressions (\ref{A2-2}) for which
$
W_{2}^{\mu;\{\nu\rho\}} = 0,
$
i.e. the summation in (\ref{A2-2}) goes only up to
$
k =1.
$
It is time again to use the Wess-Zumino equation (\ref{WZ1-2}); if we consider only the terms with two derivatives on 
$\delta$
we obtain that the completely symmetric part of
$
W^{\mu;\nu}_{2}
$
is null i.e.
$
W^{\mu;\nu}_{2} = W^{[\mu;\nu]}_{2}.
$
Now we consider the finite renormalizations
\be
T(T^{[\mu\nu]}(x_{1}),T(x_{2})) \rightarrow T(T^{[\mu\nu]}(x_{1}),T(x_{2})) 
+ \delta(x_{2} - x_{1})~U_{3}^{[\mu\nu]}(x_{1})
\label{R3c-2}
\ee
with
$
U_{3}^{[\mu\nu]} = - i~W_{2}^{[\mu;\nu]}
$
we will get a new expression (\ref{A2-2}) with only the contributions
$k = 0$
i.e. the expression (\ref{A2-2}) has the form from the statement.

It is easy to prove that the Wess-Zumino equation (\ref{WZ1-2}) is now equivalent to:
\be
d_{Q}W_{1} = - i~\partial_{\mu}W^{\mu}_{2}.
\label{WZ1-2'}
\ee

(iii) From (\ref{genericA}) we have:
\be
A^{[\mu\nu]}_{3}(x_{1},x_{2})
= \sum_{k \leq 2} 
\partial_{\rho_{1}} \dots \partial_{\rho_{k}} \delta(x_{2} - x_{1})
W^{[\mu\nu];\{\rho_{1},\dots,\rho_{k}\}}_{3}(x_{1})
\label{A3-2}
\ee
and we have the restrictions 
\be
\omega(W^{[\mu\nu];\{\rho_{1},\dots,\rho_{k}\}}_{3}) \leq 5 - k,\quad
gh(W^{[\mu\nu];\{\rho_{1},\dots,\rho_{k}\}}_{3}) = 3
\ee
for all 
$
k = 0, 1, 2.
$

We perform the finite renormalization
\be
T(T^{[\mu\nu]}(x_{1}),T^{\rho}(x_{2})) \rightarrow T(T^{[\mu\nu]}(x_{1}),T^{\rho}(x_{2})) 
+ \partial_{\sigma}\delta(x_{2} - x_{1})~U_{5}^{[\mu\nu];\rho;\sigma}(x_{1})
\label{R5a-2}
\ee
with
$
U_{5}^{[\mu\nu];\rho;\sigma} = i~W^{[\mu\nu];\{\rho\sigma\}}_{3}
$
and we eliminate the contributions corresponding to 
$k = 2$
from (\ref{A3-2}). Now we consider the finite renormalization
\be
T(T^{[\mu\nu]}(x_{1}),T^{\rho}(x_{2})) \rightarrow T(T^{[\mu\nu]}(x_{1}),T^{\rho}(x_{2})) 
+ \delta(x_{2} - x_{1})~U_{3}^{[\mu\nu];\rho}(x_{1})
\label{R5b-2}
\ee
with
$
U_{5}^{[\mu\nu];\rho} = i~W_{3}^{[\mu\nu];\rho} 
$
and we get a new expressions (\ref{A3-2}) with only the contributions
$k = 0$
i.e. the expression (\ref{A3-2}) has the form from the statement.

(iv) From (\ref{genericA}) we have:
\be
A^{\mu;\nu}_{4}(x_{1},x_{2})
= \sum_{k \leq 2} 
\partial_{\rho_{1}} \dots \partial_{\rho_{k}} \delta(x_{2} - x_{1})
W^{\mu;\nu;\{\rho_{1},\dots,\rho_{k}\}}_{4}(x_{1})
\label{A4-2}
\ee
and we have the restrictions 
\be
\omega(W^{\mu;\nu;\{\rho_{1},\dots,\rho_{k}\}}_{4}) \leq 5 - k, \quad
gh(W^{\mu;\nu;\{\rho_{1},\dots,\rho_{k}\}}_{4}) = 3
\ee
for all 
$
k = 0, 1, 2.
$

We will have to consider the (anti)symmetry (\ref{s4'-2}). From the terms with two derivatives on delta we obtain that
$
W^{\mu;\nu;\{\rho,\sigma\}}_{4}
$
is antisymmetric in the first two indices i.e. we have the writing
$
W^{\mu;\nu;\{\rho,\sigma\}}_{4} = W^{[\mu\nu];\{\rho\sigma\}}_{4}.
$

Next we consider the Wess-Zumino consistency condition (\ref{WZ2-2}). From the terms with three derivatives on delta we obtain
\be
W^{[\mu\nu];\{\rho\sigma\}}_{4} + W^{[\mu\rho];\{\sigma\nu\}}_{4} + W^{[\mu\sigma];\{\nu\rho\}}_{4} = 0.
\ee

We note now that in the finite renormalization (\ref{R5a-2}) we have used only the expression
$
U_{5}^{[\mu\nu];\{\rho;\sigma\}}
$
i.e.
$
U_{5}^{[\mu\nu];[\rho;\sigma]}
$
is still available. It is not so complicated to prove (using the preceding relation) that the choice:
$
U_{5}^{[\mu\nu];[\rho;\sigma]} = {i\over 4}~
(W^{[\mu\rho];\{\nu\sigma\}}_{4} - W^{[\nu\rho];\{\mu\sigma\}}_{4} 
- W^{[\mu\sigma];\{\nu\rho\}}_{4} + W^{[\nu\sigma];\{\mu\rho\}}_{4}) 
$
is possible i.e. it verifies the (anti)symmetry properties; moreover after this finite renormalization we get a new expression (\ref{A4-2}) for which the term corresponding to
$k = 2$
is absent. We can enforce now the (anti)symmetry property (\ref{s4'-2}): it is equivalent to:
\bea
W^{\mu;\nu;\rho}_{4} = W^{\nu;\mu;\rho}_{4}
\nonumber \\
W^{\mu;\nu}_{4} + W^{\nu;\mu}_{4} + \partial_{\rho}W^{\nu;\mu;\rho}_{4} = 0.
\eea

We also make explicit the Wess-Zumino consistency condition (\ref{WZ2-2}); it is:
\bea
d_{Q}W^{\mu}_{2} = i~\partial_{\nu}W_{3}^{[\mu\nu]}
\nonumber \\
W^{\mu;\nu}_{4} = - W^{[\mu\nu]}_{3}
\nonumber \\
W^{\mu;\nu;\rho}_{4} = - W^{\mu;\rho;\nu}_{4}.
\eea

We note immediately that we have 
$
W^{\mu;\nu;\rho}_{4} = 0
$
i.e. the expression (\ref{A4-2}) has the form from the statement. We are left from (\ref{WZ2-2}) only with 
\be
d_{Q}W^{\mu}_{2} = i~\partial_{\nu}W_{3}^{[\mu\nu]}.
\ee

(v) From (\ref{genericA}) we have:
\be
A^{[\mu\nu];\rho}_{5}(x_{1},x_{2})
= \delta(x_{2} - x_{1}) W^{[\mu\nu]}_{5}(x_{1})
+ \partial_{\sigma}\delta(x_{2} - x_{1}) W^{[\mu\nu];\rho;\sigma}_{5}(x_{1})
\label{A5-2}
\ee
and we have the restrictions 
\bea
\omega(W^{[\mu\nu];\rho}_{5})  \leq 5,\qquad \omega(W^{[\mu\nu];\rho\sigma}_{5}) \leq 4
\nonumber \\
gh(W^{[\mu\nu];\rho}_{5}) = gh(W^{[\mu\nu];\rho;\sigma}_{5}) = 4.
\eea

We consider the Wess-Zumino consistency conditions (\ref{WZ3-2}). From the terms with two derivatives on delta we obtain:
\be
W^{[\mu\nu];\rho;\sigma}_{5} = - W^{[\mu\nu];\sigma;\rho}_{5} 
\ee
i.e. we have the writing
$
W^{[\mu\nu];\rho;\sigma]}_{5} = W^{[\mu\nu];[\rho\sigma]}_{5}.
$
From the Wess-Zumino consistency conditions (\ref{WZ4-2}) we consider again the terms with two derivatives on delta and we obtain after some computations:
\be
W^{[\mu\nu];[\rho\sigma]}_{5} = W^{[\rho\sigma];[\mu\nu]}_{5}. 
\ee

We now make the finite renormalization
\be
T(T^{[\mu\nu]}(x_{1}),T^{[\rho\sigma]}(x_{2})) \rightarrow T(T^{[\mu\nu]}(x_{1}),T^{[\rho\sigma]}(x_{2})) 
+ \delta(x_{1} - x_{2})~U_{6}^{[\mu\nu];[\rho\sigma]}(x_{1})
\label{R6-2}
\ee
with
$
U_{6}^{[\mu\nu];[\rho\sigma]} = i~W^{[\mu\nu];[\rho\sigma]}_{5}
$
and we eliminate the second contributions from (\ref{A5-2}). The Wess-Zumino consistency conditions (\ref{WZ3-2}) becomes equivalent to
\bea
d_{Q}W_{3}^{[\mu\nu]} = 0
\nonumber \\
W_{5}^{[\mu\nu];\rho} = 0.
\eea

In particular we have
\be
A_{5}^{[\mu\nu];\rho} = 0.
\ee
and from (\ref{WZ3-2}) we are left with:
\be
d_{Q}W_{3}^{[\mu\nu]} = 0.
\ee

The Wess-Zumino consistency conditions (\ref{WZ4-2}) is equivalent to
\be
d_{Q}W_{4}^{\mu;\nu} = 0
\ee
which follows from the preceding relation if we remember the connection between 
$
W_{3}^{[\mu\nu]}
$
and
$
W_{4}^{\mu;\nu}
$
obtained at (iv).

(vi) From (\ref{genericA}) we have:
\be
A^{[\mu\nu];[\rho\sigma]}_{6}(x_{1},x_{2})
= \delta(x_{1} - x_{2}) W^{[\mu\nu];[\rho\sigma]}_{6}(x_{1})
\label{A6-2}
\ee
and we have the restrictions 
\be
\omega(W^{[\mu\nu];[\rho\sigma]}_{6}) \leq 5
\qquad
gh(W^{[\mu\nu];[\rho\sigma]}_{6}) = 5.
\ee

From the symmetry property (\ref{s6'-2}) we also have
\be
W^{[\mu\nu];[\rho\sigma]}_{6} = W^{[\rho\sigma];[\mu\nu]}_{6}.
\ee

However from the Wess-Zumino consistency condition (\ref{WZ6-2}) we have
\be
W^{[\mu\nu];[\rho\sigma]}_{6} = 0
\ee
so in fact:
\be
A^{[\mu\nu];[\rho\sigma]}_{6} = 0.
\ee

(vii) Finally we observe that we can make some redefinitions of the chronological products without changing the structure of the anomalies. Indeed we have
\be
T(T(x_{1}),T(x_{2})) \rightarrow T(T(x_{1}),T(x_{2})) + \delta(x_{1} - x_{2})~B(x_{1})
\ee
which makes
\be
W \rightarrow W + d_{Q}B 
\ee
and
\be
T(T^{\mu}(x_{1}),T(x_{2})) \rightarrow T(T^{\mu}(x_{1}),T(x_{2})) 
+ \delta(x_{1} - x_{2})~B^{\mu}(x_{1})
\ee
which makes
\be
W \rightarrow W + i~\partial_{\mu}B^{\mu}, \qquad 
W^{\mu} \rightarrow W^{\mu} + d_{Q}B^{\mu}.
\ee
We also observe that we can consider the finite renormalizations (\ref{R3c-2}) and
\be
T(T^{\mu}(x_{1}),T^{\nu}(x_{2})) \rightarrow T(T^{\mu}(x_{1}),T^{\nu}(x_{2})) 
+ \delta(x_{2} - x_{1})~U_{4}^{[\mu\nu]}(x_{1})
\label{R4b-2}
\ee
such that the we have the (anti)symmetry property (\ref{symmetryT}). If we take 
\be
U_{3}^{[\mu\nu]} = B^{[\mu\nu]}, \qquad U_{4}^{[\mu\nu]} = - B^{[\mu\nu]}
\ee
we have the redefinitions
\be
W^{\mu} \rightarrow W^{\mu} + i~\partial_{\nu}B^{[\mu\nu]}, \qquad 
W^{[\mu\nu]} \rightarrow W^{[\mu\nu]} + d_{Q}B^{[\mu\nu]}.
\ee
All these redefinitions do not modify the form of the anomalies from the statement and we have obtained the last assertion of the theorem.
$\qed$

As we can see one can simplify considerably the form of the anomalies if one makes convenient redefinitions of the chronological products. Moreover, the result is of purely cohomological nature i.e. we did not use the explicit form of the expressions
$
T, T^{\mu}, T^{[\mu\nu]}.
$
The main difficulty of the proof is to find a convenient way of using Wess-Zumino equations, the (anti)symmetry properties and a succession of finite renormalizations. It is a remarkable fact that the preceding result stays true for arbitrary order of the perturbation theory i.e. we have:
\begin{thm}
Suppose that we have gauge invariance up to the order $n - 1$ of the perturbation theory. Then, by convenient redefinitions of the chronological products, the anomalies from the equations (\ref{ym1}) - (\ref{ym9}) can be taken of the form: 
\bea
A_{1}(X) = \delta(X)~W(x_{1}), \qquad
A^{\mu}_{2}(X) = \delta(X)~W^{\mu}(x_{1})
\nonumber \\
A^{[\mu\nu]}_{3}(X) = \delta(X)~W^{[\mu\nu]}(x_{1}), \qquad
A^{\mu;\nu}_{4}(X) = - \delta(X)~W^{[\mu\nu]}(x_{1}),
\nonumber \\
A^{\dots}_{j}(X) = 0, \qquad j = 5, \dots,9.
\eea

The expressions
$
W, W^{\mu}
$
and
$
W^{[\mu\nu]}
$
are relative cocyles and are determined up to relative co-boundaries.
\end{thm}

{\bf Proof:} For the sake of completeness we provide a minimum number of details for the first anomaly
$
A_{1}.
$
From (\ref{genericA}) we have 
\bea
A_{1}(X) = \sum_{2 \leq l \leq n} 
\partial_{\mu}^{l}\partial_{\nu}^{l}\partial_{\rho}^{l}\partial_{\sigma}^{l}\delta(X)
W^{\{\mu\nu\rho\sigma\}}_{1}(x_{1})
+ \sum_{2 \leq k \not= l  \leq n} 
\partial_{\mu}^{l}\partial_{\nu}^{l}\partial_{\rho}^{l}\partial_{\sigma}^{k}\delta(X)
W^{\{\mu\nu\rho\};\sigma}_{1}(x_{1})
\nonumber \\
+ \sum_{2 \leq k < l \leq n} 
\partial_{\mu}^{k}\partial_{\nu}^{k}\partial_{\rho}^{l}\partial_{\sigma}^{l}\delta(X)
W^{\{\mu\nu\};\{\rho\sigma\}}_{1}(x_{1})
+ \cdots
\eea
where by $\cdots$ we mean the terms with three or less derivatives on the delta function and the symmetry property (\ref{sA1}) is true if we put some supplementary restrictions on the preceding expression. We perform the finite renormalization:
\bea
T(T^{\mu}(x_{1}),T(x_{2}),\dots,T(x_{n})) \rightarrow T(T^{\mu}(x_{1}),T(x_{2}),\dots,T(x_{n})) 
\nonumber \\
+ \sum_{2 \leq l \leq n} 
\partial_{\nu}^{l}\partial_{\rho}^{l}\partial_{\sigma}^{l}\delta(X)
U^{\mu;\{\nu\rho\sigma\}}_{21}(x_{1})
+ \sum_{2 \leq k \not= l \leq n} 
\partial_{\nu}^{k}\partial_{\rho}^{k}\partial_{\sigma}^{l}\delta(X)
U^{\mu;\{\nu\rho\};\sigma}_{22}(x_{1})
\label{R2a-n}
\eea
and if we choose it conveniently we can obtain a new expression (\ref{A1-2}) for the anomaly 
$
A_{1}
$
without terms with four derivatives on delta, i.e.
\be
A_{1}(X) = \sum_{2 \leq l \leq n}
\partial_{\mu}^{l}\partial_{\nu}^{l}\partial_{\rho}^{l}\delta(X)
W^{\{\mu\nu\rho\}}_{1}(x_{1})
+ \sum_{2 \leq k \not= l  \leq n} 
\partial_{\mu}^{k}\partial_{\nu}^{k}\partial_{\rho}^{l}\delta(X)
W^{\{\mu\nu\};\rho}_{1}(x_{1})
+ \cdots
\ee
where by $\cdots$ we mean the terms with two or less derivatives on the delta function. We impose the symmetry property (\ref{sA1}) and we can perform a finite renormalization:
\be
T(T^{\mu}(x_{1}),T(x_{2}),\dots,T(x_{n})) \rightarrow T(T^{\mu}(x_{1}),T(x_{2}),\dots,T(x_{n})) 
+ \sum_{2 \leq l \leq n} 
\partial_{\nu}^{l}\partial_{\rho}^{l}\delta(X)
U^{\mu;\{\nu\rho\}}_{2}(x_{1})
\label{R2b-n}
\ee
such that we eliminate the terms with three derivatives on delta, i.e.
\be
A_{1}(X) = \sum_{2 \leq l \leq n}
\partial_{\mu}^{l}\partial_{\nu}^{l}\delta(X) W^{\{\mu\nu\}}_{1}(x_{1})
+ \sum_{2 \leq k \not= l  \leq n} 
\partial_{\mu}^{k}\partial_{\nu}^{l}\delta(X) W^{\mu;\nu}_{1}(x_{1})
+ \cdots
\ee
where by $\cdots$ we mean the terms with one or no derivatives on the delta function.

Finally we perform a convenient finite renormalization:
\be
T(T^{\mu}(x_{1}),T(x_{2}),\dots,T(x_{n})) \rightarrow T(T^{\mu}(x_{1}),T(x_{2}),\dots,T(x_{n})) 
+ \sum_{l = 2}^{n} \partial_{\nu}^{l}\delta(X) U^{\mu;\nu}_{2}(x_{1})
\label{R2c-n}
\ee
and we get an expression for 
$
A_{1}
$
as in the statement of the theorem.
Proceeding in the same we arrive after some non-trivial combinatorics at the result from the statement for all anomalies.
$\qed$

We have proved that renormalization of gauge theories leads to some descent equations. 
We have the expressions 
$
T^{I}
$
and 
$
R^{I}
$
(with ghost numbers 
$
gh(T^{I}) = gh(R^{I}) = |I|
$ 
and canonical dimension $\leq 4$) for the interaction Lagrangian and the finite renormalizations compatible with gauge invariance; we also have the expressions 
$
W^{I}
$
(with ghost numbers 
$
gh(W^{I}) = |I| + 1
$ 
and canonical dimension $\leq 5$) for the anomalies. In the next Sections we give the most simpler way to solve in general such type of problems.
\newpage
\section{A Geometric Setting for the Gauge Invariance Problem\label{geom}}

The cohomology of the operator
$
d_{Q}
$
can be reformulated in the language of classical field theory (with Grassmann 
variables). 

The kinematical structure of a classical field theory is based on fibered 
bundle structures. Let
$
\pi: Y \mapsto X
$
be fiber bundle, where
$X$
and
$Y$
are differentiable manifolds of dimensions 
$
dim(X) = n,\quad dim(Y) = m + n
$
and
$\pi$
is the canonical projection of the fibration. Usually $X$ is interpreted as the ``space-time" manifold and the fibers of $Y$ as the field variables. An {\it adapted chart} to the fiber bundle structure is a couple
$
(V,\psi)
$
where $V$ is an open subset of $Y$ and 
$
\psi: V \rightarrow \R^{n} \times \R^{m}
$
is the so-called {\it chart map}, usually written as
$
\psi = (x^{\mu},y^{\alpha}) \quad (\mu = 1,...,n;~\alpha = 1,...,m)
$
such that
$
(\pi(V), \phi)
$
where
$
\phi = (x^{\mu}) \quad (\mu = 1,...,n)
$
is a chart on $X$ and the canonical projection has the following expression:
$
\pi(x^{\mu},y^{\alpha}) = (x^{\mu}).
$
If 
$p \in Y$
then the real numbers
$x^{\mu}(p), \quad y^{\alpha}(p)$
are called the {\it (fibered) coordinates} of $p$. For simplicity we will give up the attribute {\it adapted} in the following. Also we will refer frequently to the first entry $V$ 
of 
$
(V,\psi)
$ 
as a chart.

Next, one considers the $r$-{\it jet bundle extensions}
$
J^{r}_{n}Y \mapsto X \quad (r \in \N).
$
The construction is the following (see for instance \cite{jet}). 

\begin{thm}
Let 
$
x \in X,
$
and
$
y \in \pi^{-1}(x).
$
We denote by
$
\Gamma_{(x,y)}
$
the set of sections
$
\gamma: U \rightarrow Y
$
such that: (i) $U$ is a neighborhood of $x$; (ii) 
$
\gamma(x) = y.
$
We define on 
$
\Gamma_{(x,y)}
$
the relationship
``$
\gamma \sim \delta
$"
{\it iff} there exists a chart
$
(V,\psi)
$
on $Y$ such that 
$\gamma$ 
and 
$\delta$ 
have the same partial derivatives up to order $r$ in the given chart i.e.
\be
{\partial^{k} \over \partial x^{\mu_{1}}...\partial x^{\mu_{k}}} 
\psi \circ \gamma \circ \phi^{-1} (\phi(x)) =
{\partial^{k} \over \partial x^{\mu_{1}}...\partial x^{\mu_{k}}} 
\psi \circ \delta \circ \phi^{-1} (\phi(x)), 
\quad k \leq r.
\ee

Then this relationship is chart independent and it is an equivalence relation.
\label{equivalence}
\end{thm}

A $r$-{\it order jet with source} $x$ {\it and target} $y$
is, by definition, the equivalence class of some section
$\gamma$
with respect to the equivalence relationship defined above and it is 
denoted by
$
j^{r}_{x}\gamma.
$

Let us define
$
J^{r}_{(x.y)}\pi \equiv \Gamma_{(x,y)} / \sim
$
Then the
$r$-{\it order jet bundle extension} is, set theoretically
$
J^{r}Y \equiv \bigcup_{x} J^{r}_{(x,y)}\pi.
$
Let
$
(V,\psi), \quad \psi = (x^{\mu},y^{\sigma})
$
be a chart on $Y$. Then we define the couple
$
(V^{r},\psi^{r}),
$
where:
$
V^{r} = (\pi^{r,0})^{-1}(V)
$
and
\be
\psi = (x^{\mu},y^{\alpha},y^{\alpha}_{\mu},...,y^{\alpha}_{\mu_{1},...,\mu_{k}},...,
y^{\alpha}_{\mu_{1},...,\mu_{r}}), \quad j_{1} \leq j_{2} \leq \cdots \leq j_{k}, 
\quad k = 1,...,r
\ee
where
\bea
y^{\alpha}_{\mu_{1},...,\mu_{k}}(j^{r}_{x}\gamma) = 
\left. {\partial^{k} \over \partial x^{\mu_{1}} \cdots \partial x^{\mu_{k}}} 
y^{\alpha} \circ \gamma \circ \phi^{-1} \right|_{\phi(x)}, \quad k = 1,...,r
\nonumber \\
x^{\mu}(j^{r}_{x}\gamma) = x^{\mu}(x), \quad 
y^{\alpha}(j^{r}_{x}\gamma) = y^{\alpha}(\gamma(x)).
\eea

Then
$
(V^{r},\psi^{r})
$
is a chart on
$
J^{r}Y
$
called {\it the associated chart} of
$
(V,\psi).
$

\begin{rem}
The expressions 
$
y^{\alpha}_{\mu_{1},...,\mu_{k}}(j^{r}_{x}\gamma)
$
are defined for {\it all} indices
$
\mu_{1},...,\mu_{k} = 1,...,n,
$
and the restrictions
$
j_{1} \leq j_{2} \leq \cdots \leq j_{k}
$
in the definition of the charts are in order to avoid over-counting and are a result of the obvious symmetry property:
\be
y^{\alpha}_{\mu_{P(1)},...,\mu_{P(k)}}(j^{r}_{x}\gamma) =
y^{\alpha}_{\mu_{1},...,\mu_{k}}(j^{r}_{x}\gamma), 
\label{sym}
\ee
for any permutation
$
P \in {\cal P}_{k}, \quad k = 2,...,r.
$
\end{rem}

Now we have the following result.

\begin{thm}
If a collection of (adapted) charts
$
(V,\psi)
$
are the elements of a differentiable atlas on $Y$ then 
$
(V^{r},\psi^{r})
$
are the elements of a differentiable atlas on
$
J^{r}_{n}(Y)
$
which admits a fiber bundle structure over $Y$.
\label{smooth}
\end{thm}

To be able to use the summation convention over the dummy indices we consider
$
y^{\alpha}_{\mu_{1},...,\mu_{k}}
$
for {\it all} values of the indices
$
\mu_{1},...,\mu_{k} \in \{ 1,...,n\}
$
as smooth functions on the chart
$
V^{r}
$
defined in terms of the independent variables
$
y^{\alpha}_{\mu_{1},...,\mu_{k}}, \mu_{1} \leq \mu_{2} \leq ...\leq \mu_{k}
\quad k = 1,2,...,r
$ 
according to the formula (\ref{sym}) and we make a similar convention for the partial derivatives
$
{\partial \over \partial y^{\alpha}_{\mu_{1},...,\mu_{k}}}.
$

Then we define on the chart
$
V^{r}
$
the following vector fields:
\be
\partial^{\mu_{1},...,\mu_{k}}_{\alpha} \equiv {r_{1}!...r_{n}! \over k!} 
{\partial \over \partial y^{\alpha}_{\mu_{1},...,\mu_{k}}}, \quad k = 1,...,r
\label{partial}
\ee
for all values of the indices
$
\mu_{1},...,\mu_{k} \in \{ 1,...,n\}.
$
Here
$
r_{l}, \quad l = 1,...,n
$
is the number of times the index $l$ enters into the set
$
\{ \mu_{1},...,\mu_{k}\}.
$

One can easily verify the following formulas:
\be
\partial^{\mu_{1},...,\mu_{k}}_{\beta} y^{\alpha}_{\nu_{1},...,\nu_{l}} = 0, \quad
(k \not= l)
\ee
\be
\partial^{\mu_{1},...,\mu_{k}}_{\beta} y^{\alpha}_{\nu_{1},...,\nu_{k}}
= \delta^{\alpha}_{\beta}~{\cal S}^{+}_{\mu_{1},...,\mu_{k}}
\delta^{\mu_{1}}_{\nu_{1}} \cdots \delta^{\mu_{k}}_{\nu_{k}}
\ee
where
$
{\cal S}^{+}_{j_{1},...,j_{k}}
$
is the symmetrization projector operator in the indices
$
\mu_{1},...,\mu_{k}.
$

Also we have for any smooth function $f$ on the chart 
$
V^{r}:
$ 
\be
df = {\partial f \over \partial x^{\mu}} dx^{\mu} + \sum_{k=0}^{r} 
(\partial^{\mu_{1},...,\mu_{k}}_{\alpha}f) dy^{\alpha}_{\mu_{1},...,\mu_{k}} = 
{\partial f \over \partial x^{\mu}} dx^{\mu} + \sum_{|J| \leq r} 
(\partial^{J}_{\alpha}f) dy^{\alpha}_{J}. 
\label{df}
\ee
In the last formula we have introduced the multi-index notations in an obvious way. This formula also shows that the coefficients appearing in the definition (\ref{partial}) are exactly what is needed to use the summation convention over the dummy indices without over-counting.

We now define the expressions
\be
d_{\rho}^{r} \equiv {\partial \over \partial x^{\rho}} + \sum_{k=0}^{r-1} 
y^{\alpha}_{\rho,\mu_{1},...,\mu_{k}} \partial^{\mu_{1},...,\mu_{k}}_{\alpha}
\label{formal}
\ee
called {\it formal derivatives}. When it is no danger of confusion we denote simply 
$
d_{\mu} = d_{\mu}^{r}.
$

\begin{rem}
The formal derivatives are not vector fields on 
$
J^{r}Y.
$
\end{rem}

Next one immediately sees that
\be
d_{\mu} y^{\alpha}_{\nu_{1},...,\nu_{k}} = y^{\alpha}_{\mu,\nu_{1},...,\nu_{k}}, 
\quad k = 0,...,r-1.
\ee

From the definition of the formal derivatives it easily follows by direct computation that:
\be
\left[ \partial^{\mu_{1},...,\mu_{k}}_{\alpha} , d_{\rho} \right] = 
{1\over k} \sum_{l=1}^{k} \delta^{\mu_{l}}_{\rho} 
\partial^{\mu_{1},...,\hat{\mu}_{l},...,\mu_{k}}_{\alpha}, \quad k = 0,...,r
\label{commutator}
\ee
where we use Bourbaki conventions
$
\sum_{\emptyset} \equiv 0, \quad \prod_{\emptyset} \equiv 1.
$

The formalism presented above extends easily to the Grassmann case. We denote by
$
\epsilon_{\alpha}
$
the Grassmann parity of the variable 
$
y^{\alpha}.
$
We only have to replace commutators with graded commutators and distinguish between left and right derivatives; we will consider here only left derivatives. Then we can interpret equation
\be
d_{Q}R = 0
\ee
as an equation in classical field theory where we also suppose that the polynomials are restricted to the mass shell and we replace the derivative $\partial^{\mu}$ by $d^{\mu}$.

A final word about the notations. Because
$
y^{\alpha}_{\mu_{1}\dots\mu_{n}} = d_{\mu_{1}}\dots d_{\mu_{n}}y^{\alpha}
$
we freely use both notations. When the index 
$
\alpha
$
are downstairs
we write
$
y_{\alpha;\mu_{1}\dots\mu_{n}}.
$

We now prove a sort of Poincar\'e lemma adapted to our conditions. There are two obstacles in applying the usual Poincar\'e lemma: first our co-cycles are polynomials and second we are working on the mass shell. If only the first obstacle would be present then we could apply the so-called algebraic Poincar\'e lemma \cite{Dr}, but unfortunately this nice result breaks down if we work on shell. We make the assumption that we are on the mass shell
because the Epstein-Glaser construction is done from the very beginning in a Fock space of some free particles. We will prove below that the obstacles to the Poincar\'e lemma are easy to describe. Basically we want to find the general solution of equations of the type:
\be
d_{\mu}S^{I;\mu} = 0.
\label{div-free}
\ee
There are some trivial solutions of this equation namely of this equation
namely of the type
\be
S^{I;\mu} = d_{\nu}S^{I;\mu\nu}
\label{rotor}
\ee
where the expression
$
S^{I;\mu\nu}
$
is antisymmetric in the last two indices. We will be able to describe the
obstruction relevant to this equation i.e. solutions which are not trivial. We start first with:
\begin{prop}
Let the expression
$
S^{I;\mu}
$
be of canonical dimension
$
\omega(S^{I;\mu}) = 2
$
and verifying the relation (\ref{div-free}). Then it is of the form
\be
S^{I;\mu} = c^{I}_{\alpha}~d^{\mu}y^{\alpha} + d_{\nu}S^{I;\mu\nu}
\ee
with the expression 
$
S^{I;\mu\nu}
$
antisymmetric in the last two indices.
\end{prop}

{\bf Proof:} The generic form for
$
S^{I;\mu}
$
is:
\be
S^{I;\mu} = {1\over 2}~\sum_{\alpha,\beta}~c^{I;\mu}_{\alpha\beta}~y^{\alpha}~y^{\beta} 
+ {\rm total~divergence}
\ee
where the expressions
$
c^{I;\mu}_{\alpha\beta}
$
are constants and we note that the second contribution is linear in the fields. Also we can impose
$
c^{I;\mu}_{\alpha\beta} = \epsilon_{\alpha}~\epsilon_{\beta}~c^{I;\mu}_{\beta\alpha}.
$

Now it is easy to prove that the condition (\ref{div-free}) gives
$
c^{I;\mu}_{\alpha\beta} = 0
$
so we have 
$
S^{I;\mu} = d_{\nu}S^{I;\mu\nu}
$
with
$
\omega(S^{I;\mu\nu}) = 1.
$
We split now the expression
$
S^{I;\mu\nu}
$
in the symmetric and the antisymmetric part in the indices $\mu$ and $\nu$
denoted by
$
S^{I;\mu\nu}_{\pm}.
$
The condition (\ref{div-free}) gives
$
d_{\mu}d_{\nu}S^{I;\mu\nu}_{+} = 0
$
so we necessarily have
$
S^{I;\mu\nu}_{+} = \eta^{\mu\nu}A^{I}; 
$
obviously we must have
$
A^{I} = c^{I}_{\alpha}~y^{\alpha}
$
and we obtain the expression from the statement.
$\qed$

The case
$
\omega = 3
$
is harder.
\begin{prop}
Let the expression
$
S^{I;\mu}
$
be of canonical dimension
$
\omega(S^{I;\mu}) = 3
$
and verifying the relation (\ref{div-free}). Then it is of the form
\be
S^{I;\mu} =  \sum_{\alpha,\beta}~c^{I}_{\alpha\beta}~y^{\alpha}~d^{\mu}y^{\beta} 
+ \sum_{\alpha}~c^{I\nu}_{\alpha}~d^{\mu} d_{\nu}y^{\alpha} 
+ d_{\nu}S^{I;\mu\nu}
\ee
with 
$
c^{I}_{\alpha\beta},~c^{I\nu}_{\alpha}
$
some constants, one has
$
c^{I}_{\alpha\beta} = - \epsilon_{\alpha}~\epsilon_{\beta}~c^{I}_{\beta\alpha}
$
and
$
S^{I;\mu\nu}
$
is antisymmetric in the last two indices.
\end{prop}

{\bf Proof:} From the equation (\ref{div-free}) we get with (\ref{commutator}):
\be
d_{\mu} {\partial S^{I;\mu} \over \partial y^{\alpha}} = 0
\ee
for any
$
y^{\alpha}.
$
So we can use the preceding proposition and find out
\be 
{\partial S^{I;\mu} \over \partial y^{\alpha}} =
\sum_{\beta}~c^{I}_{\alpha\beta}~d^{\mu}y^{\beta} + d_{\nu}S^{I;\mu\nu}_{\alpha}
\ee
with the last expression antisymmetric in $\mu$ and $\nu$. Here
$
c^{I}_{\alpha\beta}
$
are constants and
$
S^{I;\mu\nu}_{\alpha}
$
have canonical dimension $\omega = 1$ so we have the generic form:
\be
S^{I;\mu\nu}_{\alpha} = \sum_{\beta}~s^{I;\mu\nu}_{\alpha\beta}~y^{\beta}
\ee
where
$
s^{I;\mu\nu}_{\alpha\beta}
$
are constants and we have antisymmetry in $\mu$ and $\nu$. So we have:
\be 
{\partial S^{I;\mu} \over \partial y^{\alpha}} =
\sum_{\beta}~c^{I}_{\alpha\beta}~d^{\mu}y^{\beta} 
+ \sum_{\beta}~s^{I;\mu\nu}_{\alpha\beta}~d_{\nu}y^{\beta}
\ee
which can be integrated: 
\be 
S^{I;\mu} = \sum_{\alpha,\beta}~c^{I}_{\alpha\beta}~y^{\alpha}~d^{\mu}y^{\beta} 
+ \sum_{\alpha,\beta}~s^{I;\mu\nu}_{\alpha\beta}~y^{\alpha}~d_{\nu}y^{\beta}
+ S^{I;\mu}_{1}
\ee
where
$
S^{I;\mu}_{1}
$
depends only on derivatives i.e. is of the form:
\be 
S^{I;\mu}_{1} = \sum_{\alpha}~c^{I\mu\nu\rho}_{\alpha}~d_{\nu}~d_{\rho}y^{\alpha}
\ee
with 
$
c^{I\mu\nu\rho}_{\alpha}
$
some constants with symmetry in $\nu$ and $\rho$. Now we obtain from (\ref{div-free}) the following equations:
\bea
c^{I}_{\alpha\beta} = - \epsilon_{\alpha}~\epsilon_{\beta}~c^{I}_{\beta\alpha}
\nonumber \\
s^{I;\mu\nu}_{\alpha\beta} = \epsilon_{\alpha}~\epsilon_{\beta}~s^{I;\mu\nu}_{\beta\alpha}
\nonumber \\
c^{I\mu\nu\rho}_{\alpha} = a_{1}~\eta^{\nu\rho}~c^{I;\mu}_{\alpha}
+ {1\over 2}~a_{2}~(\eta^{\mu\nu}~d^{I;\rho}_{\alpha} + \eta^{\mu\rho}~d^{I;\nu}_{\alpha})
\eea
and we easily obtain the expression from the statement. 
$\qed$

Now we give the main result of this Section.
\begin{thm}
Let
$
S^{I;\mu}
$
be of canonical dimension
$
\omega(S^{I;\mu}) \geq 4
$ 
at least tri-linear in the fields (and derivatives) fulfilling the relation (\ref{div-free}). Then it is of the following
generic form:
\be
S^{I;\mu} = d_{\nu}S^{I;\mu\nu}
\ee
where the expression 
$
S^{I;\mu\nu}
$
is antisymmetric in $\mu,~\nu$ i.e. it gives a trivial contribution.
\end{thm}

{\bf Proof:} (i) We first consider the case 
$
\omega(S^{I;\mu}) = 4
$
and we have from (\ref{div-free})
\be
d_{\mu} \left({\partial S^{I;\mu} \over \partial y^{\alpha}} \right) = 0;
\ee
but the expression
$
{\partial S^{I;\mu} \over \partial y^{\alpha}}
$
has the canonical dimension $3$ so we can apply the preceding proposition and obtain:
\be
{\partial S^{I;\mu} \over \partial y^{\alpha}} =
\sum_{\beta,\gamma} c^{I}_{\alpha\beta\gamma}~y^{\beta}~d^{\mu}y^{\gamma}
+ d_{\nu}S^{I;\mu\nu}_{\alpha}
\ee
with the expressions 
$
S^{I;\mu\nu}_{\alpha}
$
antisymmetric in $\mu,~\nu$; the term 
$
\sim d^{\mu}~d^{\nu}y^{\alpha}
$ 
does not appear because we have supposed the expression 
$
S^{I;\mu} 
$
at least tri-linear in the fields. We also have the generic form:
\be
S^{I;\mu\nu}_{\alpha} = {1\over 2}~\sum_{\beta,\gamma} s^{I;\mu\nu}_{\alpha\beta\gamma}~y^{\beta}~y^{\gamma}
\ee
with
$
s^{I;\mu\nu}_{\alpha\beta\gamma}
$
some constants and 
\bea
c^{I}_{\alpha\beta\gamma} = - \epsilon_{\beta}~\epsilon_{\gamma}~c^{I}_{\alpha\gamma\beta}
\nonumber \\
s^{I;\mu\nu}_{\alpha\beta\gamma} = - s^{I;\nu\mu}_{\alpha\beta\gamma}
\nonumber \\
s^{I;\mu\nu}_{\alpha\beta\gamma} =  \epsilon_{\beta}~\epsilon_{\gamma}~s^{I;\mu\nu}_{\alpha\gamma\beta}.
\label{c1}
\eea
It follows that
\be
{\partial S^{I;\mu} \over \partial y^{\alpha}} =
\sum_{\beta,\gamma} c^{I}_{\alpha\beta\gamma}~y^{\beta}~d^{\mu}y^{\gamma}
+ \sum_{\beta,\gamma} s^{I;\mu\nu}_{\alpha\beta\gamma}~y^{\beta}~d_{\nu}y^{\gamma}
\label{s1}
\ee
We impose the condition 
\be
{\partial^{2} S^{I;\mu} \over \partial y^{\beta}\partial y^{\alpha}} =
\epsilon_{\alpha}~\epsilon_{\beta}~
{\partial^{2} S^{I;\mu} \over \partial y^{\alpha}\partial y^{\beta}}
\ee
and obtain:
\be
c^{I}_{\alpha\beta\gamma} = \epsilon_{\alpha}~\epsilon_{\beta}~c^{I}_{\beta\alpha\gamma},
\qquad
s^{I;\mu\nu}_{\alpha\beta\gamma} =  \epsilon_{\alpha}~\epsilon_{\beta}~s^{I;\mu\nu}_{\beta\alpha\gamma}.
\label{c2}
\ee
From the first relations of (\ref{c1}) and (\ref{c2}) we obtain 
\be
c^{I}_{\alpha\beta\gamma} = 0.
\ee
Using the second relation (\ref{c2}) we can integrate (\ref{s1}) and get:
\be
{\partial S^{I;\mu} \over \partial y^{\alpha}} =
{1\over 2}~\sum_{\alpha,\beta,\gamma} s^{I;\mu\nu}_{\alpha\beta\gamma}~y^{\alpha}~y^{\beta}~d^{\mu}y^{\gamma}
+ S^{I;\mu}_{1}
\ee
where 
$
S^{I;\mu}_{1}
$ 
depends only on derivatives so it is null (because it must be trilinear). Now we have from (\ref{c1}) and (\ref{c2}) that the expression
$
s^{I;\mu\nu}_{\alpha\beta\gamma}
$
is completely symmetric (in the graded sense) in the indices
$
\alpha, \beta, \gamma
$
so we can integrate the preceding relation:
\be
S^{I;\mu}
= {1\over 6}~\sum_{\alpha\beta,\gamma} s^{I;\mu\nu}_{\alpha\beta\gamma}~
d_{\nu}(y^{\alpha}~y^{\beta}~y^{\gamma})
\ee
i.e. we have the expression from the statement with 
\be
S^{I;\mu\nu}
= {1\over 6}~\sum_{\alpha\beta,\gamma}~ s^{I;\mu\nu}_{\alpha\beta\gamma}~y^{\alpha}~y^{\beta}~y^{\gamma}.
\ee

(ii) Now we consider the statement of the theorem valid for 
$
\omega(S^{I;\mu}) = 4,\dots,N~(N \geq 4)
$
and we have from (\ref{div-free})
\be
d_{\mu} \left({\partial S^{I;\mu} \over \partial y^{\alpha}} \right) = 0;
\ee
we can apply the induction hypothesis and get
\be
{\partial S^{I;\mu} \over \partial y^{\alpha}} = d_{\nu}S^{I;\mu\nu}_{\alpha}.
\label{S1}
\ee
the expression
$
S^{I;\mu\nu}_{\alpha}
$
is of maximal degree $N - 1$ in 
$
y^{\alpha}
$
so we have the generic form
\be
S^{I;\mu\nu}_{\alpha_{0}} =
\sum_{k=0}^{n}~{1\over k!}~s^{I;\mu\nu}_{\alpha_{0}\dots\alpha_{k}}
y^{\alpha_{1}}\cdots y^{\alpha_{n}}
\label{Smunu}
\ee
where the expression
$
s^{I;\mu\nu}_{\alpha_{0}\dots\alpha_{k}}
$
do not depend on
$
y^{\beta}
$
are antisymmetric in $\mu,~\nu$ and (graded) antisymmetric in
$
\alpha_{1},\dots,\alpha_{n};
$
moreover
$
n \leq N - 1
$
is the maximal degree in
$
y^{\beta}
$
and
$
\omega(s^{I;\mu\nu}_{\alpha_{0}\dots\alpha_{k}}) = N - 1 - k.
$
Let us also note that we must have
$
s^{I;\mu\nu}_{\alpha_{0}\dots\alpha_{k-1}} = 0
$
because this expression has canonical dimension $1$ according to the
preceding formula but it must have at least a factor
$
d^{\rho}y^{\beta}
$
which has canonical dimension grater than $2$. We have two cases:

(a)
$
n = N - 1.
$

In this case the expression
$
s^{I;\mu\nu}_{\alpha_{0}\dots\alpha_{n}}
$
are in fact constants. It is easy to prove from Frobenius condition of
integrability that this expression is completely antisymmetric
(in the graded sense) in all indices
$
\alpha_{0},\dots,\alpha_{n};
$
now we can integrate (\ref{S1}) with respect to the variables
$
y^{\beta}
$
and we have
\be
S^{I;\mu} = {1\over (N-1)!}~s^{I;\mu\nu}_{\alpha_{0}\dots\alpha_{N-1}}
y^{\alpha_{0}}\cdots y^{\alpha_{N-2}}~d_{\nu}y^{\alpha_{N-1}}
+ \cdots
\ee
where by $\cdots$ we mean terms of degree
$
< N -1
$
in
$
y^{\beta}.
$
From here
\be
S^{I;\mu} = {1\over N!}~d_{\nu}
(s^{I;\mu\nu}_{\alpha_{0}\dots\alpha_{N-1}}
y^{\alpha_{0}}\cdots y^{\alpha_{N-1}}) + \cdots
\ee

The first term is a trivial solution and can be eliminated. The new
$
S^{I;\mu}
$
will be of degree
$
< N - 1
$
in the variables
$
y^{\beta};
$
the new
$
S^{I;\mu}
$
verifies again (\ref{S1}) and (\ref{Smunu}) with
$
n = N - 3.
$

(b)
$
n \leq N - 3.
$

In this case Frobenius condition of integrability shows that the expression
$
d_{\nu}s^{I;\mu\nu}_{\alpha_{0}\dots\alpha_{n}}
$
is completely antisymmetric (in the graded sense) in all indices
$
\alpha_{0},\dots,\alpha_{n};
$
again we can integrate the system (\ref{S1}) and get
\be
S^{I;\mu} = {1\over (n+1)!}~(d_{\nu}s^{\mu\nu I}_{\alpha_{0}\dots\alpha_{n}})
y^{\alpha_{0}}\cdots y^{\alpha_{n}}
+ \cdots
\ee
where by $\cdots$ we mean terms of degree
$
< n -1
$
in
$
y^{\beta}.
$
From here
\be
S^{I;\mu} = {1\over (n+1)!}~d_{\nu}
(s^{I;\mu\nu}_{\alpha_{0}\dots\alpha_{n}}
y^{\alpha_{0}}\cdots y^{\alpha_{n}}) + \cdots
\ee

The first term is a trivial solution and can be eliminated. The new
$
S^{I;\mu}
$
will again verify (\ref{S1}) and (\ref{Smunu}). Because
$
s^{I;\mu\nu}_{\alpha_{0}\dots\alpha_{n-1}} = 0
$
we will now obtain from Frobenius condition of integrability
that the expression
$
s^{I;\mu\nu}_{\alpha_{0}\dots\alpha_{n}}
$
is completely antisymmetric (in the graded sense) in all indices
$
\alpha_{0},\dots,\alpha_{n}
$
and we can repeat the argument from case (a).  As a result we obtain a new
$
S^{I;\mu}
$
verifying (\ref{S1}) and (\ref{Smunu}) with
$
n \rightarrow n - 1.
$

(iii) By recursion we end up with an expressions
$
S^{I;\mu}_{\alpha}
$
and
$
S^{I;\mu\nu}_{\alpha}
$
independent of the variables
$
y^{\beta}.
$
Because the expressions are at least tri-linear in the fields they can be non-zero only for 
$
N \geq 2.3 = 6. 
$ 
We can repeat the line of argument with
$
y^{\alpha}~~\rightarrow ~~y^{\alpha}_{\mu}
$
because
$
\omega({\partial S^{I;\mu} \over \partial y^{\alpha}_{\mu}}) = N - 2 \geq 4
$
and we will eliminate the dependence on the first order derivatives. After a finite number of steps we get
$
S^{I;\mu} = 0.
$
$\qed$

Let us denote by
$
y^{A}
$
any of the variables
$
y^{\alpha}
$
and their derivatives. We also denote by
$
\epsilon_{A}
$
the Grassmann parity of
$
y^{A}.
$
Then we have the following simple corollary:
\begin{cor}
Suppose that in the preceding theorem we renounce at the hypothesis of tri-linearity. Then the solutions of the equation (\ref{div-free}) are of the form:
\be
S^{I;\mu} =  \sum_{A,B}~c^{I}_{AB}~y^{A}~d^{\mu}y^{B} 
+ \sum_{A}~c^{I}_{A}~d^{\mu}y^{A} + d_{\nu}S^{I;\mu\nu}
\ee
where
$
c^{I}_{AB}, c^{I}_{A}
$
are constants verifying 
\be
c^{I}_{AB} = - \epsilon_{A}~\epsilon_{B}~c^{I}_{BA}
\ee
and the last contribution is the trivial solution.
\end{cor}
\newpage
\section{The Cohomology of the Gauge Charge Operator\label{q}}

We consider a vector space 
$
{\cal H}
$
of Fock type generated (in the sense of Borchers theorem) by the vector field 
$
v_{\mu}
$ 
(with Bose statistics) and the scalar fields 
$
u, \tilde{u}
$
(with Fermi statistics). The Fermi fields are usually called {\it ghost fields}. We suppose that all these (quantum) fields are of null mass. Let $\Omega$ be the vacuum state in
$
{\cal H}.
$
In this vector space we can define a sesquilinear form 
$<\cdot,\cdot>$
in the following way: the (non-zero) $2$-point functions are by definition:
\be
<\Omega, v_{\mu}(x_{1}) v_{\mu}(x_{2})\Omega> =i~\eta_{\mu\nu}~D_{0}^{(+)}(x_{1} - x_{2}),
\qquad
<\Omega, u(x_{1}) \tilde{u}(x_{2})\Omega> =- i~D_{0}^{(+)}(x_{1} - x_{2})
\ee
and the $n$-point functions are generated according to Wick theorem. Here
$
\eta_{\mu\nu}
$
is the Minkowski metrics (with diagonal $1, -1, -1, -1$) and 
$
D_{0}^{(+)}
$
is the positive frequency part of the Pauli-Villars distribution
$
D_{0}
$
of null mass. To extend the sesquilinear form to
$
{\cal H}
$
we define the conjugation by
\be
v_{\mu}^{\dagger} = v_{\mu}, \qquad 
u^{\dagger} = u, \qquad
\tilde{u}^{\dagger} = - \tilde{u}.
\ee

Now we can define in 
$
{\cal H}
$
the operator $Q$ according to the following formulas:
\bea
~[Q, v_{\mu}] = i~\partial_{\mu}u,\qquad
[Q, u] = 0,\qquad
[Q, \tilde{u}] = - i~\partial_{\mu}v^{\mu}
\nonumber \\
Q\Omega = 0
\label{Q-0}
\eea
where by 
$
[\cdot,\cdot]
$
we mean the graded commutator. One can prove that $Q$ is well defined. Indeed, we have the causal commutation relations 
\be
~[v_{\mu}(x_{1}), v_{\mu}(x_{2}) ] =i~\eta_{\mu\nu}~D_{0}(x_{1} - x_{2})~\cdot I,
\qquad
[u(x_{1}), \tilde{u}(x_{2})] = - i~D_{0}(x_{1} - x_{2})~\cdot I
\ee
and the other commutators are null. The operator $Q$ should leave invariant these relations, in particular 
\be
[Q, [ v_{\mu}(x_{1}),\tilde{u}(x_{2})]] + {\rm cyclic~permutations} = 0
\ee
which is true according to (\ref{Q-0}). It is useful to introduce a grading in 
$
{\cal H}
$
as follows: every state which is generated by an even (odd) number of ghost fields and an arbitrary number of vector fields is even (resp. odd). We denote by 
$
|f|
$
the ghost number of the state $f$. We notice that the operator $Q$ raises the ghost number of a state (of fixed ghost number) by an unit. The usefullness of this construction follows from:
\begin{thm}
The operator $Q$ verifies
$
Q^{2} = 0.
$ 
The factor space
$
Ker(Q)/Ran(Q)
$
is isomorphic to the Fock space of particles of zero mass and helicity $1$ (photons). 
\end{thm}
{\bf Proof:} (i) The fact that $Q$ squares to zero follows easily from (\ref{Q-0}): the operator 
$
Q^{2} = 0
$
commutes with all field operators and gives zero when acting on the vacuum. 

(ii) The generic form of a state 
$
\Psi \in {\cal H}^{(1)} \subset {\cal H}
$
from the one-particle Hilbert subspace is
\be
\Psi = \left[ \int f_{\mu}(x) v^{\mu}(x) + \int g_{1}(x) u(x) + \int g_{2}(x) \tilde{u}(x) \right] \Omega
\ee
with test functions
$
f_{\mu}, g_{1}, g_{2}
$
verifying the wave equation equation. We impose the condition 
$
\Psi \in Ker(Q) \quad \Longleftrightarrow \quad Q\Psi = 0;
$
we obtain 
$
\partial^{\mu}f_{\mu} = 0
$
and
$
g_{2} = 0
$
i.e. the generic element
$
\Psi \in {\cal H}^{(1)} \cap Ker(Q)
$
is
\be
\Psi = \left[ \int f_{\mu}(x) v^{\mu}(x) + \int g(x) u(x) \right] \Omega
\label{kerQ-0}
\ee
with $g$ arbitrary and 
$
f_{\mu}
$
constrained by the transversality condition 
$
\partial^{\mu}f_{\mu} = 0;
$
so the elements of
$
{\cal H}^{(1)} \cap Ker(Q)
$
are in one-one correspondence with couples of test functions
$
(f_{\mu}, g)
$
with the transversality condition on the first entry. Now, a generic element
$
\Psi^{\prime} \in {\cal H}^{(1)} \cap Ran(Q)
$
has the form 
\be
\Psi^{\prime} = Q\Phi = \left[\int \partial_{\mu}g^{\prime}(x) v^{\mu}(x) 
- \int \partial^{\mu}f^{\prime}_{\mu}(x) u(x) \right] \Omega
\label{ranQ-0}
\ee
so if
$
\Psi \in {\cal H}^{(1)} \cap Ker(Q)
$
is indexed by the couple 
$
(f_{\mu}, g)
$
then 
$
\Psi + \Psi^{\prime}
$
is indexed by the couple
$
(f_{\mu} + \partial_{\mu}g^{\prime}, g - \partial^{\mu}f^{\prime}_{\mu}).
$
If we take 
$
f^{\prime}_{\mu}
$
conveniently we can make 
$
g = 0.
$
We introduce the equivalence relation 
$
f_{\mu}^{(1)} \sim f_{\mu}^{(2)} \quad \Longleftrightarrow 
f_{\mu}^{(1)} - f_{\mu}^{(2)} = \partial_{\mu}g^{\prime}
$
and it follows that the equivalence classes from
$
({\cal H}^{(1)} \cap Ker(Q))/({\cal H}^{(1)} \cap Ran(Q))
$ 
are indexed by equivalence classes of wave functions
$
[f_{\mu}];
$
it remains to prove that the sesquilinear form 
$<\cdot,\cdot>$ 
induces a positively defined form on
$
({\cal H}^{(1)} \cap Ker(Q))/({\cal H}^{(1)} \cap Ran(Q))
$ 
and we have obtained the usual one-particle Hilbert space for the photon.

(iii) We  go now to the $2$-particle space. We borrow an argument from the proof of K\"unneth formula \cite{Dr}. Any $2$-particle state is generated by states of the form: 
\be
\Psi = \sum_{j=1}^{n} f_{j} \otimes g_{j} 
\label{2-particle}
\ee
with 
$
f_{j}, g_{j}
$
one-particle states. We impose the condition 
$
\Psi \in Ker(Q)
$
and observe that it is sufficient to take 
$
f_{j}, g_{j}
$
states of fixed ghost number. Moreover, we can take 
$
f_{j}
$ 
such that their span does not intersect  
$
Ran(Q).
$
Indeed if we have constants
$
\beta_{j}
$
not all null such that
$
\sum_{j=1}^{n} \beta_{j}~f_{j} \in Ran(Q)
$
then by a redefinition of the vectors
$
f_{j}
$
we can arrange such that
$
f_{1} = \sum_{j=2}^{n} \beta^{\prime}_{j}~f_{j} + Qh.
$
We substitute this in the formula for $\Psi$ and get:
$
\Psi = \sum_{j=2}^{n} f_{j} \otimes (\beta^{\prime}_{j} g_{1} + g_{j}) +
Q(h \otimes g_{1}) - (-1)^{|h|}~h \otimes Qg_{1}
$
so if we eliminate the co-boundary we can replace the state 
$\Psi$
by an equivalent one in which 
$
f_{1} \rightarrow h.
$
In this way we replace the expression (\ref{2-particle}) by an equivalent expression for which
$
\sum_{j=1}^{n} |f_{j}|
$
decreases by an unit. Recursively we obtain another expression (\ref{2-particle}) modulo 
$
Ran(Q)
$
for which 
$
Span~(f_{j})_{j=1}^{n} \cap Ran(Q) = \{0\}.
$
Now the condition 
$
Q \Psi = 0
$
writes
$
\sum_{j=1}^{n} (Qf_{j} \otimes g_{j} + (-1)^{|f_{j}|}~f_{j} \otimes Qg_{j}) = 0 
$
and it easily follows that both sums must be separately null i.e. we must have 
$
Qg_{j} = 0
$
and 
$
Qf_{j} = 0
$
for all 
$j = 1,\dots,n.$
It means that we have the canonical isomorphism
$
({\cal H}^{(2)} \cap Ker(Q))/({\cal H}^{(2)} \cap Ran(Q)) \cong
({\cal H}^{(1)} \cap Ker(Q))/({\cal H}^{(1)} \cap Ran(Q)) \otimes
({\cal H}^{(1)} \cap Ker(Q))/({\cal H}^{(1)} \cap Ran(Q)).
$

Now we can proceed by induction to the general $n$-particle states. 
$\qed$

We see that the condition 
$
[Q, T] = i~\partial_{\mu}T^{\mu}
$
means that the expression $T$ leaves invariant the physical Hilbert space (at least in the adiabatic limit).

Now we have the physical justification for solving another cohomology problem namely to determine the cohomology of the operator 
$
d_{Q} = [Q,\cdot]
$
induced by $Q$ in the space of Wick polynomials. To solve this problem it is convenient to use the formalism from the preceding Section. We consider that the (classical) fields
$
y^{\alpha}
$
are
$
v_{\mu}, u, \tilde{u}
$
of null mass and we consider the set 
$
{\cal P}
$
of polynomials in these fields and their derivatives. We note that on
$
{\cal P}
$
we have a natural grading. We introduce by convenience the notation:
\be
B \equiv d_{\mu}v^{\mu}
\ee
and define the graded derivation 
$
d_{Q}
$
on
$
{\cal P}
$
according to
\bea
d_{Q}v_{\mu} = i d_{\mu}u, \qquad d_{Q}u = 0, \qquad d_{Q}\tilde{u} = - i~B
\nonumber \\
~[d_{Q}, d_{\mu} ] = 0.
\eea
Then one can easily prove that 
$
d_{Q}^{2} = 0
$
and the cohomology of this operator is isomorphic to the cohomology of the preceding operator (denoted also by $d_{Q}$) and acting in the space of Wick monomials. The operator 
$
d_{Q}
$
raises the grading and the canonical dimension by an unit. To determine the cohomology of 
$
d_{Q}
$
it is convenient to introduce the {\it field strength}
\be
F_{\mu\nu} \equiv d_{\mu}v_{\nu} - d_{\nu}v_{\mu} = v_{\nu;\mu} - v_{\mu;\nu}
\ee 
and observe that 
\bea
d_{Q}F_{\mu\nu} = 0, 
\nonumber \\
d_{\nu}F^{\mu\nu} = d^{\mu}B,
\nonumber \\
F_{\mu\nu;\rho} + F_{\nu\rho;\mu} + F_{\rho\mu;\nu} = 0;
\eea
the last relation is called {\it Bianchi identity}. Next we prove that the tensor
\be
F^{(0)}_{\mu\nu;\rho_{1},\dots,\rho_{n}} \equiv F_{\mu\nu;\rho_{1},\dots,\rho_{n}}
+ {1\over n + 1}~\sum_{l=1}^{n} [\eta_{\mu\rho_{l}}~
B_{\rho_{1},\dots,\hat{\rho}_{l},\dots,\rho_{n}} - (\mu \leftrightarrow \nu) ]
\label{f-fb}
\ee 
is traceless in all indices and the expressions
$
F^{(0)}_{\mu\nu;\rho}
$
also verify the Bianchi identities. Now we define 
\be
g_{\mu_{1},\dots,\mu_{n}} \equiv {1\over n}~\sum_{l=1}^{n} 
v_{\mu_{l};\mu_{1},\dots,\hat{\mu}_{l},\dots,\mu_{n}}
\ee
which is the completely symmetric part of the derivative
$
v_{\mu_{1};\mu_{2},\dots,\mu_{n}} 
$
and prove that 
\be
v_{\mu_{1};\mu_{2},\dots,\mu_{n}} = g_{\mu_{1},\dots,\mu_{n}} + {1\over n}~
\sum_{l=2}^{n}~d_{\mu_{2}}\dots\hat{d}_{\mu_{l}}\dots d_{\mu_{n}}F_{\mu_{l}\mu_{1}}.
\label{v-gf}
\ee
Finally we define
\be
g^{(0)}_{\mu_{1},\dots,\mu_{n}} \equiv g_{\mu_{1},\dots,\mu_{n}}
- {2\over n(2n + 1)}~\sum_{1 \leq p < q \leq n} \eta_{\mu_{p}\mu_{q}}~
B_{\mu_{1},\dots,\hat{\mu}_{p},\dots,\hat{\mu}_{q},\dots,\mu_{n}}
\label{g-gb}
\ee
which is completely symmetric and traceless.

We will use repeatedly the K\"unneth theorem:
\begin{thm}
Let 
$
{\cal P}
$
be a graded space of polynomials and $d$ an operator verifying
$
d^{2} = 0
$ 
and raising the grading by an unit. Let us suppose that
$
{\cal P}
$
is generated by two subspaces
$
{\cal P}_{1}, {\cal P}_{2}
$
such that
$
{\cal P}_{1} \cap {\cal P}_{2} = \{0\}
$
and
$
d{\cal P}_{j} \subset {\cal P}_{j}, j = 1,2.
$
We define by 
$
d_{j}
$
the restriction of $d$ to
$
{\cal P}_{j}.
$
Then there exists the canonical isomorphism
$
H(d) \cong H(d_{1}) \times H(d_{2})
$ 
of the associated cohomology spaces. 
\label{kunneth}
\end{thm}

The proof goes in a similar way to the preceding theorem (see \cite{Dr}). Now we can prove an important result describing the cohomology of the operator 
$
d_{Q};
$
we denote by
$
Z_{Q}
$
and 
$
B_{Q}
$
the co-cyles and the co-boundaries of this operator.
\begin{thm}
Let 
$
p \in Z_{Q}.
$
Then $p$ is cohomologous to a polynomial in $u$ and
$
F^{(0)}_{\mu\nu;\rho_{1},\dots,\rho_{n}}.
$
If we factorize the space 
$
{\cal P}_{0} \subset {\cal P}
$
of such polynomials to the Bianchi identities we obtain a space which is isomorphic to the cohomology space 
$
H_{Q}
$
of
$
d_{Q}.
$
\label{m=0}
\end{thm}
{\bf Proof:} (i) The idea is to define conveniently two subspaces 
$
{\cal P}_{1}, {\cal P}_{2}
$
and apply K\"unneth theorem. First we use on
$
{\cal P}
$
new variables. We eliminate the variables
$
v_{\mu_{1};\mu_{2},\dots,\mu_{n}}~(n \geq 2) 
$
in terms of 
$
g_{\mu_{1},\dots,\mu_{n}}~(n \geq 2) 
$
and
$
F_{\mu\nu;\rho_{1},\dots,\rho_{n-2}}
$
using (\ref{v-gf}). Next we eliminate
$
F_{\mu\nu;\rho_{1},\dots,\rho_{n-2}}
$
in terms of
$
F^{(0)}_{\mu\nu;\rho_{1},\dots,\rho_{n-2}}
$
and
$
B_{\rho_{1},\dots,\rho_{n-2}}
$
using (\ref{f-fb}). Finally we eliminate
$
g_{\mu_{1},\dots,\mu_{n}}~(n \geq 2) 
$
in terms of
$
g^{(0)}_{\mu_{1},\dots,\mu_{n}}~(n \geq 2) 
$
and
$
B_{\mu_{1},\dots,\mu_{n-2}}
$
according to (\ref{g-gb}).

(ii) Now we can take in K\"unneth theorem
$
{\cal P}_{1} = {\cal P}_{0}
$
from the statement and
$
{\cal P}_{2}
$
the subspace generated by the variables
$
B_{\mu_{1},\dots,\mu_{n}}~(n \geq 0),~
g^{(0)}_{\mu_{1},\dots,\mu_{n}}~(n \geq 2),~
\tilde{u}_{\mu_{1},\dots,\mu_{n}}~(n \geq 0),~
u_{\mu_{1},\dots,\mu_{n}}(n > 0)
$
and
$
v_{\mu}.
$
We have
$
d_{Q}{\cal P}_{1} = \{0\}
$
and
\bea
d_{Q}u_{\mu_{1},\dots,\mu_{n}} = 0
\nonumber \\
d_{Q}g^{(0)}_{\mu_{1},\dots,\mu_{n}} = i~u_{\mu_{1},\dots,\mu_{n}}
\nonumber \\
d_{Q}\tilde{u}_{\mu_{1},\dots,\mu_{n}} = - i~B_{\mu_{1},\dots,\mu_{n}}
\nonumber \\
d_{Q}B_{\mu_{1},\dots,\mu_{n}} = 0
\nonumber \\
d_{Q}v_{\mu} = iu_{\mu}
\eea
so we meet the conditions of K\"unneth theorem. Let us define in
$
{\cal P}_{2}
$
the graded derivation $h$ by:
\bea
hu_{\mu} = - i~v_{\mu} 
\nonumber \\
hu_{\mu_{1},\dots,\mu_{n}} = - i~g^{(0)}_{\mu_{1},\dots,\mu_{n}}~(n \geq 2)
\nonumber \\
hB_{\mu_{1},\dots,\mu_{n}} = i~\tilde{u}_{\mu_{1},\dots,\mu_{n}}~(n \geq 0)
\eea
and zero on the other variables from 
$
{\cal P}_{2}.
$
It is easy to prove that $h$ is well defined: the condition of tracelessness is essential to avoid conflict with the equations of motion. Then one can prove that
\be
[d_{Q},h] = Id
\ee
on polynomials of degree one in the fields and because the left hand side is a derivation operator we have
\be
[d_{Q},h] = n \cdot Id
\ee 
on polynomials of degree $n$ in the fields. It means that $h$ is a homotopy for 
$
d_{Q}
$
restricted to 
$
{\cal P}_{2}
$
so the the corresponding cohomology is trivial: indeed, if 
$
p \in {\cal P}_{2}
$
is a co-cycle of degree $n$ in the fields then it is a co-boundary
$
p = {1\over n} d_{Q}hp.
$

According to K\"unneth formula if $p$ is an arbitrary cocycle from 
$
{\cal P}
$
it can be replaced by a cohomologous polynomial from
$
{\cal P}_{0}
$
and this proves the theorem.
$\qed$

We repeat the whole argument for the case of massive photons i.e. particles of spin $1$ and positive mass. 

We consider a vector space 
$
{\cal H}
$
of Fock type generated (in the sense of Borchers theorem) by the vector field 
$
v_{\mu},
$ 
the scalar field 
$
\Phi
$
(with Bose statistics) and the scalar fields 
$
u, \tilde{u}
$
(with Fermi statistics). We suppose that all these (quantum) fields are of mass
$
m > 0.
$
In this vector space we can define a sesquilinear form 
$<\cdot,\cdot>$
in the following way: the (non-zero) $2$-point functions are by definition:
\bea
<\Omega, v_{\mu}(x_{1}) v_{\mu}(x_{2})\Omega> =i~\eta_{\mu\nu}~D_{m}^{(+)}(x_{1} - x_{2}),
\quad
<\Omega, u(x_{1}) \tilde{u}(x_{2})\Omega> =- i~D_{m}^{(+)}(x_{1} - x_{2}),
\nonumber \\
<\Omega, \Phi(x_{1}) \Phi(x_{2})\Omega> =- i~D_{m}^{(+)}(x_{1} - x_{2})
\eea
and the $n$-point functions are generated according to Wick theorem. Here
$
D_{m}^{(+)}
$
is the positive frequency part of the Pauli-Villars distribution
$
D_{m}
$
of mass $m$. To extend the sesquilinear form to
$
{\cal H}
$
we define the conjugation by
\be
v_{\mu}^{\dagger} = v_{\mu}, \qquad 
u^{\dagger} = u, \qquad
\tilde{u}^{\dagger} = - \tilde{u},
\qquad \Phi^{\dagger} = \Phi.
\ee

Now we can define in 
$
{\cal H}
$
the operator $Q$ according to the following formulas:
\bea
~[Q, v_{\mu}] = i~\partial_{\mu}u,\qquad
[Q, u] = 0,\qquad
[Q, \tilde{u}] = - i~(\partial_{\mu}v^{\mu} + m~\Phi)
\qquad
[Q,\Phi] = i~m~u,
\nonumber \\
Q\Omega = 0.
\label{Q-m}
\eea
One can prove that $Q$ is well defined. We have a result similar to the first theorem of this Section:
\begin{thm}
The operator $Q$ verifies
$
Q^{2} = 0.
$ 
The factor space
$
Ker(Q)/Ran(Q)
$
is isomorphic to the Fock space of particles of mass $m$ and spin $1$ (massive photons). 
\end{thm}
{\bf Proof:} (i) The fact that $Q$ squares to zero follows easily from (\ref{Q-m}).

(ii) The generic form of a state 
$
\Psi \in {\cal H}^{(1)} \subset {\cal H}
$
from the one-particle Hilbert subspace is
\be
\Psi = \left[ \int f_{\mu}(x) v^{\mu}(x) + \int g_{1}(x) u(x) + \int g_{2}(x) \tilde{u}(x) + \int h(x) \Phi(x) \right] \Omega
\ee
with test functions
$
f_{\mu}, g_{1}, g_{2}, h
$
verifying the wave equation equation. We impose the condition 
$
\Psi \in Ker(Q) \quad \Longleftrightarrow \quad Q\Psi = 0;
$
we obtain 
$
h = {1\over m}~\partial^{\mu}f_{\mu}
$
and
$
g_{2} = 0
$
i.e. the generic element
$
\Psi \in {\cal H}^{(1)} \cap Ker(Q)
$
is
\be
\Psi = \left[ \int f_{\mu}(x) v^{\mu}(x) + \int g(x) u(x) 
+ {1\over m}~\int \partial^{\mu}f_{\mu}(x) \Phi(x) \right] \Omega
\label{kerQ-m}
\ee
with $g$ arbitrary and 
$
f_{\mu}
$
so the elements of
$
{\cal H}^{(1)} \cap Ker(Q)
$
are in one-one correspondence with couples of test functions
$
(f_{\mu}, g).
$
Now, a generic element
$
\Psi^{\prime} \in {\cal H}^{(1)} \cap Ran(Q)
$
has the form 
\be
\Psi^{\prime} = Q\Phi = \left\{\int \partial_{\mu}g^{\prime}(x) v^{\mu}(x)
+ \left[ m h^{\prime}(x) - \int \partial^{\mu}f^{\prime}_{\mu}(x)\right] u(x)
- m g^{\prime}(x) \Phi(x) \right\} \Omega
\label{ranQ-m}
\ee
so if  
$
\Psi \in {\cal H}^{(1)} \cap Ker(Q)
$
is indexed by the couple 
$
(f_{\mu}, g)
$
then 
$
\Psi + \Psi^{\prime}
$
is indexed by the couple
$
(f_{\mu} + \partial_{\mu}g^{\prime}, 
g + m~h^{\prime} - \partial^{\mu}f^{\prime}_{\mu}).
$
If we take 
$
h^{\prime}
$
conveniently we can make 
$
g = 0
$
and if we take 
$
g^{\prime}
$
conveniently we can make 
$
f^{\mu}
$
of null divergence; it follows that the equivalence classes from
$
({\cal H}^{(1)} \cap Ker(Q))/({\cal H}^{(1)} \cap Ran(Q))
$ 
are indexed by wave functions
$
f_{\mu}
$
constrained by the transversality condition
$
\partial^{\mu}f_{\mu} =  0;
$
it remains to prove that the sesquilinear form 
$<\cdot,\cdot>$ 
induces a positively defined form on
$
({\cal H}^{(1)} \cap Ker(Q))/({\cal H}^{(1)} \cap Ran(Q))
$ 
and we have obtained the usual one-particle Hilbert space for the massive photon.

(iii) We  go now to the $n$-particle space as in the first theorem.
$\qed$

Now we determine the cohomology of the operator 
$
d_{Q} = [Q,\cdot]
$
induced by $Q$ in the space of Wick polynomials. As before, it is convenient to use the formalism from the preceding Section. We consider that the (classical) fields
$
y^{\alpha}
$
are
$
v_{\mu}, u, \tilde{u}, \Phi
$
of mass $m$ and we consider the set 
$
{\cal P}
$
of polynomials in these fields and their derivatives. We introduce by convenience the notation:
\be
C \equiv d_{\mu}v^{\mu} + m \Phi
\ee
and define the graded derivation 
$
d_{Q}
$
on
$
{\cal P}
$
according to
\bea
d_{Q}v_{\mu} = i d_{\mu}u, \qquad d_{Q}u = 0, \qquad d_{Q}\tilde{u} = - i~C, \qquad
d_{Q}\Phi = i~m~u,
\nonumber \\
~[d_{Q}, d_{\mu} ] = 0.
\eea
Then one can prove that 
$
d_{Q}^{2} = 0
$
and the cohomology of this operator is isomorphic to the cohomology of the preceding operator (denoted also by $d_{Q}$) and acting in the space of Wick monomials. To determine the cohomology of 
$
d_{Q}
$
it is convenient to introduce the field strength
$
F_{\mu\nu}
$ 
as before and also
\bea
\phi_{\mu} \equiv d_{\mu}\Phi - m~v_{\mu},
\nonumber \\
\phi_{\mu_{1},\dots,\mu_{n}} \equiv d_{\mu_{1}}\dots d_{\mu_{n}}\Phi 
- m~g_{\mu_{1},\dots,\mu_{n}}~(n \geq 2).
\eea
Observe that we have
\bea
d_{Q}F_{\mu\nu} = 0, 
\nonumber \\
d^{\nu}F_{\mu\nu} = d_{\mu}C - m \phi_{\mu},
\nonumber \\
F_{\mu\nu;\rho} + F_{\nu\rho;\mu} + F_{\rho\mu;\nu} = 0,
\nonumber \\
d_{Q}\phi_{\mu_{1},\dots,\mu_{n}} = 0,
\nonumber \\
d^{\mu}\phi_{\mu} = - m~C = i~m~d_{Q}\tilde{u}.
\eea 

In the massive case we do not have explicit formulas for the traceless parts of the various tensors; we even do not know if such a traceless parts do exists! However, due to a theorem proved in the Appendix, such traceless parts 
$
F^{(0)}_{\mu\nu;\rho_{1},\dots,\rho_{n}},
\phi^{(0)}_{\mu_{1},\dots,\mu_{n}} 
$
and
$
g^{(0)}_{\mu_{1},\dots,\mu_{n}} 
$
do exists; moreover they are linear combinations of
$
F_{\mu\nu;\rho_{1},\dots,\rho_{n}},
\phi_{\mu_{1},\dots,\mu_{n}} 
$
and
$
g_{\mu_{1},\dots,\mu_{n}} 
$
and traces of these tensors respectively. Now we can describe the cohomology of the operator 
$
d_{Q}
$
in the massive case.
\begin{thm}
Let 
$
p \in Z_{Q}.
$
Then $p$ is cohomologous to a polynomial in 
$
F^{(0)}_{\mu\nu;\rho_{1},\dots,\rho_{n}}
$
and
$
\phi^{(0)}_{\mu_{1},\dots,\mu_{n}}
$ 
If we factorize the space 
$
{\cal P}_{0} \subset {\cal P}
$
of such polynomials to the Bianchi identities we obtain a space which is isomorphic to the cohomology space 
$
H_{Q}
$
of
$
d_{Q}.
$
\label{m>0}
\end{thm}
{\bf Proof:} (i) As before, we use on
$
{\cal P}
$
new variables. In the first step, we eliminate the variables
$
v_{\mu_{1};\mu_{2},\dots,\mu_{n}} 
$
in terms of 
$
g_{\mu_{1},\dots,\mu_{n}} 
$
and
$
F_{\mu\nu;\rho_{1},\dots,\rho_{n-2}};
$
and we eliminate the variables
$
\Phi_{\mu_{1},\dots,\mu_{n}}
$
in terms of 
$
\phi_{\mu_{1},\dots,\mu_{n}}
$
and
$
g_{\mu_{1},\dots,\mu_{n}}.
$

In the second step we eliminate
$
F_{\mu\nu;\rho_{1},\dots,\rho_{n}}
$
in terms of
$
F^{(0)}_{\mu\nu;\rho_{1},\dots,\rho_{n}},~
C_{\rho_{1},\dots,\rho_{n}}
$
and we eliminate
$
g_{\mu_{1},\dots,\mu_{n}} 
$
in terms of
$
g^{(0)}_{\mu_{1},\dots,\mu_{n}},~ 
C_{\mu_{1},\dots,\mu_{n}}
$
and
$
\phi_{\mu_{1},\dots,\mu_{n}}.
$

In the final step we note that the traces of 
$
u_{\mu_{1},\dots,\mu_{n}}, \tilde{u}_{\mu_{1},\dots,\mu_{n}}, C_{\mu_{1},\dots,\mu_{n}}
$
and
$
\phi_{\mu_{1},\dots,\mu_{n}}
$
are functions of derivatives of lower order so they can be recursively expressed in terms of the traceless variables:
$
u^{(0)}_{\mu_{1},\dots,\mu_{n}}, \tilde{u}^{(0)}_{\mu_{1},\dots,\mu_{n}}, C^{(0)}_{\mu_{1},\dots,\mu_{n}}
$
and
$
\phi^{(0)}_{\mu_{1},\dots,\mu_{n}}.
$

(ii) Now we can take in K\"unneth theorem
$
{\cal P}_{1} = {\cal P}_{0}
$
from the statement and
$
{\cal P}_{2}
$
the subspace generated by the variables
$
C^{(0)}_{\mu_{1},\dots,\mu_{n}},~g^{(0)}_{\mu_{1},\dots,\mu_{n}},
\tilde{u}^{(0)}_{\mu_{1},\dots,\mu_{n}},
u^{(0)}_{\mu_{1},\dots,\mu_{n}}
$
and
$
v_{\mu}, \Phi.
$
We have
$
d_{Q}{\cal P}_{1} = \{0\}
$
and
\bea
d_{Q}u^{(0)}_{\mu_{1},\dots,\mu_{n}} = 0,
\nonumber \\
d_{Q}g^{(0)}_{\mu_{1},\dots,\mu_{n}} = i~u^{(0)}_{\mu_{1},\dots,\mu_{n}},
\nonumber \\
d_{Q}\tilde{u}^{(0)}_{\mu_{1},\dots,\mu_{n}} = - i~C^{(0)}_{\mu_{1},\dots,\mu_{n}},
\nonumber \\
d_{Q}C^{(0)}_{\mu_{1},\dots,\mu_{n}} = 0,
\nonumber \\
d_{Q}v_{\mu} = iu_{\mu}, \qquad d_{Q}\Phi = i~m~u
\eea
so we meet the conditions of K\"unneth theorem. Let us define in
$
{\cal P}_{2}
$
the graded derivation $h$ by:
\bea
hu = - {i\over m}~\Phi, \qquad hu_{\mu} = - i~v_{\mu},
\nonumber \\
hu^{(0)}_{\mu_{1},\dots,\mu_{n}} = - i~g^{(0)}_{\mu_{1},\dots,\mu_{n}}~(n \geq 2),
\nonumber \\
hC^{(0)}_{\mu_{1},\dots,\mu_{n}} = i~\tilde{u}^{(0)}_{\mu_{1},\dots,\mu_{n}}
\eea
and zero on the other variables from 
$
{\cal P}_{2}.
$
It is easy to prove that $h$ is well defined due to the condition of tracelessness. Then one can prove as before that we have
\be
[d_{Q},h] = n \cdot Id
\ee 
on polynomials of degree $n$ in the fields. It means that $h$ is a homotopy for 
$
d_{Q}
$
restricted to 
$
{\cal P}_{2}
$
so the the corresponding cohomology is trivial.

According to K\"unneth formula if $p$ is an arbitrary cocycle from 
$
{\cal P}
$
it can be replaced by a cohomologous polynomial from
$
{\cal P}_{0}
$
and this proves the theorem.
$\qed$

We note that in the case of null mass the operator 
$
d_{Q}
$
raises the canonical dimension by one unit and this fact is not true anymore in the massive case. We are lead to another cohomology group. Let us take as the space of co-chains the space 
$
{\cal P}^{(n)}
$
of polynomials of canonical dimension
$
\omega \leq n;
$
then 
$
Z_{Q}^{(n)} \subset {\cal P}^{(n)}
$
and
$
B_{Q}^{(n)} \equiv d_{Q}{\cal P}^{(n-1)}
$
are the co-cyles and the co-boundaries respectively. It is possible that a polynomial is a co-boundary as an element of
$
{\cal P}
$
but not as an element of
$
{\cal P}^{(n)}.
$
The situation is described by the following generalization of the preceding theorem.
\begin{thm}
Let 
$
p \in Z^{(n)}_{Q}.
$
Then $p$ is cohomologous to a polynomial of the form  
$
p_{1} + d_{Q}p_{2}
$
where
$
p_{1} \in {\cal P}_{0}
$ 
and
$
p_{2} \in {\cal P}^{(n-1)}.
$
If we factorize the space of such polynomials to the Bianchi identities we obtain a space which is isomorphic to the cohomology space 
$
H^{(n)}_{Q}
$
of
$
d_{Q}
$
in
$
{\cal P}^{(n)}.
$
\label{Q-cohomology}
\end{thm}

We will call the cocyles of the type
$
p_{1}
$
(resp.
$
d_{Q}p_{2})
$
{\it primary} (resp. {\it secondary}).

The situations described above (of massless and massive photons) are susceptible of the following generalizations. We can consider a system of 
$
r_{1}
$ 
species of particles of null mass and helicity $1$ if we use in the first part of this Section 
$
r_{1}
$ 
triplets
$
(v^{\mu}_{a}, u_{a}, \tilde{u}_{a}), a \in I_{1}
$
of massless fields; here
$
I_{1}
$
is a set of indices of cardinal 
$
r_{1}.
$
All the relations have to be modified by appending an index $a$ to all these fields. If we repeatedly apply K\"unneth theorem we end up with a generalization of theorem \ref{m=0}: the space 
$
{\cal P}_{0}
$
is generated by 
$
u_{a}
$ and
$
F^{(0)}_{a\mu\nu;\rho_{1},\dots,\rho_{n}}.
$

In the massive case we have to consider 
$
r_{2}
$ 
quadruples
$
(v^{\mu}_{a}, u_{a}, \tilde{u}_{a}, \Phi_{a}),  a \in I_{2}
$
of fields of mass 
$
m_{a}
$; here
$
I_{2}
$
is a set of indices of cardinal 
$
r_{2}.
$
We also have a generalization of theorem \ref{m>0}: the space 
$
{\cal P}_{0}
$
is generated
$
F^{(0)}_{a\mu\nu;\rho_{1},\dots,\rho_{n}}
$
and
$
\phi^{(0)}_{a;\mu_{1},\dots,\mu_{n}}.
$ 

We can consider now the most general case involving fields of spin not greater that $1$.
We take 
$
I = I_{1} \cup I_{2} \cup I_{3}
$
a set of indices and for any index we take a quadruple
$
(v^{\mu}_{a}, u_{a}, \tilde{u}_{a},\Phi_{a}), a \in I
$
of fields with the following conventions:
(a) For
$
a \in I_{1}
$
we impose 
$
\Phi_{a} = 0
$
and we take the masses to be null
$
m_{a} = 0;
$
(b) For
$
a \in I_{2}
$
we take the all the masses strictly positive:
$
m_{a} > 0;
$
(c) For 
$
a \in I_{3}
$
we take 
$
v_{a}^{\mu}, u_{a}, \tilde{u}_{a}
$
to be null and the fields
$
\Phi_{a} \equiv \phi^{H}_{a} 
$
of mass 
$
m^{H}_{a} \geq 0.
$
The fields
$
\phi^{H}_{a} 
$
are called {\it Higgs fields}.

If we define
$
m_{a} = 0, \forall a \in I_{3}
$
then we can define in 
$
{\cal H}
$
the operator $Q$ according to the following formulas for all indices
$
a \in I:
$
\bea
~[Q, v^{\mu}_{a}] = i~\partial^{\mu}u_{a},\qquad
[Q, u_{a}] = 0,
\nonumber \\
~[Q, \tilde{u}_{a}] = - i~(\partial_{\mu}v^{\mu}_{a} + m_{a}~\Phi_{a})
\qquad
[Q,\Phi_{a}] = i~m_{a}~u_{a},
\nonumber \\
Q\Omega = 0.
\label{Q-general}
\eea

Then the space 
$
{\cal P}_{0}
$
is generated by
$
u_{a}, a \in I_{1},
$
$
F^{(0)}_{a\mu\nu;\rho_{1},\dots,\rho_{n}}, a \in I_{1} \cup I_{2}
$
and
$
\phi^{(0)}_{a;\mu_{1},\dots,\mu_{n}}, a \in I_{2} \cup I_{3}.
$ 
If we consider matter fields also i.e some set of Dirac fields with Fermi statistics:
$
\Psi_{A}, A \in I_{4}
$ 
then we impose
\be
d_{Q}\Psi_{A} = 0 
\ee
and the space 
$
{\cal P}_{0}
$
is generated by
$
\Psi_{A}
$
and
$
\bar{\Psi}_{A}
$
also.
\newpage
\section{The Relative Cohomology of the Operator $d_{Q}$\label{relative}}
A polynomial 
$
p \in {\cal P}
$
verifying the relation
\be
d_{Q}p = i~d_{\mu}p^{\mu}
\label{rel-co}
\ee
for some polynomials
$
p^{\mu}
$
is called a {\it relative cocycle} for 
$
d_{Q}.
$
The expressions of the type
\be
p = d_{Q}b + i~d_{\mu}b^{\mu}, \qquad (b, b^{\mu} \in {\cal P})
\ee
are relative co-cyles and are called {\it relative co-boundaries}. We denote by
$
Z_{Q}^{\rm rel}, B_{Q}^{\rm rel} 
$
and
$
H_{Q}^{\rm rel}
$
the corresponding cohomological spaces. In (\ref{rel-co}) the expressions
$
p_{\mu}
$
are not unique. It is possible to choose them Lorentz covariant? The next proposition gives a positive answer in a quite general case. The proof will illustrate the descent technique.
\begin{thm}
Let us suppose that the relative cocycle $p$ is at least tri-linear in the fields and Lorentz covariant. Then the expressions
$
p^{\mu}
$
from (\ref{rel-co}) can be chosen to be Lorentz covariant also.
\end{thm}
{\bf Proof:} Let us denote by
$
\delta_{g}
$
the action of the Lorentz transformation 
$
g \in G = SL(2,\C)
$ 
in the space 
$
{\cal P}^{(k)}.
$
It is clear that
$
\delta_{g}
$
commutes with 
$
d_{\mu}.
$
If we denote by
$
C^{n}(G,{\cal P}^{(k)})~(n \geq 0)
$
the space of maps
$
p: G^{\times n} \rightarrow {\cal P}^{(k)}
$
with the convention that for $n = 0$ the functions $p$ are independent of $g$ then we have the co-chain operator
$
d:C^{n}(G,{\cal P}^{(k)}) \rightarrow C^{n+1}(G,{\cal P}^{(k)}) 
$
\bea
(d\cdot p)(g_{1},\dots,g_{n+1}) \equiv \delta_{g_{1}}\cdot p(g_{2},\dots,g_{n+1}) 
\nonumber \\
+ \sum_{j=1}^{n} (-1)^{j} p(g_{1},\dots,g_{j}g_{j+1},\dots,g_{n+1})
+ (-1)^{n+1} p(g_{1},\dots,g_{n}).
\eea
Because
$
d^{2} = 0
$
we can define the corresponding cohomology spaces
$
Z^{n}(G,{\cal P}^{(k)}), B^{n}(G,{\cal P}^{(k)})
$
and
$
H^{n}(G,{\cal P}^{(k)})
$
\cite{Gui}. By hypothesis we have
\be
\delta_{g}\cdot p = p
\ee
which can be written as
\be
d\cdot p = 0.
\ee
Then we have from (\ref{rel-co}):
\be
d_{\mu}(\delta_{g}\cdot p^{\mu} - p^{\mu}) = 0
\ee
so with the Poincar\'e lemma we have:
\be
\delta_{g}\cdot p^{\mu} - p^{\mu} = d_{\nu}p^{\mu\nu}(g)
\ee
for some polynomials
$
p^{\mu\nu}(g)
$
antisymmetric in 
$
\mu, \nu;
$
the preceding identity can be written as
\be
d\cdot p^{\mu} = d_{\nu}p^{\mu\nu}.
\ee
Proceding in the same way we obtain the expressions
$
p^{\mu\nu\rho}(g_{1},g_{2})
$
and
$
p^{\mu\nu\rho\sigma}(g_{1},g_{2},g_{3})
$
which are completely antisymmetric and we have
\bea
d\cdot p^{\mu\nu} = d_{\rho}p^{\mu\nu\rho}
\nonumber \\
d\cdot p^{\mu\nu\rho} = d_{\sigma}p^{\mu\nu\rho\sigma}
\nonumber \\
d\cdot p^{\mu\nu\rho\sigma} = 0.
\eea
We have obtained that
$
p^{\mu\nu\rho\sigma} \in H^{3}(G,{\cal P}^{(k)}).
$
But $G$ is a connected simply connected Lie group and in this case the study of group cohomology can be reduced to the study of the corresponding Lie algebra cohomology. Because $G$ is also simple we can apply one of the Whitehead lemmas (see \cite{Gui} ch. II, \$ 11, cor. 11.1) and conclude that 
$
H^{n}(Lie(G),{\cal P}^{(k)})
$
are trivial for
$
n \geq 0;
$
we obtain that 
$
p^{\mu\nu\rho\sigma}
$
is a trivial cocycle i.e. it is of the form:
\be
p^{\mu\nu\rho\sigma} = d\cdot q^{\mu\nu\rho\sigma}
\ee
where we can take the co-chain
$
q^{\mu\nu\rho\sigma}
$
to be completely antisymmetric. If we make the redefinition
\be
p^{\mu\nu\rho} \rightarrow p^{\mu\nu\rho} - d_{\sigma}q^{\mu\nu\rho\sigma}
\ee
then we have
$
d\cdot p^{\mu\nu\rho} = 0
$
i.e. 
$
p^{\mu\nu\rho} \in H^{2}(G,{\cal P}^{(k)}),
$
etc. In the end we can obtain
$
d\cdot p^{\mu} = 0
$
i.e.
\be
\delta_{g} \cdot p^{\mu} = p^{\mu}
\ee
and this is the invariance property we claimed in the statement.
$\qed$

Now we consider the framework and notations from the end of the preceding Section. Then we have the following result which describes the most general form of the Yang-Mills interaction. Summation over the dummy indices is used everywhere. We will need the following notation:
\be
m^{*}_{a} \equiv \left\{\begin{array}{rcl} 
m_{a} & \mbox{for} & m_{a} \not= 0 \\
m^{H}_{a} & \mbox{for} & m_{a} = 0.\end{array}\right.
\ee
\begin{thm}
Let $T$ be a relative cocycle for 
$
d_{Q}
$
which is as least tri-linear in the fields and is of canonical dimension
$
\omega(T) \leq 4
$
and ghost number
$
gh(T) = 0.
$
Then:
(i) $T$ is (relatively) cohomologous to a non-trivial co-cycle of the form:
\bea
T = f_{abc} \left( {1\over 2}~v_{a\mu}~v_{b\nu}~F_{c}^{\nu\mu}
+ u_{a}~v_{b}^{\mu}~d_{\mu}\tilde{u}_{c}\right)
\nonumber \\
+ f^{\prime}_{abc} (\Phi_{a}~\phi_{b}^{\mu}~v_{c\mu} + m_{b}~\Phi_{a}~\tilde{u}_{b}~u_{c})
\nonumber \\
+ {1\over 3!}~f^{\prime\prime}_{abc}~\Phi_{a}~\Phi_{b}~\Phi_{c}
+ {1\over 4!}~\sum_{a,b,c,d \in I_{1}}~g_{abcd}~\Phi_{a}~\Phi_{b}~\Phi_{c}~\Phi_{d}
+ j^{\mu}_{a}~v_{a\mu} + j_{a}~v_{a}
\eea
where we can take the constants
$
f_{abc} = 0
$
if one of the indices is in
$
I_{3};
$
also
$
f^{\prime}_{abc} = 0
$
if 
$
c \in I_{3}
$
or one of the indices $a$ and $b$ are from
$
I_{1}.
$

Moreover we have:

(a) The constants
$
f_{abc}
$
are completely antisymmetric
\be
f_{abc} = f_{[abc]}.
\label{anti-f}
\ee

(b) The expressions
$
f^{\prime}_{abc}
$
are antisymmetric  in the indices $a$ and $b$:
\be
f^{\prime}_{abc} = - f^{\prime}_{bac}
\label{anti-f'}
\ee
and are connected to 
$f_{abc}$
by:
\be
f_{abc}~m_{c} = f^{\prime}_{cab} m_{a} - f^{\prime}_{cba} m_{b}.
\label{f-f'}
\ee

(c) The (completely symmetric) expressions 
$f^{\prime\prime}_{abc} = f^{\prime\prime}_{\{abc\}}$
verify
\be
f^{\prime\prime}_{abc}~m_{c} = 
f'_{abc}~\left[(m^{*}_{a})^{2} - (m^{*}_{b})^{2} - m_{a}^{2} + m_{b}^{2}\right].
\label{f"}
\ee

(d) the expressions
$
j^{\mu}_{a}
$
and
$
j_{a}
$
are bilinear in the Fermi matter fields: in tensor notations;
\bea
j_{a}^{\mu} = \sum_{\epsilon}~
\overline{\psi} t^{\epsilon}_{a} \otimes \gamma^{\mu}\gamma_{\epsilon} \psi
\nonumber \\
j_{a} = \sum_{\epsilon}~
\overline{\psi} s^{\epsilon}_{a} \otimes \gamma_{\epsilon} \psi
\label{current}
\eea
where  for every
$
\epsilon = \pm
$
we have defined the chiral projectors of the algebra of Dirac matrices
$
\gamma_{\epsilon} \equiv {1\over 2}~(I + \epsilon~\gamma_{5})
$
and
$
t^{\epsilon}_{a},~s^{\epsilon}_{a}
$
are 
$
|I_{4}| \times |I_{4}|
$
matrices. If $M$ is the mass matrix
$
M_{AB} = \delta_{AB}~M_{A}
$
then we must have
\be
d_{\mu}j^{\mu}_{a} = m_{a}~j_{a} 
\qquad \Leftrightarrow \qquad
m_{a}~s_{a}^{\epsilon} = i(M~t^{\epsilon}_{a} - t^{-\epsilon}_{a}~M).
\label{conserved-current}
\ee

(ii) The relation 
$
d_{Q}T = i~d_{\mu}T^{\mu}
$
is verified by:
\be
T^{\mu} = f_{abc} \left( u_{a}~v_{b\nu}~F^{\nu\mu}_{c} -
{1\over 2} u_{a}~u_{b}~d^{\mu}\tilde{u}_{c} \right)
+ f^{\prime}_{abc}~\Phi_{a}~\phi_{b}^{\mu}~u_{c}
+ j^{\mu}_{a}~u_{a}
\label{Tmu}
\ee

(iii) The relation 
$
d_{Q}T^{\mu} = i~d_{\nu}T^{\mu\nu}
$
is verified by:
\be
T^{\mu\nu} \equiv {1\over 2} f_{abc}~u_{a}~u_{b}~F_{c}^{\mu\nu}.
\ee
\label{T1}
\end{thm}
{\bf Proof:}
(i) By hypothesis we have
\be
d_{Q}T = i~d_{\mu}T^{\mu}.
\label{descent-t0}
\ee
If we apply 
$
d_{Q}
$
we obtain
$
d_{\mu}d_{Q}~T^{\mu} = 0
$
so with the Poincar\'e lemma there must exist the polynomials
$
T^{\mu\nu}
$
antisymmetric in $\mu, \nu$ such that
\be
d_{Q}T^{\mu} = i~d_{\nu}T^{\mu\nu}.
\label{descent-t}
\ee
Continuing in the same way we find
$
T^{\mu\nu\rho},~T^{\mu\nu\rho\sigma}
$
which are completely antisymmetric and we also have
\bea
d_{Q}T^{\mu\nu} = i~d_{\rho}T^{\mu\nu\rho}
\nonumber \\
d_{Q}T^{\mu\nu\rho} = i~d_{\sigma}T^{\mu\nu\rho\sigma}
\nonumber \\
d_{Q}T^{\mu\nu\rho\sigma} = 0.
\label{descent-T}
\eea
According to the preceding theorem one can choose the expressions
$
T^{I}
$
to be Lorentz covariant; we also have
\be
gh(T^{I}) = |I|.
\ee 

From the last relation we find, using Theorem \ref{Q-cohomology} that
\be
T^{\mu\nu\rho\sigma} = d_{Q}B^{\mu\nu\rho\sigma} + T_{0}^{\mu\nu\rho\sigma}
\ee
with
$
T_{0}^{\mu\nu\rho\sigma} \in {\cal P}_{0}^{(4)}.
$
The generic form of such an expression is:
\be
T_{0}^{\mu\nu\rho\sigma} = {1\over 4!}~\epsilon^{\mu\nu\rho\sigma}~f_{[abcd]}~u_{a}~u_{b}~u_{c}~u_{d};
\ee
the contributions corresponding to
$
a, b, c, d \in I_{1}
$
are primary co-cyles and the contributions for which at least one of the indices is in
$
I_{2}
$
are secondary co-cyles. 

If we substitute the preceding expression in the second relation (\ref{descent-T}) we find out
\be
d_{Q}(T^{\mu\nu\rho} - i~d_{\sigma}B^{\mu\nu\rho\sigma}) = i~d_{\sigma}T_{0}^{\mu\nu\rho\sigma}.
\ee 
The right hand side can be written as a co-boundary: we define
\be
B^{\mu\nu\rho}_{0} \equiv {1\over 3!}~\epsilon^{\mu\nu\rho\sigma}~f_{[abcd]}~u_{a}~u_{b}~u_{c}~v_{d\sigma}
\ee
and we have in fact;
\be
d_{Q}(T^{\mu\nu\rho} - i~d_{\sigma}B^{\mu\nu\rho\sigma} - B^{\mu\nu\rho}_{0}) = 0.
\ee 
We apply again Theorem \ref{Q-cohomology} and obtain
\be
T^{\mu\nu\rho} = B^{\mu\nu\rho} + i~d_{\sigma}B^{\mu\nu\rho\sigma} + T^{\mu\nu\rho}_{0}
\ee 
where
$
T_{0}^{\mu\nu\rho} \in {\cal P}_{0}^{(4)}.
$

We substitute the last relation into the first relation (\ref{descent-T}) and obtain
\be
d_{Q}(T^{\mu\nu} - i~d_{\rho}B^{\mu\nu\rho}) = i~d_{\rho}T_{0}^{\mu\nu\rho}.
\ee 
The right hand side must be a co-boundary. But it is not hard to prove that this is not possible, so we have in fact
$
f_{[abcd]} = 0 \qquad  \Leftrightarrow \qquad T_{0}^{\mu\nu\rho\sigma} = 0 
\qquad \Leftrightarrow \qquad B^{\mu\nu\rho}_{0} = 0
$
so 
\be
T^{\mu\nu\rho} = B^{\mu\nu\rho} + i~d_{\sigma}B^{\mu\nu\rho\sigma}
\ee
and
\be
d_{Q}(T^{\mu\nu} - i~d_{\rho}B^{\mu\nu\rho}) = 0.
\ee 

(ii) We use again Theorem \ref{Q-cohomology} and obtain
\be
T^{\mu\nu} - i~d_{\rho}B^{\mu\nu\rho} = d_{Q}B^{\mu\nu} + T^{\mu\nu}_{0}
\ee 
where
$
T_{0}^{\mu\nu} \in {\cal P}_{0}^{(4)}.
$
The generic form of such an expression is:
\be
T_{0}^{\mu\nu} = {1\over 2}~f^{(1)}_{[ab]c}~u_{a}~u_{b}~F_{c}^{\mu\nu}
+ {1\over 2}~f^{(2)}_{[ab]c}~\epsilon^{\mu\nu\rho\sigma}~u_{a}~u_{b}~F_{c\rho\sigma};
\ee
the contributions corresponding to
$
a, b \in I_{1}
$
are primary co-cyles and the contributions for which at least one of the indices $a, b$ is in
$
I_{2}
$
are secondary co-cyles. We substitute this in (\ref{descent-t}) and get:
\be
d_{Q}(T^{\mu} - i~d_{\nu}B^{\mu\nu}) = i~d_{\nu}T_{0}^{\mu\nu}.
\label{descent-t'}
\ee
The right hand side must be a co-boundary. But one can easily obtain that
\be
d_{\nu}T_{0}^{\mu\nu} = - i~d_{Q}B^{\mu}_{1} 
- {i\over 2}~f^{(1)}_{[ab]c}~m_{c}~u_{a}~u_{b}~\phi_{c}^{\mu}
\ee
where
\be
B^{\mu}_{1} \equiv f^{(1)}_{[ab]c}~\left(u_{a}~v_{b\nu}~F_{c}^{\nu\mu}
- {1\over 2}~u_{a}~u_{b}~d^{\mu}\tilde{u}_{c}\right)
- f^{(2)}_{[ab]c}~\epsilon^{\mu\nu\rho\sigma}~u_{a}~v_{b\nu}~F_{c\rho\sigma}.
\ee
The term 
$
u u \phi^{\mu}
$ 
must be a co-boundary and there is only the possibility:
\be
B^{\mu}_{2} \equiv f^{\prime}_{cab}~\Phi_{a}~\phi_{c}^{\mu}~u_{b}
\ee
where we can take 
$
f^{\prime}_{cab} = 0
$
if one of the indices $a, c$ is from 
$
I_{1}.
$
Now the relation
\be
- {1\over 2}~f^{(1)}_{[ab]c}~m_{c}~u_{a}~u_{b}~\phi_{c}^{\mu} = i~d_{Q}B^{\mu}_{2} 
\ee
gives the restriction:
\be
f^{(1)}_{[ab]c}~m_{c} = f^{\prime}_{cab} m_{a} - f^{\prime}_{cba} m_{b}.
\ee
If this is true then we have
\be
i~d_{\nu}T_{0}^{\mu\nu} = d_{Q}B^{\mu}_{0} 
\ee
where
\be
B^{\mu}_{0} = B^{\mu}_{1} - B^{\mu}_{2}
\ee
and (\ref{descent-t'}) becomes:
\be
d_{Q}(T^{\mu} - i~d_{\nu}B^{\mu\nu} - B^{\mu}_{0}) = 0.
\ee

(iii) Now it is again time we use Theorem \ref{Q-cohomology} and obtain
\be
T^{\mu} - B^{\mu}_{0} - i~d_{\nu}B^{\mu\nu} = d_{Q}B^{\mu} + T^{\mu}_{0}
\ee 
where
$
T_{0}^{\mu} \in {\cal P}_{0}^{(4)}.
$
The generic form of such an expression is:
\bea
T_{0}^{\mu} = u_{a}~j_{a}^{\mu} 
+ \sum_{a \in I_{3}}~\tilde{f}_{abc}~\Phi_{a}~\phi_{c}^{\mu}~u_{b}
\nonumber 
\eea
where 
$
j_{a}^{\mu}
$
has the form from the statement; but the last term can be eliminated if we redefine the expressions
$
f^{\prime}_{cab}
$
so in fact we can take:
\be
T_{0}^{\mu} = u_{a}~j_{a}^{\mu}.
\ee
It means that we have
\be
T^{\mu} = d_{Q}B^{\mu} + i~d_{\nu}B^{\mu\nu} + T^{\mu}_{1}
\ee
where
\be
T^{\mu}_{1} \equiv B^{\mu}_{0} + T_{0}^{\mu}.
\ee
Now we get from (\ref{descent-t0}) 
\be
d_{Q}(T - i~d_{\mu}B^{\mu}) = i~d_{\mu}T_{1}^{\mu}
\label{descent-0t'}
\ee
The right hand side must be a co-boundary. But one can easily obtain that
\bea
d_{\nu}T_{1}^{\mu\nu} = - i~d_{Q}B_{0} 
- {1\over 2}~f^{(1)}_{[ab]c}~u_{a}~F_{b\mu\nu}~F_{c}^{\mu\nu}
- {1\over  2}~f^{(2)}_{[ab]c}~
\epsilon_{\mu\nu\rho\sigma}~u_{a}~F^{\mu\nu}_{b}~F^{\rho\sigma}_{c} 
\nonumber \\
- m_{b}~m_{c}~f^{\prime}_{cba}~u_{a}~v_{b}^{\mu}~v_{c\mu}
+ m_{b}~(f^{\prime}_{cba} + f^{\prime}_{bca})~u_{a}~v_{b}^{\mu}~d_{\mu}\Phi_{c}
- f^{\prime}_{cab}~d_{\mu}\Phi_{a}~d^{\mu}\Phi_{c}~u_{b}
\nonumber \\
- f^{\prime}_{cab}~[ m_{c}^{2} - (m_{c}^{*})^{2}]~\Phi_{a}~\Phi_{c}~u_{b}
+ u_{a}~d_{\mu}j^{\mu}_{a}
\eea
where
\bea
B_{0} \equiv f^{(1)}_{[ab]c} \left( {1\over 2}~v_{a\mu}~v_{b\nu}~F_{c}^{\nu\mu}
+ u_{a}~v_{b}^{\mu}~ d_{\mu}\tilde{u}_{c}\right)
- f^{\prime}_{cab} (\Phi_{a}~\phi_{c}^{\mu}~v_{b\mu} + m_{c}~\Phi_{a}~\tilde{u}_{c}~u_{b})
\nonumber \\
- {1\over  2}~f^{(2)}_{[ab]c}~
\epsilon_{\mu\nu\rho\sigma}~v_{a}^{\mu}~v^{\nu}_{b}~F^{\rho\sigma}_{c}. 
\label{b0}
\eea
It means that the expression
\bea
- {i\over 2}~f^{(1)}_{[ab]c}~u_{a}~F_{b\mu\nu}~F_{c}^{\mu\nu}
- {i\over  2}~f^{(2)}_{[ab]c}~
\epsilon_{\mu\nu\rho\sigma}~u_{a}~F^{\mu\nu}_{b}~F^{\rho\sigma}_{c} 
\nonumber \\
- m_{b}~m_{c}~f^{\prime}_{cba}~u_{a}~v_{b}^{\mu}~v_{c\mu}
+ m_{b}~(f^{\prime}_{cba} + f^{\prime}_{bca})~u_{a}~v_{b}^{\mu}~d_{\mu}\Phi_{c}
- f^{\prime}_{cab}~d_{\mu}\Phi_{a}~d^{\mu}\Phi_{c}~u_{b}
\nonumber \\
- f^{\prime}_{cab}~[ m_{c}^{2} - (m_{c}^{*})^{2}]~\Phi_{a}~\Phi_{c}~u_{b}
+ u_{a}~d_{\mu}j^{\mu}_{a}
\eea
must be a co-boundary. It is easy to argue that the terms
$
u F F
$
and
$
u d\Phi d\Phi
$
cannot be written as co-boundaries so we necessarily have
\bea
f^{(1)}_{[ab]c} = - f^{(1)}_{[ac]b}, \qquad f^{(2)}_{[ab]c} = - f^{(2)}_{[ac]b},
\nonumber \\
f^{\prime}_{cab} = - f^{\prime}_{acb}.
\nonumber
\eea
It means that the constants
$
f^{(1)}_{abc}
$
and
$
f^{(2)}_{abc}
$
are completely antisymmetric and
$
f^{\prime}_{abc}
$
are antisymmetric in the first two indices. We are left with the condition:
\be
- f^{\prime}_{cab}~[ m_{c}^{2} - (m_{c}^{*})^{2}]~\Phi_{a}~\Phi_{c}~u_{b}
+ u_{a}~d_{\mu}j^{\mu}_{a} = - i~d_{Q}B_{1}
\ee
so necessarily we must have:
\be
B_{1} = \Phi_{a}~j_{a} + {1\over 3!}~f^{\prime\prime}_{\{abc\}}~\Phi_{a}~\Phi_{b}~\Phi_{c}
\ee
with 
$
j_{a}
$
as in the statement. We easily obtain (\ref{f"}) and (\ref{conserved-current}) from the statement.

(iv) If we denote 
\be
T_{1} \equiv B_{0} + B_{1}
\ee
then we have from (\ref{descent-0t'})
\be
d_{Q}(T - i~d_{\mu}B^{\mu} - T_{1}) = 0
\ee
so a last use of Theorem \ref{Q-cohomology} gives
\be
T - T^{\mu}_{1} - i~d_{\mu}B^{\mu} = d_{Q}B + T_{0}
\ee 
where
$
T_{0} \in {\cal P}_{0}^{(4)}.
$
The generic form of such an expression is:
\be
T_{0} = {1\over 3!}~\sum_{a,b,c \in I_{3}}~
\tilde{f}^{\prime\prime}_{abc}~\Phi_{a}~\Phi_{b}~\Phi_{c}
+ {1\over 4!}~\sum_{a,b,c,d \in I_{3}}~g_{\{abcd\}}~\Phi_{a}~\Phi_{b}~\Phi_{c}~\Phi_{d}
\nonumber 
\ee
but we can get rid of the first term if we redefine the expressions
$
f^{\prime\prime}_{\{abc\}}.
$
It is easy to prove that the expression
$
f^{(2)}_{[abc]}~\epsilon_{\mu\nu\rho\sigma}~v_{a}^{\mu}~v^{\nu}_{b}~F^{\rho\sigma}_{c}
$
from (\ref{b0}) is in fact a total divergence so it can be eliminated and we obtain the expression $T$ from the statement. 

(v) We prove now that $T$ from the statement is not a trivial (relative) cocycle. Indeed, if this would be true i.e.
$
T = d_{Q}B + i~d_{\mu}B^{\mu}
$
then we get 
$
d_{\mu}(T^{\mu} - d_{Q}B^{\mu}) = 0
$
so with Poincar\'e lemma we have
$
T^{\mu} = d_{Q}B^{\mu} + i~d_{\nu}B^{[\mu\nu]}.
$
In the same way we obtain from here:
$
T^{[\mu\nu]} = d_{Q}B^{[\mu\nu]} + i~d_{\rho}B^{[\mu\nu\rho]}.
$
But it is easy to see that there is no such an expression
$
B^{[\mu\nu\rho]}
$
with the desired antisymmetry property in ghost number $3$ so we have in fact
$
T^{[\mu\nu]} = d_{Q}B^{[\mu\nu]}.
$
This relation contradicts the fact that
$
T^{[\mu\nu]}
$
is a non-trivial cocycle for
$
d_{Q}
$
as it follows from Theorem \ref{m=0}.
$\qed$

If $T$ is bilinear in the fields we cannot use the Poincar\'e lemma but we can make a direct analysis. The result is the following.
\begin{thm}
Let $T$ be a relative cocycle for 
$
d_{Q}
$
which is bilinear in the fields, of canonical dimension
$
\omega(T) \leq 4
$
and ghost number
$
gh(T) = 0.
$
Then:
(i) $T$ is (relatively) cohomologous to an expression of the form:
\be
T = \sum_{a \in I_{1}}~f_{ab} ( v_{a\mu}~\phi_{b}^{\mu} - m_{b}~u_{a}~\tilde{u}_{b})
+ f^{\prime}_{\{ab\}} \phi_{a\mu}~\phi_{b}^{\mu} 
+ \sum_{a,b \in I_{3}}~f^{\prime\prime}_{\{ab\}} \Phi_{a}~\Phi_{b}.  
\ee

(ii) The relation 
$
d_{Q}T = i~d_{\mu}T^{\mu}
$
is verified with 
\be
T^{\mu} = \sum_{a \in I_{1}}~f_{ab} u_{a}~\phi_{b}^{\mu}
\ee
and we also have
$
d_{Q}T^{\mu} = 0.
$
\end{thm}

The first theorem gives us the generic form of the interaction Lagrangian for Yang-Mills models. Both theorems can be used to describe the finite renormalizations
$
R^{I}
$
(see the end of Section \ref{ggt}) which preserve gauge invariance. The expression from the first theorem produces a renormalization of the coupling constant and the expression from the second theorem produces renormalization of the propagators (or wave functions). 

In the same way one can analyze the descent equations (\ref{W}) and provide the general form of the anomalies for Yang-Mills models. We give only the result.
\begin{thm}
Let $W$ be a relative cocycle for 
$
d_{Q}
$
which is as least tri-linear in the fields, of canonical dimension
$
\omega(W) \leq 5
$
and ghost number
$
gh(W) = 1.
$
Then:
(i) $W$ is (relatively) cohomologous to a non-trivial co-cycle of the form:
\bea
W = {1\over 2}~f_{abcd} (u_{a}~v_{b\mu}~v_{c\nu}~F_{d}^{\mu\nu}
- u_{a}~u_{b}~v_{c}^{\mu}~\partial_{\mu}\tilde{u}_{d}),
\nonumber \\
- f^{\prime}_{abcd} \left(u_{a}~v_{b\mu}~\Phi_{c}~\phi_{d}^{\mu} 
- {1\over 2}~m_{d}~u_{a}~u_{b}~\Phi_{c}~\tilde{u}_{d}\right)
\nonumber \\
+ \sum_{a,b \in I_{1}}~g_{abc} \left(u_{a}~v_{b\mu}~\phi_{c}^{\mu} 
- {1\over 2}~m_{c}~u_{a}~u_{b}~\tilde{u}_{c}\right)
\nonumber \\
+ {1\over 3!}~f^{\prime\prime}_{a\{bcd\}}~u_{a}~\Phi_{b}~\Phi_{c}~\Phi_{d}
+ {1\over 4!}~\sum_{b,c,d,e \in I_{3}}~
g_{a\{bcde\}}~u_{a}~\Phi_{b}~\Phi_{c}~\Phi_{d}~\Phi_{e}
\nonumber \\
+ j^{\mu}_{ab}~u_{a}~v_{b\mu} + j_{ab}~u_{a}~\Phi_{b}
+ \sum_{a\in I_{1}}~k_{a}~u_{a}
\nonumber \\
+ h^{(1)}_{a\{bc\}}~u_{a}~F_{b}^{\mu\nu}~F_{c\mu\nu}
+ h^{(2)}_{a\{bc\}}~\epsilon_{\mu\nu\rho\sigma}~u_{a}~F_{b}^{\mu\nu}~F^{c\rho\sigma}
\nonumber \\
+ h^{(3)}_{a\{bc\}}~u_{a}~\phi_{b\mu}~\phi_{c\mu}
+ \sum_{a \in I_{1}~b,c \in I_{3}}~h^{(4)}_{a\{bc\}}~u_{a}~\Phi_{b}~\Phi_{c}.
\label{w}
\eea
We can take the constants
$
f_{abcd} = 0
$
if one of the indices is in
$
I_{3};
$
we can take
$
f^{\prime}_{abcd} = 0
$
if one of the indices $a$ and $b$ is in
$
I_{3}
$
or one of the indices $c$ and $d$ are from
$
I_{1};
$
also we can take 
$
g_{abc} = 0
$
if 
$
c \in I_{3}
$
and
$
h^{(4)}_{abc} = 0
$
if 
$
b, c \in I_{3}.
$
Moreover we have:
(a) The constants
$
f_{abcd}
$
are completely antisymmetric;
\be
f_{abcd} = f_{[abcd]}.
\ee

(b) The expressions
$
f^{\prime}_{abcd}
$
is antisymmetric in $a,b$ and in $c,d$:
\be
f^{\prime}_{abcd} = f^{\prime}_{[ab][cd]}
\ee
and verifies
\be
f_{abcd}~m_{d} = f^{\prime}_{abcd} m_{c} + f^{\prime}_{bcad} m_{a} + f^{\prime}_{cabd} m_{b}.
\ee

(c) For 
$
a \in I_{2}
$
we can write 
$f^{\prime\prime}_{abcd} = m_{a}~\tilde{f}_{abcd}$
and eliminate the completely symmetric part 
$
\tilde{f}_{\{abcd\}};
$
we also have:
\be
f^{\prime\prime}_{abcd}~m_{b} - f^{\prime\prime}_{bacd}~m_{a}
= f^{\prime}_{abcd}~\left[(m^{*}_{d})^{2} - (m^{*}_{c})^{2} - m_{c}^{2} + m_{d}^{2}\right];
\ee

(d) The expressions
$
j^{\mu}_{ab}, j_{ab}
$
and
$
k_{a}
$
are bilinear in the Fermi matter fields: in tensor notations;
\bea
j_{ab}^{\mu} = \sum_{\epsilon}~
\overline{\psi} t^{\epsilon}_{ab} \otimes \gamma^{\mu}\gamma_{\epsilon} \psi
\nonumber \\
j_{ab} = \sum_{\epsilon}~
\overline{\psi} s^{\epsilon}_{ab} \otimes \gamma_{\epsilon} \psi
\nonumber \\
k_{a} = \sum_{\epsilon}~
\overline{\psi} k^{\epsilon}_{a} \otimes \gamma_{\epsilon} \psi
\eea
and we have the relations
\be
m_{b}~s_{ab}^{\epsilon} - m_{a}~s_{ba}^{\epsilon} 
= i(M~t^{\epsilon}_{ab} - t^{-\epsilon}_{ab}~M).
\ee

(ii) The relation 
$
d_{Q}W = - i~d_{\mu}W^{\mu}
$
is verified by:
\bea
W^{\mu} = f_{abcd} \left( {1\over 2}~u_{a}u_{b}~v_{c\nu}~F^{\mu\nu}_{d}
+ {1\over 3!}~u_{a}~u_{b}~u_{c}~d^{\mu}\tilde{u}_{d} \right)
- {1\over 2}~f^{\prime}_{abc}~u_{a}~u_{b}~\Phi_{c}~\phi_{d}^{\mu}
\nonumber \\
+ {1\over 2}~\sum_{a,b \in I_{1}}~g_{abc}~u_{a}~u_{b}~\phi_{c}^{\mu}
+ {1\over 2}~j^{\mu}_{ab}~u_{a}~u_{b}.
\eea

(iii) The relation 
$
d_{Q}W^{\mu} = i~d_{\nu}W^{\mu\nu}
$
is verified by:
\be
W^{\mu\nu} \equiv {1\over 3!}~f_{abcd}~u_{a}~u_{b}~u_{c}~F_{d}^{\mu\nu}.
\ee

(iv) If we have 
$
W = 0
$
i.e. the equation (\ref{ym1}) does not have anomalies, then we also have
$
W^{\mu} = 0,\quad W^{\mu\nu} = 0.
$
\label{W1}
\end{thm}

If the expression $W$ is bilinear in the fields we can make a direct analysis:
\begin{thm}
Let $W$ be a relative cocycle for 
$
d_{Q}
$
which is bilinear in the fields, of canonical dimension
$
\omega(W) \leq 5
$
and ghost number
$
gh(W) = 1.
$
Then $W$ is (relatively) cohomologous to an expression of the form:
\be
W = \sum_{a \in I_{1}, b \in I_{3}}~g_{ab} u_{a}~\Phi_{b} 
\ee
and we have 
$
d_{Q}W = 0.
$
\end{thm} 

As a matter of terminology, if in the generic scheme presented above we have 
$
I_{2} = I_{3} = \emptyset
$ 
we say that we have a {\it pure gauge model}. The physically relevant cases are quantum electro-dynamics and quantum chromo-dynamics. If 
$
I_{2} \not= \emptyset
$ 
we say that the theory is {\it spontaneously broken}. In this case it can be proved that we must necessarily have
$
I_{3} \not= \emptyset;
$ 
without Higgs fields gauge invariance is not valid already in the second order of perturbation theory. The physically relevant case is the electro-weak interaction (the standard model).

Using Wick expansion property (\ref{wick-chrono2}) one can prove that the tree graphs give anomalies only for 
$
n = 2, 3.
$
\newpage
\section{Yang-Mills Models in Higher Orders of Perturbation Theory\label{yang}}

The theory is gauge invariant in all orders {\it iff} we can prove that 
$
W = 0
$
in an arbitrary order. This is possible in some simple cases like quantum electro-dynamics.
We have to take in the generic scheme presented in the preceding Section 
$
|I_{1}| = |I_{4}| = 1, \quad I_{2} = I_{3} = \emptyset.
$
So we have a triplet 
$
(v_{\mu},u,\tilde{u})
$
of null mass fields (
$
v_{\mu}
$
is called the {\it electromagnetic potential}) and one Dirac field of mass $M$ with the interaction Lagrangian
\be
T = :v_{\mu} \overline{\psi} \gamma^{\mu} \psi:
\ee
and
\be
T^{\mu} = :u \overline{\psi} \gamma^{\mu} \psi:
\ee
An important observation is the following one. Let us define the so-called 
{\it charge conjugation} operator according to
\bea
U_{c}~v_{\mu}~U_{c}^{-1} = - v_{\mu}, \qquad 
U_{c}~u~U_{c}^{-1} = - u, \qquad
U_{c}~\tilde{u}~U_{c}^{-1} = - \tilde{u},
\nonumber \\
U_{c}~\psi~U_{c}^{-1} = - C~\gamma_{0}~\psi^{\dagger},
\nonumber \\
U_{c}\Omega = 0
\eea
where $C$ is the {\it charge conjugation matrix}. Then we can easily prove that 
\be
U_{c}~T~U_{c}^{-1} = T, \qquad U_{c}~T^{\mu}~U_{c}^{-1} = T^{\mu}.
\ee
The result (sometimes called Furry theorem) is then:
\begin{thm}
The chronological products can be chosen such that the theory is gauge invariant in all orders of perturbation theory.
\end{thm}
{\bf Proof:}
(i) First we can define the chronological products such that they are charge conjugation invariant in all orders of perturbation theory by induction. We suppose that the assertion is true up to order $n - 1$ i.e. 
\bea
U_{c}~T^{I_{1},\dots,I_{k}}~U_{c}^{-1} = T^{I_{1},\dots,I_{k}}, \qquad k < n.
\nonumber
\eea
If
$
T^{I_{1},\dots,I_{n}}
$
do not verify this relation we simply replace:
\be
T^{I_{1},\dots,I_{n}}  \rightarrow 
{1\over 2}~(T^{I_{1},\dots,I_{n}} + U_{c}~T^{I_{1},\dots,I_{n}}~U_{c}^{-1}).
\ee

So we can suppose that we have 
\be
U_{c}~T^{I_{1},\dots,I_{k}}~U_{c}^{-1} = T^{I_{1},\dots,I_{k}}, \qquad \forall n.
\ee

(ii) Suppose now that the theory is gauge invariant up to order $n - 1$. Then in order $n$ we might have the anomaly $W$. From the preceding relation we have however:
\be
U_{c}~W~U_{c}^{-1} = W.
\ee
In our particular case the relation (\ref{w}) considerably simplifies:
\be
W = u~\overline{\psi}\psi + u~\overline{\psi}\gamma_{5}\psi
+ h^{(1)}~u~F^{\mu\nu}~F_{\mu\nu}
+ h^{(2)}~\epsilon_{\mu\nu\rho\sigma}~u~F^{\mu\nu}~F^{\rho\sigma}.
\ee
If we substitute this generic expression in the preceding relation we obtain 
$
W = 0
$
which proves gauge invariance in order $n$.
$\qed$

In the similar way one can treat other models for which a charge conjugation operator do exists e.g. $SU(n)$ invariant models without spontaneously broken symmetry.

Now we consider again the generic case from the preceding Section. One can compute explicitly the expression of the anomaly $W$ in the second order of the perturbation theory. Imposing 
$
W = 0
$
one finds out new restrictions on the various constants. The computations are given in \cite{standard}, \cite{fermi} and \cite{ano} so we give only the results. Computing 
$
A_{3}^{[\mu\nu]}
$
we find 
\be
f_{abcd} = 2 i~(f_{abe}~f_{cde} + f_{bce}~f_{ade} + f_{cae}~f_{bde})
\ee
so if we impose 
$
f_{abcd} = 0
$
we find out that the constants
$
f_{abc}
$
verify Jacobi identities. Computing 
$
A_{2}^{\mu}
$
we find the same expression for 
$
f_{abcd}
$
and moreover
\be
f^{\prime}_{abcd} = 2 i~(f_{abe}~f^{\prime}_{cde} 
+ f^{\prime}_{cae}~f^{\prime}_{edb} - f^{\prime}_{ceb}~f^{\prime}_{eda})
\ee
\be
t_{ab}^{\epsilon} = 2~([t_{a}^{\epsilon}~t_{b}^{\epsilon}] - i~f_{abc}~t_{c}^{\epsilon})
\ee
so the cancellationn of this anomaly tells us that 
$
t_{a}^{\epsilon}
$
and
$
(T_{c})_{ab} = - f^{\prime}_{abc}
$
are representations of the Lie algebra with structure constants
$
f_{abc}.
$

Finally, computing 
$
A_{1}
$
we find the same expressions for 
$
f_{abcd},~f^{\prime}_{abcd},~t_{ab}^{\epsilon}
$
and moreover
\be
s_{ab}^{\epsilon} = 2~
(t_{a}^{-\epsilon}~s_{b}^{\epsilon} - s_{b}^{\epsilon}~t_{a}^{\epsilon}
+ i~f^{\prime}_{cba}~s_{c}^{\epsilon})
\ee
\be
f^{\prime\prime}_{abcd} = 2 i~H_{abcd}, \qquad a \in I_{1}
\ee
\be
f^{\prime}_{abcd} = i~m_{a}~(F_{abcd} - F_{\{abcd\}}), \qquad a \in I_{2}
\ee
where
\be 
H_{abcd} = f^{\prime}_{eba}~f^{\prime\prime}_{ecd} + f^{\prime}_{eca}~f^{\prime\prime}_{ebd}
+ f^{\prime}_{eda}~f^{\prime\prime}_{ebc}
\ee
and 
\be
F_{abcd} \equiv \left\{\begin{array}{rcl} 
{2 \over m_{a}}~H_{abcd} & \mbox{for} &  a \in I_{2}\\
0 & \mbox{for} & a \in I_{1} \cup I_{3}\end{array}\right.
\ee
We also have
\be
g_{ab_{1}\dots b_{4}} = 8 i~{\cal S}_{b_{1},\dots,b_{4}}~
(f^{\prime}_{eb_{1}a}~g_{eb_{2}b_{3}b_{4}})
\ee
and all other possible pieces of the anomaly (\ref{w}) are null. The explicit expressions for the finite renormalizations which must be used to put $W$ in such a form are: 
\bea
T(T^{\mu\nu}(x_{1}),T(x_{2})) \rightarrow T(T^{\mu\nu}(x_{1}),T(x_{2}))
+ \delta(x_{1} - x_{2})~N^{\mu\nu}(x_{1})
\nonumber \\
T(T^{\mu\nu}(x_{1}),T^{\rho}(x_{2})) \rightarrow T(T^{\mu\nu}(x_{1}),T^{\rho}(x_{2}))
+ \delta(x_{1} - x_{2})~N^{\mu\nu;\rho}(x_{1})
\nonumber \\
T(T^{\mu}(x_{1}),T(x_{2})) \rightarrow T(T^{\mu}(x_{1}),T(x_{2}))
+ \delta(x_{1} - x_{2})~N^{\mu}(x_{1})
\nonumber \\
T(T^{\mu}(x_{1}),T^{\nu}(x_{2})) \rightarrow T(T^{\mu}(x_{1}),T^{\nu}(x_{2}))
+ \delta(x_{1} - x_{2})~\tilde{N}^{\mu\nu}(x_{1})
\nonumber \\
T(T(x_{1}),T(x_{2})) \rightarrow T(T(x_{1}),T(x_{2}))
+ \delta(x_{1} - x_{2})~N(x_{1})
\eea
where:
\bea
N^{\mu\nu} \equiv {1\over 2}~f_{abe}~f_{cde}~u_{a}~u_{b}~v_{c}^{\mu}~v_{d}^{\nu}
\nonumber \\
N^{\mu\nu;\rho}\equiv - {1\over 2}~
f_{abe}~f_{cde}~[\eta^{\mu\rho}~u_{a}~u_{b}~u_{c}~v_{d}^{\nu} - (\mu \leftrightarrow \nu)]
\nonumber \\
N^{\mu} \equiv f_{abe}~f_{cde}~u_{a}~v_{b}^{\mu}~v_{c}^{\nu}~v_{d\nu}
+ f^{\prime}_{cea}~f^{\prime}_{edb}~u_{a}~v_{b}^{\mu}~\Phi_{c}~\Phi_{d}
\nonumber \\
\tilde{N}^{\mu\nu} \equiv f_{abe}~f_{cde}~u_{a}~v_{b}^{\nu}~u_{c}~v_{d}^{\mu}
\nonumber \\
N \equiv {1\over 2}~f_{abe}~f_{cde}~v_{a}^{\mu}~v_{b}^{\nu}~v_{c\mu}~v_{d\nu}
+ f^{\prime}_{cea}~f^{\prime}_{edb}~v_{a\mu}~v_{b}^{\mu}~\Phi_{c}~\Phi_{d}
\nonumber \\
+ {1\over 2}~\sum_{a \in I_{2}}~{1\over m_{a}}~
f^{\prime}_{eba}~f^{\prime\prime}_{ecd}~v_{a\mu}~v_{b}^{\mu}~\Phi_{c}~\Phi_{d}
\label{n}
\eea
If we go to the third order of perturbation theory and use the Wick expansion property (\ref{wick-chrono2}) we obtain a much simpler expression for the generic anomaly:
\bea
W = \sum_{a,b \in I_{1}}~g_{abc} \left(u_{a}~v_{b\mu}~\phi_{c}^{\mu} 
- {1\over 2}~m_{c}~u_{a}~u_{b}~\tilde{u}_{c}\right)
+ \sum_{a \in I_{1}}~k_{a}~u_{a}
\nonumber \\
+ h^{(1)}_{abc}~u_{a}~F_{b}^{\mu\nu}~F_{c\mu\nu}
+ h^{(2)}_{abc}~\epsilon_{\mu\nu\rho\sigma}~u_{a}~F_{b}^{\mu\nu}~F^{c\rho\sigma}
+ h^{(3)}_{abc}~u_{a}~\phi_{b\mu}~\phi_{c\mu}
+ \sum_{a \in I_{1}~b,c \in I_{3}}~h^{(4)}_{abc}~u_{a}~\Phi_{b}~\Phi_{c}
\nonumber \\
+ {1\over 3!}~f^{\prime\prime}_{a\{bcd\}}~u_{a}~\Phi_{b}~\Phi_{c}~\Phi_{d}
+ {1\over 4!}~\sum_{b,c,d,e \in I_{3}}~
g_{a\{bcde\}}~u_{a}~\Phi_{b}~\Phi_{c}~\Phi_{d}~\Phi_{e}
\eea 

Explicit computations gives non-null expressions for 
$
h^{(2)}_{abc}
$
(the so-called {\it axial anomaly}) and 
$
g_{a\{bcde\}}
$
which gives the value of the quadri-linear Higgs coupling i.e. a supplementary term in the last relation (\ref{n}).

Let us provide as a particular case the standard model of the electro-weak interactions.
We have to take in the general scheme:
$
I_{1} = I_{\rm ph} \cup I_{g}
$
where
$
|I_{1}| = 1,~|I_{2}| = 3,~|I_{3}| = 1;
$ 
we denote the corresponding indices by $0$, $1,2,3$, $H$ and 
$
j \in I_{g}
$ 
respectively. The vector fields corresponding to 
$
I_{\rm ph}, I_{2}
$
and
$
I_{q}
$
are the {\it photon}, the {\it heavy Bosons} and the {\it gluons}. The field 
$
\phi_{H}
$
is called the {\it Higgs} field. We also have:
$ 
|I_{4}| = 8 {\cal N}
$
where 
$
{\cal N}
$
is called the number of {\it generations}. Then the non-zero constants
$
f_{abc}
$
for the values 
$
I_{1} \cup I_{2}
$
are:
\be
f_{210} = g~sin \theta, \quad f_{321} = g cos~\theta, 
\quad f_{310} = 0, \quad f_{320} = 0
\ee
with
$cos \theta > 0, \quad g > 0$
and the other constants determined through the anti-symmetry property. The expressions
$
f_{jkl}, j,k,l \in I_{g}
$
are the structure constants of the Lie algebra
$
su(3)
$
and this means that
$
|I_{g}| = 8. 
$

The Jacobi identity is verified and the corresponding Lie algebra is isomorphic to
$
u(1) \times su(2) \times su(3).
$
The angle 
$\theta$, 
determined by the condition
$cos~\theta > 0$
is called the {\it Weinberg angle}. The masses of the heavy Bosons are constrained by:
\be
m_{1} = m_{2} = m_{3} cos~\theta;
\ee

The non-zero constants
$f^{\prime}_{abc}$
are completely determined by the antisymmetry property in the first two 
indices and:
\bea
f^{\prime}_{H11} = f'_{H22} = {\varepsilon~g\over 2}, \quad
f^{\prime}_{H33} = {\epsilon~g \over 2cos~\theta}, \quad
f^{\prime}_{21H} = g~sin~\theta,
\nonumber \\
f^{\prime}_{321} = - f'_{312} = {g \over 2}, \quad
f^{\prime}_{123} = - g~{cos~2\theta \over 2cos~\theta},
\eea
the rest of them being zero. Here 
$\varepsilon = \pm$
but if
$\epsilon = -1$
we can make the redefinition
$
\phi_{H} \rightarrow - \phi_{H}
$
and make
$\epsilon = 1$.

The non-zero constants
$f^{\prime\prime}_{abc}$
are determined by:
\be
f^{\prime\prime}_{H11} = f^{\prime\prime}_{H22} = f^{\prime\prime}_{H33} = 
{g\over 2 m_{1}} m_{H}^{2},
\qquad 
f^{\prime\prime}_{HHH} = {3 m_{H}^{2} \over 2}
\ee
and we also have
\be
g_{HHHH} = 0.
\ee
Moreover, we must have a supplementary term in the last relation from (\ref{n}) such that the known form of the Higgs potential is obtained.

The Dirac fields are considered with values in
$
\C^{2} \otimes \C^{4 {\cal N}}
$
so use a matrix notation i.e. we put
\be
\psi = \left( \matrix{ \psi_{1} \cr \psi_{2}} \right)
\ee
with
$
\psi_{1}, \psi_{2} \in \C^{4 {\cal N}}.
$
Then 
\bea
t^{+}_{1} = {1\over 2} g \left( \matrix{ 0 & C^{-1} \cr C & 0} \right) \quad 
t^{+}_{2} = {1\over 2} g \left( \matrix{ 0 & -i C^{-1} \cr i C & 0} \right) 
\nonumber \\
t^{+}_{3} = {1\over 2}
\left[ -g cos~\theta \left( \matrix{ I & 0 \cr 0 & - I} \right) 
+ g' sin~\theta {\bf 1}\right]
\nonumber \\
t^{+}_{0} = - {1\over 2}
\left[ g sin~\theta \left( \matrix{ I & 0 \cr 0 & - I} \right) 
+ g' cos~\theta {\bf 1}\right]
\eea
\be
t^{-}_{1} = t^{-}_{2} =  0, \quad 
t^{-}_{3} = - tg~\theta~t^{+}_{0}, \quad t^{-}_{0} = t^{+}_{0}
\ee
with $C$ a 
$4~{\cal N} \times 4~{\cal N}$
unitary matrix, $I$ the
$4~{\cal N} \times 4~{\cal N}$
unit matrix and
\be
g' = g \left( \matrix{ D & 0 \cr 0 & - I} \right) 
\label{g'}
\ee
with $D$ a diagonal, traceless
$
Tr(D) = 0
$
and Hermitian
$4~{\cal N} \times 4~{\cal N}$
matrix which commutes with $C$.  The matrix $C$ is called the {\it Cabibbo-Kobayashi-Maskawa (CKM)} matrix.
Because every Fermi fields can be redefined by multiplication with a phase factor without changing the physics (i.e. the expressions
$
T^{I}
$
) one can use this freedom to put this matrix in a preferred form \cite{Sc2}. 
It seems that there are only 
$
{\cal N} = 3
$ 
generations and the corresponding fields 
$
\psi_{1j}, \psi_{2j},~j = 1,\dots,12
$
are
\bea
\psi_{1} = \nu_{e},\nu_{\mu},\nu_{\tau},u_{p},c_{p},t_{p}
\nonumber \\
\psi_{2} = {\bf e},\mu,\tau,d_{p},s_{p},b_{p}.
\eea
Here the Dirac fields
$
{\bf e},\mu,\tau
$ 
are the {\it leptons} (producing the electron and the particles $\mu$ and $\tau$),
$
\nu_{e},\nu_{\mu},\nu_{\tau}
$
the associated neutrinos and the Dirac fields
$
u_{p},c_{p},t_{p},d_{p},s_{p},b_{p}
$
are the {\it quarks (up, charm, top, down, strange, bottom)} each with 
$
p = 1,2,3
$
{\it colors}. 

All the preceding conditions are compatible with gauge invariance conditions up to the third order of perturbation theory.

One can introduce the {\it electric charge} operator according
$
Q_{e}
$
to
\bea
Q_{e}\Omega
\nonumber \\
~[Q_{e}, v^{\mu}_{1}] = ie~v^{\mu}_{2}, \quad [Q_{e}, v^{\mu}_{2}] = - ie~v^{\mu}_{1},
\nonumber \\
~[Q_{e}, \Phi_{1}] = ie~\Phi_{2}, \quad [Q_{e}, \Phi_{2}] = - ie~\Phi_{1},
\nonumber \\
~[Q_{e}, u_{1}] = ie~u_{2}, \quad [Q_{e}, u_{2}] = - ie~u_{1},
\nonumber \\
~[Q_{e}, \tilde{u}_{1}] = ie~u_{2}, \quad [Q_{e}, \tilde{u}_{2}] = - ie~u_{1},
\nonumber \\
~[Q_{e}, \psi] = i~t_{0}^{+}~\psi
\label{electric}
\eea
and the rest of the fields are commuting with
$
Q_{e};
$
here $e$ is a positive numbere (the {\it electric charge}). Then one can prove that the electric charge is leaving invariant the expressions
$
T^{I}:
$
\be
~[Q_{e},T^{I}] = 0.
\ee
If one takes the matrix $D$ from the expression (\ref{g'}) to be proportional to
$
- tan(\theta)
$
in the lepton sector and
$
{1\over 3}~tan(\theta)
$
in the quark sector, then we have the condition of tracelessness for $D$; moreover, the lepton states will have charge 
$
- e,
$ 
the quarks 
$
u, c, t
$
will have charge 
$
{2e\over 3}
$
and the quarks 
$
d, s, b
$
will have charge 
$
- {e\over 3}.
$
\section{Conclusions}

The cohomological methods presented in this paper leads to the most simple understanding of quantum gauge models in lower orders of perturbation theory and extract completely the information from the consistency Wess-Zumino equations. We have illustrate the methods for the case of Yang-Mills models. In a subsequent paper we will consider the same methods for case of gravity considered as a perturbative theory of particles of helicity (spin) $2$.

{\bf Acknowledgements:} The author had many interesting discussions on the topics of this paper with prof. G. Scharf
\newpage
\section{Appendix}
In this Appendix we prove a {\it trace decomposition} result:
\begin{thm}
Let 
$
t_{\mu_{1},\dots,\mu_{n}}
$
be a Lorentz covariant tensor and also parity invariant. Then one can write this tensor in the following form:
\be
t_{\mu_{1},\dots,\mu_{n}} = \sum_{P}~\eta_{I_{1}}\dots \eta_{I_{k}}~t^{P}_{I_{0}}
\ee
where the sum goes over the partitions
$
P = \{I_{0},\dots,I_{k}\}
$
of the set 
$
\{\mu_{1},\dots,\mu_{n}\}
$
such that 
$
|I_{1}| = \cdots = |I_{k}| = 2
$
and the tensors
$
t^{P}_{I_{0}}
$
are Lorentz covariant, parity invariant and also traceless. These tensors can be obtained from various traces of the tensor
$
t_{\mu_{1},\dots,\mu_{n}}.
$
\end{thm}
{\bf Proof:}
(i) As it is usual in such sort of problems it is convenient to consider instead of
$
t_{\mu_{1},\dots,\mu_{n}}.
$
the associated 
$
SL(2,\C)-
$
covariant tensor:
\be
t_{a_{1},\dots,a_{n};\bar{b}_{1},\dots,\bar{b}_{n}} \equiv
\sigma^{\mu_{1}}_{a_{1}\bar{b}_{1}} \dots \sigma^{\mu_{n}}_{a_{n}\bar{b}_{n}}~
t_{\mu_{1},\dots,\mu_{n}}.
\label{t}
\ee
Here
$
\sigma^{\mu} = (I, \sigma_{1},\sigma_{2},\sigma_{3})
$
are the Pauli matrices and we use dotted and undotted Weyl indices
$
a, \bar{b} = 1,2.
$
We will use in the following a number of formulas involving Pauli matrices. We find convenient to list them. First we define:
\be
\sigma^{\mu\nu}_{ab} \equiv {i\over 4}~
[\sigma^{\mu}_{a\bar{b}}~\epsilon^{\bar{b}\bar{d}}~\sigma^{\mu}_{b\bar{d}} 
- ( \mu \leftrightarrow \nu)]
\qquad
\bar{\sigma}^{\mu\nu}_{\bar{c}\bar{d}} \equiv - {i\over 4}~
[\sigma^{\mu}_{a\bar{c}}~\epsilon^{ab}~\sigma^{\mu}_{b\bar{d}} 
- ( \mu \leftrightarrow \nu)].
\ee
The first expression is symmetric in $a,b$ and the second is symmetric in 
$
\bar{c},\bar{d}.
$
Then:
\be
\sigma^{\mu}_{a\bar{b}}~\epsilon^{\bar{b}\bar{d}}~\sigma^{\nu}_{c\bar{d}} = 
\epsilon_{ca}~g^{\mu\nu} - 2 i~\sigma^{\mu\nu}_{ac}, 
\label{sigma1}
\ee
\be
\eta_{\mu\nu} \sigma^{\mu}_{a\bar{b}} \sigma^{\nu}_{c\bar{d}} = 
2 \epsilon_{ac}~\epsilon_{\bar{b}\bar{d}},
\label{sigma2}
\ee
\be
\eta_{\alpha\rho}\sigma^{\alpha\beta}_{ab}~\sigma^{\rho}_{c\bar{d}} = - {i\over 2}~
(\epsilon_{ac}~\sigma^{\beta}_{b\bar{d}} + \epsilon_{bc}~\sigma^{\beta}_{a\bar{d}}),
\label{sigma3}
\ee
\be
\eta_{\alpha\beta}~\sigma^{\mu\alpha}_{ab}~\sigma^{\nu\beta}_{cd} = 
- {1\over 4}~(\epsilon_{ac}~\epsilon_{bd} + \epsilon_{ad}\epsilon_{bc}) \eta^{\mu\nu}
- {i\over 2}~( \epsilon_{ac}~\sigma^{\mu\nu}_{bd} + \epsilon_{ad}~\sigma^{\mu\nu}_{bc} + 
\epsilon_{bc}~\sigma^{\mu\nu}_{ad} + \epsilon_{bd}~\sigma^{\mu\nu}_{ac} ),
\label{sigma4}
\ee
\be
\eta_{\alpha\beta}~
\bar{\sigma}^{\mu\alpha}_{\bar{a}\bar{b}}~\bar{\sigma}^{\nu\beta}_{\bar{c}\bar{d}} = 
- {1\over 4}~(\epsilon_{\bar{a}\bar{c}}~\epsilon_{\bar{b}\bar{d}} + \epsilon_{\bar{a}\bar{d}}\epsilon_{\bar{b}\bar{c}}) \eta^{\mu\nu}
+ {i\over 2}~( \epsilon_{\bar{a}\bar{c}}~\bar{\sigma}^{\mu\nu}_{\bar{b}\bar{d}} 
+ \epsilon_{\bar{a}\bar{d}}~\bar{\sigma}^{\mu\nu}_{\bar{b}\bar{c}} + 
\epsilon_{\bar{b}\bar{c}}~\bar{\sigma}^{\mu\nu}_{\bar{a}\bar{d}} 
+ \epsilon_{\bar{b}\bar{d}}~\bar{\sigma}^{\mu\nu}_{\bar{a}\bar{c}} ),
\label{sigma5}
\ee
\be
\eta_{\alpha\beta}~\sigma^{\mu\alpha}_{ab}~\bar{\sigma}^{\nu\beta}_{\bar{c}\bar{d}} = 
- {1\over 8}~[ \sigma^{\mu}_{a\bar{c}}~\sigma^{\nu}_{b\bar{d}}
+ (a \leftrightarrow b) + (\bar{c} \leftrightarrow \bar{d}) 
+ (a \leftrightarrow b, \bar{c} \leftrightarrow \bar{d})],
\label{sigma6}
\ee
\bea
\sigma^{\mu\nu}_{ab}~\epsilon^{bd}~\sigma^{\alpha\beta}_{cd} = 
- {1\over 4}~\epsilon_{ac}~(\eta^{\mu\beta}~\eta^{\nu\alpha} - \eta^{\nu\beta}~g^{\mu\alpha}
+ i~\epsilon^{\mu\nu\alpha\beta} )
\nonumber \\
+ {i\over 4}~(\eta^{\mu\beta}~\sigma^{\alpha\nu}_{ac} - \eta^{\nu\beta}~\sigma^{\alpha\mu}_{ac}
- \eta^{\mu\alpha}~\sigma^{\beta\nu}_{ac} + \eta^{\nu\alpha}~\sigma^{\beta\mu}_{ac} )
+ {i\over 4}~({\epsilon^{\mu\nu\beta}}_{\rho}~\sigma^{\alpha\rho}_{ac} 
- {\epsilon^{\mu\nu\alpha}}_{\rho}~\sigma^{\beta\rho}_{ac}),
\label{sigma7}
\eea
\be
\sigma^{\mu\nu}_{ab}~\epsilon^{bd}~\sigma^{\alpha}_{d\bar{c}} = 
{i\over 2}~(\eta^{\mu\alpha}~\sigma^{\nu}_{a\bar{c}} - \eta^{\nu\alpha}~\sigma^{\mu}_{a\bar{c}}
- i~{\epsilon^{\mu\nu\alpha}}_{\beta}~\sigma^{\beta}_{a\bar{c}} ).
\label{sigma8}
\ee

(ii) The correspondence between 
$
t_{\mu_{1},\dots,\mu_{n}}
$
and
$
t_{a_{1},\dots,a_{n};\bar{b}_{1},\dots,\bar{b}_{n}}
$
is one-one because we have the formulas (\ref{sigma1}) and (\ref{sigma2}). We have:
\be
t^{\mu_{1},\dots,\mu_{n}} = {1\over 2^{n}}~
\sigma^{\mu_{1}}_{a_{1}\bar{b}_{1}} \dots \sigma^{\mu_{n}}_{a_{n}\bar{b}_{n}}~
t^{a_{1},\dots,a_{n};\bar{b}_{1},\dots,\bar{b}_{n}}
\label{tt}
\ee
where the Weyl indices are raised and lowered with the metric 
$
\epsilon_{ab}
$
and
$
\epsilon_{\bar{a}\bar{b}}
$
e.g.
$
t^{a} = \epsilon^{ab}~t_{b}.
$

(iii) We consider an arbitrary tensor 
$
t_{a_{1},\dots,a_{n}}
$
and we decompose it with respect to the first two indices, into the symmetric and antisymmetric part:
\be
t_{a_{1},\dots,a_{n}} = t_{\{a_{1},a_{2}\},a_{3},\dots,a_{n}}
+ \epsilon_{a_{1}a_{2}}~t_{a_{3},\dots,a_{n}}
\ee 
Now we have by direct computation:
\bea
t_{\{a_{1},a_{2}\},a_{3},\dots,a_{n}} = t_{\{a_{1},a_{2},a_{3}\},\dots,a_{n}}
\nonumber \\
+ {1\over 3}~(t_{\{a_{1},a_{2}\},a_{3},\dots,a_{n}} - t_{\{a_{1},a_{3}\},a_{2},\dots,a_{n}})
+ {1\over 3}~(t_{\{a_{1},a_{2}\},a_{3},\dots,a_{n}} - t_{\{a_{2},a_{3}\},a_{1},\dots,a_{n}})
\eea
and the second (third) term is antisymmetric in 
$
a_{2}, a_{3}
$
(resp. in
$
a_{1}, a_{3}).
$
It means that we have in fact a decomposition:
\be
t_{a_{1},\dots,a_{n}} = 
\epsilon_{a_{1},a_{2}}~t^{(3)}_{a_{3},\dots,a_{n}}
+ \epsilon_{a_{2}a_{3}}~t^{(1)}_{a_{1},a_{4},\dots,a_{n}}
+ \epsilon_{a_{3}a_{1}}~t^{(2)}_{a_{2},a_{4},\dots,a_{n}}
+ t_{\{a_{1},a_{2},a_{3}\},\dots,a_{n}}.
\ee
We continue by recursion and we find out
\be
t_{a_{1},\dots,a_{n}} = \sum_{P}~\epsilon_{I_{1}}~t^{(P)}_{I_{0}}
+ t_{\{a_{1},a_{2},a_{3},\dots,a_{n}\}}
\ee 
where 
$
P = \{I_{0},I_{1}\}
$
is a partition of the set 
$
A \equiv \{a_{1},\dots,a_{n}\}
$
such that
$
|I_{1}| = 2.
$
We apply the same argument to every tensor 
$
t^{(P)}_{I_{0}}
$
and at the very end we get the decomposition formula:
\be
t_{a_{1},\dots,a_{n}} = \sum_{P}~\epsilon_{I_{1}}\dots \epsilon_{I_{k}}~t^{(P)}_{I_{0}}
\ee 
where 
$
P = \{I_{0},\dots,I_{k}\}
$
is a partition of the set 
$
A \equiv \{a_{1},\dots,a_{n}\}
$
such that
$
|I_{1}| = \dots = |I_{k}| = 2
$
and the tensors 
$
t^{(P)}_{I_{0}}
$
are completely symmetric. In the same way we have:
\be
t_{a_{1},\dots,a_{n};\bar{b}_{1},\dots,\bar{b}_{n}} = \sum_{P,Q}~
\epsilon_{I_{1}}\dots \epsilon_{I_{k}}~\epsilon_{\bar{J}_{1}}\dots \epsilon_{\bar{J}_{l}}~
t^{(P,Q)}_{I_{0},\bar{J}_{0}}
\label{ttt}
\ee 
where 
$
P = \{I_{0},\dots,I_{k}\}
$
is a partition of the set 
$
A \equiv \{a_{1},\dots,a_{n}\}
$
and 
$
Q = \{\bar{J}_{0},\dots,\bar{J}_{l}\}
$
is a partition of the set 
$
B \equiv \{\bar{b}_{1},\dots,\bar{b}_{n}\}
$
such that
$
|I_{1}| = \dots = |I_{k}| = |\bar{J}_{1}| = \dots = |\bar{J}_{l}| = 2;
$
the tensors 
$
t^{(P,Q)}_{I_{0},\bar{J}_{0}}
$
are completely symmetric in the dotted and undotted indices. The preceding formula is in fact the decomposition in irreducible tensors: the tensor
$
t^{(P,Q)}_{I_{0},\bar{J}_{0}}
$
transforms according to the irreducible representation
$
D^{\left(|I_{0}|/2,|\bar{J}_{0}|\right)}.
$

(iv) We consider all possible terms from (\ref{ttt}) and the contributions they are producing in (\ref{tt}). The term without $\epsilon$ factors from (\ref{ttt}) is producing in (\ref{tt}) a traceless contribution because of (\ref{sigma2}). We consider a term with at least one factor
$
\epsilon_{\bar{J}}
$
and use the formula (\ref{sigma1}) to eliminate all such factors. Because the representation is irreducible only one of the 
$
2^{l}
$
resulting contributions can be non-zero. So we must have either a contribution with at least 
$\eta$ factor or a contribution only with factors
$
\sigma^{\mu\nu}_{ab}
$
i.e. of the type:
\be
\sigma^{\alpha_{1}\beta_{1}}_{a_{1}b_{1}}\dots \sigma^{\alpha_{p}\beta_{p}}_{a_{p}b_{p}}~
\sigma^{\rho_{1}}_{c_{1}\bar{d}_{1}}\dots \sigma^{\rho_{q}}_{c_{q}\bar{d}_{q}}~
t^{a_{1},\dots,a_{p};b_{1},\dots,b_{p};c_{1},\dots,c_{q};\bar{d}_{1},\dots,\bar{d}_{q}}
\label{sigma'}
\ee
and we must prove that these contributions are producing in (\ref{tt}) either traceless terms or terms with one factor $\eta$. We have two cases: if the contribution is without factors
$
\epsilon_{I}
$
then the tensor 
$
t^{a_{1},\dots,a_{p};b_{1},\dots,b_{p};c_{1},\dots,c_{q};\bar{d}_{1},\dots,\bar{d}_{q}}
$
must be completely symmetric in the dotted and undotted indices. But in this case one can show that the contribution induced in (\ref{tt}) it is traceless: if we take the trace of two indices of type $\rho$ we use (\ref{sigma2}),  if we take the trace between an index of type $\alpha$ and an index of type $\rho$ we use (\ref{sigma3}) and if we take the trace between two indices of type $\alpha$ we use (\ref{sigma4}).

So in the preceding formula it remains to consider the case when we have at least one factor
$
\epsilon^{I}.
$
Again we have two subcases: if the factor
$
\epsilon^{I}
$
is of the type
$
\epsilon^{b_{j}b_{k}}
$
we use the formulas (\ref{sigma7}) and if it is of the type
$
\epsilon^{b_{j}c_{k}}
$
we use the formulas (\ref{sigma8}) to obtain in (\ref{tt}) a contribution with a factor $\eta$ or
$
\epsilon^{\mu\nu\rho\sigma}.
$
So we still have to consider the case when we have in (\ref{sigma'}) only factors of the type
$
\epsilon^{c_{j}c_{k}}.
$
In this case we use the formula (\ref{sigma1}). The first term from (\ref{sigma1}) is giving a null contribution in (\ref{sigma'}) so by recursion we obtain a contribution of the type
\be
\sigma^{\alpha_{1}\beta_{1}}_{a_{1}b_{1}}\dots \sigma^{\alpha_{p}\beta_{p}}_{a_{p}b_{p}}~
\bar{\sigma}^{\rho_{1}\lambda_{1}}_{\bar{c}_{1}\bar{d}_{1}}\dots 
\bar{\sigma}^{\rho_{q}\lambda_{q}}_{\bar{c}_{q}\bar{d}_{q}}~
t^{\{a_{1},\dots,a_{p}\};\{b_{1},\dots,b_{p}\};
\{\bar{c}_{1},\dots,\bar{c}_{q}\};\{\bar{d}_{1},\dots,\bar{d}_{q}\}}
\label{sigma''}
\ee
where 
$
t^{\{a_{1},\dots,a_{p}\};\{b_{1},\dots,b_{p}\};
\{\bar{c}_{1},\dots,\bar{c}_{q}\};\{\bar{d}_{1},\dots,\bar{d}_{q}\}}
$
is completely symmetric in 
$
a_{1},\dots,a_{p},
$
etc. In this case we can use (\ref{sigma4}) - (\ref{sigma6}) to prove that the resulting contribution in (\ref{tt}) is traceless. 

(v) In the end we obtain in (\ref{tt}) a traceless part and terms with  at least one factor 
$\eta$ or
$
\epsilon^{\mu\nu\rho\sigma}.
$
If there are two factors off the type $\varepsilon$ we use the formula:
\be
\epsilon^{\mu\nu\rho\sigma}~\epsilon^{\alpha\beta\gamma\delta}
= {\cal A}_{\mu\nu\rho\sigma}~
(\eta^{\mu\alpha}~\eta^{\nu\beta}~\eta^{\rho\gamma}~\eta^{\sigma\delta})
\ee 
so we have in (\ref{tt}) a traceless part, terms with at least one factor 
$\eta$ and no
$
\varepsilon^{\mu\nu\rho\sigma}
$ 
factor and terms with one
$
\varepsilon^{\mu\nu\rho\sigma}
$ 
factor. The last contribution must be zero because of parity invariance.

(vi) Let us denote by
$
T^{n}_{k}
$
the space of tensors of rank $n$ of the type
$
\eta_{I_{1}}\dots\eta_{I_{k}}~t_{I_{0}}
$
with 
$
t_{I_{0}}
$
a traceless tensor. According to (v) we have the following decomposition for the space of parity invariant tensors of rank $n$:
\be
T^{n}_{+} = \sum_{k}~T^{n}_{k}.
\ee
We introduce on 
$
T^{n}_{+}
$
the sesquilinear non-degenerate form:
\be
<t,s> \equiv \eta^{\mu_{1}\nu_{1}} \dots \eta^{\mu_{n}\nu_{n}}~
t_{\mu_{1},\dots,\mu_{n}}~s_{\nu_{1},\dots,\nu_{n}}
\ee
and observe that for $k$ different from $l$ we have: 
\be
<T^{n}_{k},T^{n}_{l}> = 0.
\label{TT}
\ee
Indeed, we may take 
$
l > k.
$
But
$
t \in T^{n}_{k}
$
is of the form
$
t = \eta\dots\eta~t_{0}
$
with $k$ factors $\eta$ and
$
t_{0} \in T^{n-2k}_{0}.
$
We eliminate all $\eta$'s and we get
\be
<t,s> \sim <t_{0},s_{0}>
\ee
where 
$
s_{0} \in T^{n-2k}_{l-k}
$
so we have at least one factor $\eta$ in 
$
s_{0}.
$
By contraction with the traceless tensor
$
t_{0}
$
we get zero i.e. we have (\ref{TT}). 

Now we choose a basis
$
e^{(k)}_{\alpha}
$
in
$
T^{n}_{k}
$
and we remark that we must have
\be
det(<e^{(k)}_{\alpha},e^{(k)}_{\beta}>) \not= 0.
\label{det}
\ee
Indeed, if this would not be true that we would have a non-null
$
t \in T^{n}_{k}
$
such that 
\be
<t,e^{(k)}_{\alpha}> = 0, \quad \forall \alpha \qquad \Leftrightarrow\qquad
t \bot T^{n}_{k}.
\ee
If we use (\ref{TT}) we find out that 
$
t \in T^{n}_{+}
$
and because 
$<\cdot,\cdot>$
is non-degenerate 
it follows that 
$
t = 0.
$
The contraction proves (\ref{det}). 

We write any 
$
t \in T^{n}_{+}
$
in the form 
\be
t = \sum_{k,\alpha}~t^{(k)}_{\alpha}~e^{(k)}_{\alpha}
\ee
and we have from here
\be
<t,e^{(k)}_{\alpha}> = \sum_{\beta}~<e^{(k)}_{\alpha},e^{(k)}_{\beta}>~t^{(k)}_{\beta}.
\ee
If we take into account (\ref{det}) it means that we can express the tensors
$
t^{(k)}_{\beta}
$
as linear combinations of
$
<t,e^{(k)}_{\alpha}>.
$
But it is easy to see that these expressions are some traces of the tensor $t$. This proves the last assertion from the statement.
$\qed$

\end{document}